\documentclass[twocolumn,iop,numberedappendix]{emulateapj}

\usepackage{amsmath,amssymb}
\usepackage{mathrsfs}

\newcommand{\qdot}{\dot{q}}
\newcommand{\mdot}{\dot{M}}
\newcommand{\lnu}{L_{\nu}}
\newcommand{\lcore}{L_{\nu,\,{\rm core}}}
\newcommand{\msun}{M_\odot}
\newcommand{\beq}{\begin{equation}}
\newcommand{\eeq}{\end{equation}}
\newcommand{\lcrit}{\lcore^{\rm crit}}
\newcommand{\lacc}{\ensuremath{\lnu^{\rm acc}}}
\newcommand{\vff}{v_{\rm ff}}
\newcommand{\vesc}{v_{\rm esc}}
\newcommand{\rnu}{r_\nu}
\newcommand{\rs}{\ensuremath{r_{\rm S}}}
\newcommand{\rgain}{\ensuremath{r_{\rm gain}}}
\newcommand{\ye}{Y_{\rm e}}
\newcommand{\intd}{{\rm d}}
\newcommand{\invs}{{\rm s^{-1}}}
\newcommand{\erg}{{\rm ergs}}
\newcommand{\adv}{\ensuremath{_{\rm adv}}}
\newcommand{\heat}{\ensuremath{_{\rm heat}}}

\newcommand{\ct}{\ensuremath{c_T}}

\newcommand{\nue}{\ensuremath{\nu_{\rm e}}}
\newcommand{\nuebar}{{\ensuremath{\overline{\nu}_{\rm e}}}}
\newcommand{\elec}{\ensuremath{e^-}}
\newcommand{\posi}{\ensuremath{e^+}}
\newcommand{\ergsec}{\erg\ s$^{-1}$}

\shorttitle{Neutrino Mechanism of Core-Collapse Supernovae}
\shortauthors{Pejcha \& Thompson}

\begin{document}

\title{The Physics of the Neutrino Mechanism of Core-Collapse Supernovae}

\author{Ond\v{r}ej Pejcha\altaffilmark{1} and Todd A. Thompson\altaffilmark{1,2,3}}
\altaffiltext{1}{Department of Astronomy, The Ohio State University, 140 West 18th Avenue, Columbus, OH 43210, USA}
\altaffiltext{2}{Center for Cosmology and Astroparticle Physics, The Ohio State University, 191 West Woodruff Avenue, Columbus, OH 43210, USA}
\altaffiltext{3}{Alfred P. Sloan Fellow}
\email{pejcha@astronomy.ohio-state.edu} 

\begin{abstract}
Although it is known that the stalled accretion shock in models of core-collapse supernovae turns into explosion when the neutrino luminosity from the proto-neutron star (PNS) exceeds a critical value ($\lcrit$) (the ``neutrino mechanism''), the physics of $\lcrit$ has never been systematically explored. We solve the accretion problem between the proto-neutron star (PNS) surface and the accretion shock. We quantify the deep connection between the general problem
of accretion flows with bounding shocks and the neutrino mechanism. In particular, we show that there is a maximum, critical sound speed above which the shock jump conditions cannot be satisfied and steady-state accretion is impossible. This physics is general and does not depend on a specific heating mechanism.  For the simple model of pressure-less free-fall onto a shock bounding an isothermal accretion flow, we show that shock solutions are possible only for sound speed $c_T < \ct^{\rm crit}$ and that $c_T^2/v_{\rm esc}^2 = 3/16 = 0.1875$ at $\ct^{\rm crit}$. We generalize this result to the supernova problem, showing that the same physics determines $\lcrit$.  The critical condition for explosion can be written as $c_S^2/v^2_{\rm esc}\simeq0.19$, where
$c_S$ is the adiabatic sound speed.  This ``antesonic'' condition describes $\lcrit$ over a broad range of parameters, and other criteria proposed in the literature fail to capture this physics. We show that the accretion luminosity reduces $\lcrit$ non-trivially.  A larger PNS radius decreases $\lcrit$, implying that a stiff high-density equation of state may be preferred. Finally, using an analytic model, we provide evidence that the reduction of $\lcrit$ seen in recent multi-dimensional simulations results from reduced cooling efficiency, rather than an increased heating rate.
\end{abstract}

\section{Introduction}

When the core of a massive star collapses at the end of its life as a result of the Chandrasekhar instability, the central matter reaches nuclear densities, stiffens dramatically as a result of the hard core repulsion of the strong force, and drives a shockwave into the super-sonically collapsing outer mantle.  The outward progress of the shockwave is halted by a combination of energy losses from the dissociation of nuclei across the shock, neutrino emission as the shockwave moves from the optically-thick inner core to the optically-thin surrounding envelope, and the ram pressure of the overlying infalling outer iron core. A quasi-static accretion phase then ensues, which lingers for many dynamical times before eventual explosion. It consists of a hot, optically-thick, accreting proto-neutron star (PNS) with neutrinosphere radius $\sim\!30$ to $60$\,km, radiating its gravitational binding energy in neutrinos of all flavors, surrounded by a standoff accretion shock with radius of $\sim\! 200$\,km.

Despite decades of modelling effort, the mechanism responsible for reviving the shockwave to positive velocities has not been conclusively identified. The ``delayed neutrino mechanism'' relies on neutrino heating in the semi-transparent subsonic accretion flow to drive explosion. A fraction of the neutrinos diffusing out of the collapsed PNS deposit their energy in the accretion flow, and for a sufficiently large heating rate explosion results.  This mechanism was originally discussed by \citet{colgate66}, and then developed by \citet{bethewilson85}.  However, the most detailed one-dimensional models generically fail to explode, except for the lowest mass progenitors \citep{rampp00,bruenn01,liebendorfer01,mezzacappa01,thompson03,kitaura06,janka08}. These models clearly neglect multi-dimensional effects such as convection, both in the PNS interior \citep{keil96,mezzacappa98,dessart06}, and in the heating region between the neutrinosphere and the shock \citep{herant92,herant94,janka96,bhf95,fryer00,fryer02,fryer04,buras06a}, the standing accretion shock instability (SASI) and/or vortical-acoustic instability \citep{foglizzo02,blondin03,blondin06,ohnishi06,iwakami08,marek09,fernandez10}, and the potential for other multi-dimensional non-linear phenomena such as the ``acoustic'' mechanism of \citet{burrows06,burrows07b}, which was critically evaluated by \citet{weinberg08}. Indeed, many of these works have claimed that multi-dimensional effects are crucial for explosion because convection between the PNS and the shock allows the matter to stay in the region of net energy deposition longer, and an increase in the average shock radius as a result of the SASI and convection makes the heating region larger, while putting more matter higher in the gravitational potential well.

An approach complimentary to these multi-dimensional modelling efforts was followed by \citet{bg93} who calculated the steady-state structure of the region between the PNS and the accretion shock, during the quasi-time-steady epoch following collapse, bounce, and shock formation. For a fixed mass accretion rate, they increased the PNS core neutrino luminosity by hand until no steady-state solution was possible. The critical core neutrino luminosity ($\lcrit$) that bounds steady-state accretion solutions from above has been identified with the initiation of the supernova explosion. With this knowledge in hand, the task of getting a supernova simulation to explode via the neutrino mechanism can be understood as either an effort to increase the neutrino flux (or effective heating rate) up to the critical value, {\it or} to decrease $\lcrit$ so that it can be reached with the available luminosities.  Examples of the former include the work by \citet{wilson88}, which appealed to salt-finger (doubly-diffusive) convection within the PNS to increase the core luminosity at early times after collapse, or \citet{thompson05}, who found that viscous heating in models of rotating collapse could enable explosion.  Examples of the latter include \citet{yamasaki06}, who showed that convection in the heating region generically decreases $\lcrit$, or \citet{murphy08} and \citet{nordhaus10}, who found the same in 2D and 3D simulations \citep[see also][]{hanke11}.

However, it is to a great extent unknown why there is a critical neutrino luminosity at all, what determines its existence, and how it scales with the parameters of the problem. Indeed, the question of the physics of $\lcrit$ is particularly paradoxical because during the stalled accretion shock phase, the size of the PNS, its neutrino luminosity, and the mass accretion rate through the shock change significantly, yet the simulations show that the position of the shock moves only very slowly. This implies there are many possible near-equilibrium configurations that yield similar shock radii for very different global parameters.  For example, in low mass progenitors, the incoming mass accretion rate can decrease by a factor of 10 in the first second after bounce, but the neutrino luminosities vary by less than a factor of 2, and yet the shock radius is essentially constant.  Given this early-time behavior, if the neutrino luminosity were to increase very slowly at a later time in the evolution, why would one expect a catastrophic change from steady accretion to dynamical outward expansion?  One might instead expect that an increase in the net energy deposition would simply result in the shock just getting pushed to some different equilibrium position with larger shock radius. Contrary to this expectation, the existence of $\lcrit$ implies instead that if the energy deposition below the shock reaches a critical value, the shock can no longer exist together with steady-state accretion, the shockwave expands dynamically, and the star explodes as a supernova \citep[e.g.,][]{bg93,herant94,bhf95,janka96}.

Within the framework of time-steady spherically symmetric models of supernovae, this paper provides an answer to two questions connected to these problems, and a possible explanation for a third. First, what is the physics of $\lcrit$? In Section~\ref{sec:isot_acc}, we investigate isothermal and polytropic accretion flows with a bounding shockwave and we show that even in such a simplistic setting there exists a critical value of the controlling parameter --- the isothermal sound speed --- that separates steady-state accretion solutions from wind solutions, which are identified with explosions \citep{burrows87,bg93,bhf95,yamasaki05}. This critical value is determined by conservation of momentum and energy across the standoff accretion shock as was found by \citet{yamasaki05,yamasaki06}; above the critical value it is not possible to satisfy these conditions with a steady-state accretion solution, and instead the flow must rearrange itself into an outgoing wind. In Section~\ref{sec:accretion_shock}, we calculate the steady-state structure of the accretion flow between the neutrinosphere and the shock in the supernova problem. We explicitly show that the critical neutrino luminosity discovered by \citet{bg93} is equivalent to the critical condition from the isothermal model. {\it That is, the neutrino mechanism of supernovae as formulated by \citet{bg93} is identical to the statement that the shock jump conditions cannot be satisfied together with the Euler equations for steady-state accretion when the sound speed of the flow exceeds the critical value.}  Furthermore, we discuss the effects of radiation transport and the accretion luminosity on the absolute value of $\lcrit$, and we show how the latter depends on the mass and radius of the PNS and the mass accretion rate.

The second question this paper answers is ``What criterion does $\lcrit$ correspond to in terms of variables within the accretion flow itself?'' For example, it has been claimed that when the advection time in the heating region becomes longer than the heating timescale, explosion must result \citep{janka01,thompson_murray01,thompson05,buras06b,schecketal08,murphy08}. Is this heuristic criterion identical to the critical neutrino luminosity?  The answer is no. In fact, we show in Section~\ref{sec:conditions} that $\lcrit$ corresponds to a virtually constant ratio of the sound speed to the escape velocity in the accretion flow --- the ``antesonic\footnote{\label{foot:ante}We call this condition ``antesonic'' because the required sound speed is significantly below the local escape velocity, and the critical radius at which this point occurs is smaller than the radius of sonic point in Bondi accretion. This condition is reached in supernovae at the time of explosion, before the flow arranges itself into a super-sonic neutrino-driven wind (see Section~\ref{sec:isot_acc} for details).} condition:'' $c_S^2\simeq 0.19 v_{\rm esc}^2$, where $v_{\rm esc}$ is the escape speed --- and that this condition is a consequence of the physics of the critical luminosity itself.

Third, an intriguing observation was made by \citet{ohnishi06}, \citet{murphy08} and \citet{nordhaus10} who found that the mere fact of increasing the dimension of the simulation lowers $\lcrit$. While going from 1D to 2D allows for the effects of convection and the SASI, it is not clear what is gained by going from 2D to 3D, or why $\lcrit$ should decrease monotonically with dimension. In Section~\ref{sec:toy_model}, we provide evidence that the reduction of $\lcrit$ observed in 2D and 3D simulations occurs because the flow can organize itself to cool less efficiently and yet still maintain hydrostatic equilibrium, giving an overall lower cooling efficiency, lower critical luminosities, larger entropy, and larger shock radii. Thus, the reduction of $\lcrit$ in 2D and 3D may be due to less efficient cooling, rather than more efficient heating.

\section{Isothermal accretion bounded by a shock}
\label{sec:isot_acc}

Here we review the basic physics of spherically-symmetric isothermal accretion with a shock. We find that this idealized model problem is a key to understanding the mechanism of supernovae. We also discuss here the analogous problem of polytropic accretion flows with a constant adiabatic index $\Gamma$ before solving the more complete supernova problem in Section 3.

\subsection{Topology of solutions}
\label{sec:isot_topology}

\begin{figure*}
\center{\includegraphics[width=0.8\textwidth]{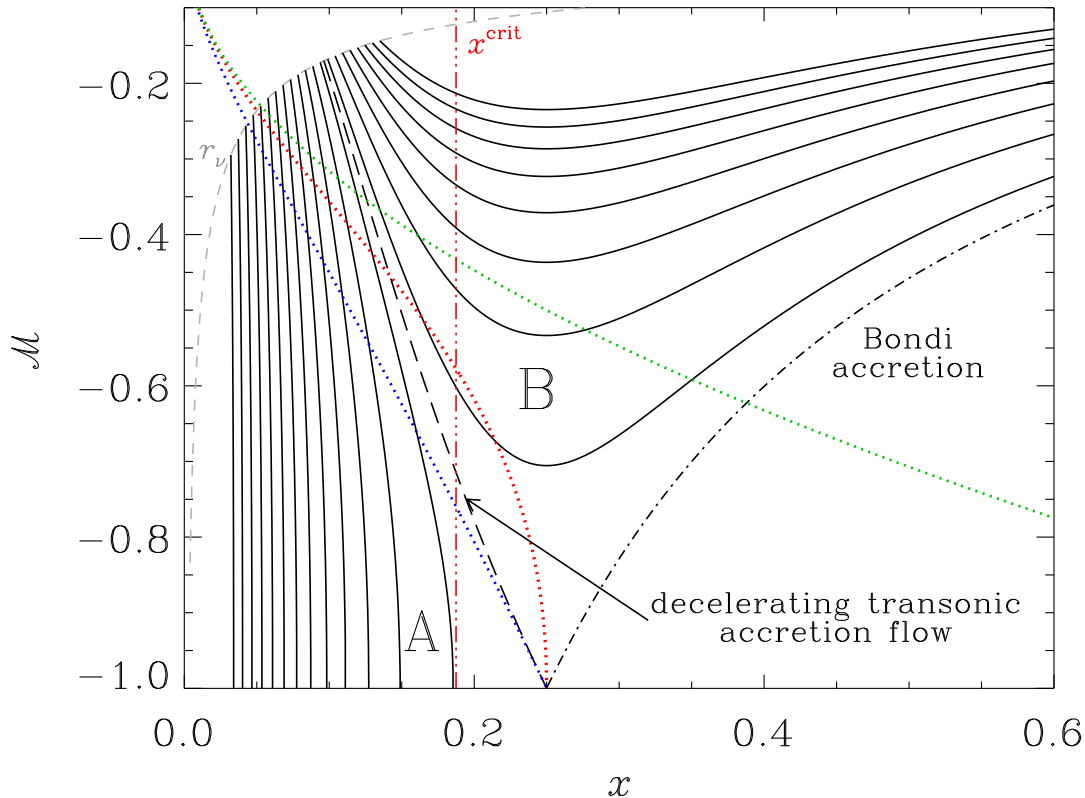}}
\caption{Isothermal accretion plotted in the space of Mach number $\mathscr{M}$ and rescaled radial coordinated $x=r\ct^2/(2GM)$. Solid black lines show solutions to eq.~(\ref{eq:isot_acc}) with $\mdot = -1\,\msun$ s$^{-1}$ and $M=1.4\,\msun$ starting from  $\rnu = 30$\,km (grey dashed line), and with fixed velocity $v_\nu = \mdot/(4\pi \rnu^2 \rho_\nu)$, where $\rho_\nu = 3\times 10^{10}$\,g cm$^3$. The value of $\ct^2$ increases from $4\times 10^{18}$ (black solid line starting at lowest $x$) to $1.68\times 10^{19}$\,cm$^2$ s$^{-2}$ (highest line) in the steps of $6.4 \times 10^{17}$\,cm$^2$ s$^{-2}$. The dashed line is the decelerating transonic solution going through the sonic point $\mathscr{M}_{\rm sonic} = -1$ at $x_{\rm sonic}=0.25$. The dash-dot line is Bondi accretion flow. Dotted lines are velocities just downstream of a shock positioned at any $x$, assuming that the upstream flow is either Bondi accretion flow (blue bottom line), in free fall (green top line), or in pressure-less free fall (red middle line). A viable accretion solution with a shock starts at $\rnu$ and follows any of the black lines until it crosses a dotted line, where it jumps to the assumed upstream velocity profile. Above a certain $\ct^{\rm crit}$ there is no accreting solution with a steady-state shock. The critical value $x^{\rm crit}$, where this happens, is shown with a vertical red dash-dot-dot line for the pressure-less free fall upstream of the shock. }
\label{fig:isot_acc}
\end{figure*}

The velocity structure of spherical steady-state isothermal flows is described by the equation
\beq
\left(\mathscr{M} - \frac{1}{\mathscr{M}}\right)\frac{\intd \mathscr{M}}{\intd x} = \frac{2}{x} - \frac{1}{2x^2},
\label{eq:isot_acc}
\eeq
where $\mathscr{M} = v/\ct$ is the Mach number, $v$ is the fluid velocity, $\ct$ is the isothermal sound speed, $x = r \ct^2/(2GM)$ is the rescaled radial coordinate $r$, and $M$ is the mass of the central object. It is possible for a standing shock wave in the flow to exist at a point that satisfies the two Rankine-Hugoniot shock jump conditions
\begin{eqnarray}
\rho^- \mathscr{M}^-  &=& \rho^+ \mathscr{M}^+,\label{eq:isot_masscons}\\
\rho^- (\mathscr{M}^-)^2 + \rho^-  &=& \rho^+ (\mathscr{M}^+)^2 + \rho^+  ,\label{eq:isot_momcons}
\end{eqnarray}
which express conservation of mass and momentum, respectively. Here, $\rho$ is the mass density and the $+$ and $-$ superscripts correspond to the quantities evaluated just upstream and downstream of the shock, respectively. The physically relevant solution to equations~(\ref{eq:isot_masscons}--\ref{eq:isot_momcons}) is $\mathscr{M}^+\mathscr{M}^- = 1$.

In Figure~\ref{fig:isot_acc}, we show with black solid lines Mach-number profiles of flows with a single constant mass accretion rate $\mdot=-1\,\msun\ {\rm s}^{-1}$ obtained by solving equation~(\ref{eq:isot_acc}). The integration starts at a fixed radius $\rnu$ corresponding to the surface of the star, and with a fixed velocity $v_\nu = \mdot/(4\pi \rnu^2 \rho_\nu)$, where $\rho_\nu$ is the prescribed mass density at $\rnu$. Different lines correspond to different values of $\ct^2$. The dashed line shows the decelerating transonic solution, which goes through the sonic point\footnote{In this paper we assign the name ``sonic point'' to the position of the flow where the numerator and denominator of the fluid momentum equation simultaneously vanish. Although this point is generally called the ``critical point'', we avoid this name to prevent confusion with the critical neutrino luminosity discussed later in the paper.} located at $x_{\rm sonic} = 0.25$ and $\mathscr{M}_{\rm sonic} = -1$. It has high velocity at large radii and $\mathscr{M} \rightarrow 0$ as $x\rightarrow 0$, unlike Bondi accretion flow, which starts with zero velocity at infinity, goes through the sonic point and has $\mathscr{M} \rightarrow -\infty$ as $x\rightarrow 0$ (shown with dash-dotted line). The decelerating transonic solution separates accretion ``breezes'', which are always subsonic ($-1 < \mathscr{M} \leq 0$) and denoted as B, from solutions going through $\mathscr{M} = -1$ and denoted as A. Although the A-type solutions lying below the decelerating transonic solution are considered unphysical in standard Bondi accretion theory, the parts of these solutions that span from $\rnu$ to the shock are viable in this setting \citep{mccrea56}.

The dotted lines in Figure~\ref{fig:isot_acc} show the velocities just downstream of the shock, $\mathscr{M}^-$, corresponding to three different assumed profiles of the upstream velocities $\mathscr{M}^+$: (1) Bondi flow (blue), (2) free fall with $\mathscr{M}^+ = -x^{-1/2}$ (green), and (3) pressure-less free fall (red), that is a good approximation in more realistic supernova calculations \citep{colgate66} and that requires modification of equation~(\ref{eq:isot_momcons}) by assuming that $\mathscr{M}^+ = -x^{-1/2} \ll -1$, and
\beq
\rho^- (\mathscr{M}^-)^2 + \rho^- = \rho^+x^{-1}.\tag{\ref{eq:isot_momcons}a}
\label{eq:isot_momconsa}
\eeq
The physically relevant solution to equations~(\ref{eq:isot_masscons}) and (\ref{eq:isot_momconsa}) is $\mathscr{M}^- = (\sqrt{x^{-1}-4}-x^{-1/2})/2$. For the sake of clarity, we do not show the upstream profiles in Figure~\ref{fig:isot_acc} except for Bondi flow (dash-dotted line).

The velocity profile of the shocked accretion flow for a given $\ct$ is constructed by following one of the black lines from $\rnu$ until it crosses a dotted line that corresponds to the appropriate shock jump conditions. At this intersection, the solution jumps to whatever velocity profile $\mathscr{M}^+$ was assumed to be present upstream. 

The essential point of Figure~\ref{fig:isot_acc} is that for a fixed $\mdot$ the presence of a shock is not guaranteed for all values of $\ct$. For example, consider the blue dotted line, which corresponds to Bondi accretion flow upstream of the shock. For the smallest values of $\ct$, the required shock radius --- the intersection of the black solid line with blue dotted line --- would be below the radius of the star, an unphysical situation. As we step to higher $\ct$, the shock appears at $\rnu$ and moves outward. At a critical value $\ct^{\rm crit}$ corresponding to the decelerating transonic accretion flow (dashed line), the shock can only coincide with the sonic point at $x^{\rm crit} = x_{\rm sonic} = 0.25$. Here there is no jump in the velocity, and the shock degenerates into a jump in the derivative of the velocity \citep{velli94,delzanna98}. A shock is not possible for $\ct > \ct^{\rm crit}$, because the blue dotted line lies below the decelerating transonic solution (dashed line) in the range of radii of interest and hence B-type solutions are not viable solutions inside the shock -- that is, solutions of type B do not intersect the blue dotted line.

The situation is somewhat different when the flow upstream of the shock is in free fall, as shown with the red and green dotted lines in Figure~\ref{fig:isot_acc}. In these cases, the shock jump conditions allow part of the B-type solutions as viable solutions downstream of the shock. Furthermore, for values of $\ct$ higher than that corresponding to the decelerating transonic solution (dashed line) two shock radii are possible for a single value of $\ct$, because the curvature of B-type solutions and shock jump conditions yields two intersections. When stepping to higher values of $\ct$, the two shock radii come closer together and finally coincide at a critical value $\ct^{\rm crit}$. Contrary to the case of Bondi flow above the shock, there is no relation between the critical sound speed $\ct^{\rm crit}$ and the sonic point. The shock radius at $\ct^{\rm crit}$ also does not coincide with the velocity minimum of any of the B-type solutions. Above the critical sound speed $\ct^{\rm crit}$, a shock in the flow is not possible, qualitatively similar to the case of Bondi flow (blue dotted line) described above.

Finally, we briefly discuss the ``upper'' solutions that appear when $\ct$ is higher than the value corresponding to the decelerating transonic solution (dashed line in Fig.~\ref{fig:isot_acc}), but still $\ct < \ct^{\rm crit}$. We see that the maximum shock radius for these solutions is identical to the sonic point radius. The shock radius decreases with increasing $\ct$ until it merges with the normal solution branch at $\ct^{\rm crit}$. These upper solutions were found by \citet{yamasaki05} also in the full problem, but these authors show that the upper solutions are unstable to perturbations and thus most likely do not occur in a real physical system. Because of this, we do not discuss the upper solutions further except in Sections~\ref{sec:prop_full} and \ref{sec:isot_corres}.

\subsection{The critical sound speed}

\begin{figure*}
\plotone{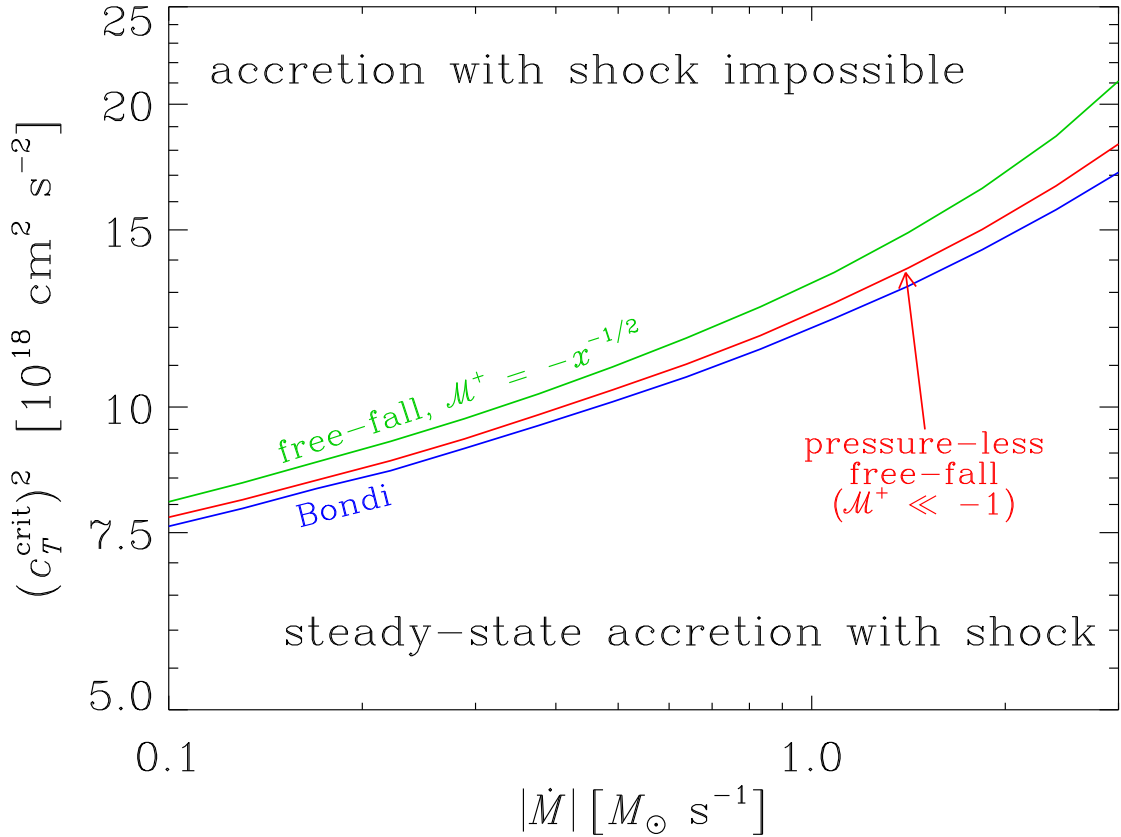}
\caption{The maximum isothermal sound speed $\ct^{\rm crit}$ that allows for a shock in the flow at a given $\mdot$. The PNS parameters --- $M$, $\rnu$, $\rho_\nu$ --- and color coding --- green, red, blue --- are the same as in Figure~\ref{fig:isot_acc}.}
\label{fig:isot_crit}
\end{figure*}

In Figure~\ref{fig:isot_crit} we show the critical values $\ct^{\rm crit}$ as a function of mass accretion rate $\mdot$ for the three sets of shock jump conditions shown in Figure~\ref{fig:isot_acc}. The three curves are similar over a significant range of $\mdot$ despite the relatively stark differences in the shock jump conditions. Specifically, the top and the middle curves closely resemble the bottom blue line, which is, in fact, the relation between the sound speed and the mass-accretion rate for isothermal transonic accretion \citep[p.~68]{lamcas99}. The curvature of this relation in the log-log plane of Figure~\ref{fig:isot_crit} arises from exponential near-hydrostatic density profile of the atmosphere, which is a good approximation to the decelerating transonic accretion flow below the sonic point. 

From Figure~\ref{fig:isot_acc} it is obvious that for pressure-less free fall upstream of the shock the solution described by $\ct^{\rm crit}$ does not correspond to any special radial point in the flow. However, we also see that the shock position $x^{\rm crit}$ at $\ct^{\rm crit}$ is constant when expressed in the dimensionless coordinates since this is the point where the dotted line just meets the flow profile. Changes in $\mdot$, $\rho_\nu$ or $\rnu$ only move the starting inner point of the flow, but one can always find a value of $\ct = \ct^{\rm crit}(\mdot, \rho_\nu, \rnu)$ to get the critical solution at $x^{\rm crit}$. By plugging the solution of $\mathscr{M}^-$ from the shock jump condition in equation~(\ref{eq:isot_momconsa}) into equation~(\ref{eq:isot_acc})\footnote{We note that this procedure is valid only for determining the position of the critical point in isothermal accretion. As seen from Figure~\ref{fig:isot_acc}, the critical point occurs when the line of all possible shock positions (solution to eqs.~[\ref{eq:isot_masscons}] and [\ref{eq:isot_momconsa}], dotted red line) just touches the flow profiles (solutions to eq.~[\ref{eq:isot_acc}], solid black lines). Thus, at this point only, the tangents and values are the same, which we then use to calculate the position of the critical point, equation~(\ref{eq:isot_cond}).}, we obtain exactly $x^{\rm crit} = 3/16$ implying that the critical condition for existence of steady-state accretion shock in isothermal accretion is
\beq
\frac{(\ct^{\rm crit})^2}{v_{\rm esc}^2} = \frac{3}{16} = 0.1875,
\label{eq:isot_cond}
\eeq
where $v_{\rm esc} = \sqrt{2GM/r}$ is the escape velocity. This result is valid when the flow upstream of the shock is in free fall and has negligible thermal pressure. We note that equation~(\ref{eq:isot_cond}) is valid only at $\ct^{\rm crit}$. There are indeed shock solutions with $x_{\rm shock} > x^{\rm crit}$ that correspond to the upper solution branch as discussed in Sections~\ref{sec:isot_topology} and \ref{sec:prop_full} and by \citet{yamasaki05}, but these solutions still have $\ct < \ct^{\rm crit}$. The actual value of $\ct^{\rm crit}$ for given $\mdot$, $M$, $\rnu$, and $\rho_\nu$ is calculated numerically as is demonstrated in Figure~\ref{fig:isot_crit}.

As the ratio in equation~(\ref{eq:isot_cond}) is lower than for the sonic point, we call this condition ``antesonic'' (see footnote~\ref{foot:ante}). For the case of free fall with non-negligible pressure shown with green line in Figure~\ref{fig:isot_acc}, the numerical factor is $(5-\sqrt{21})/2 \simeq 0.2087$.
Furthermore, we emphasize that equation~(\ref{eq:isot_cond}) does not imply any connection to the Parker point\footnote{The Parker point occurs when the right-hand side of equation~(\ref{eq:isot_acc}) vanishes, and in the case of isothermal flow the Parker point coincides with the sonic point \citep[p.~63]{lamcas99}.} or sonic point. Thus, one cannot assume that the existence of a shock is equivalent to the existence of a sonic point in the context of supernova stalled accretion shocks \citep[as in][]{shigeyama95}. However, if the flow upstream of the shock is a Bondi accretion flow with the same sound speed, then the critical condition is exactly the same as the expression for the  position of the sonic point: $(\ct^{\rm crit}/v_{\rm esc})^2 = 0.25$.

Our calculation of the critical condition in equation~(\ref{eq:isot_cond}) also explains why the critical curves for different shock jump conditions in Figure~\ref{fig:isot_crit} are so similar: their values can be estimated in the same way as for the position of the sonic point \citep[p.~68]{lamcas99}, except that for the shock jump conditions relevant in the supernova context the value of $x^{\rm crit}$ is lower than $0.25$ (eq.~[\ref{eq:isot_cond}]).

\begin{figure}
\plotone{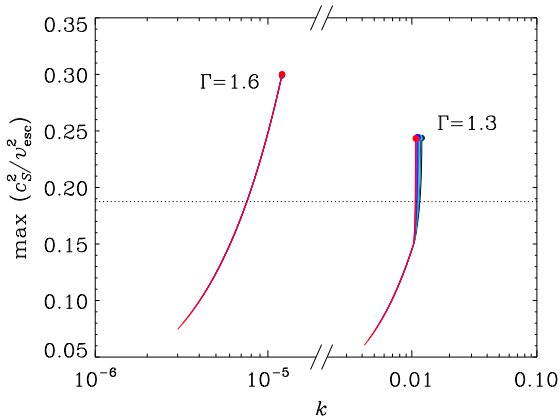}
\caption{Antesonic factor for polytropic equation of state. For each $\Gamma$, we plot $\max\,(c_S^2/v_{\rm esc}^2)$ as a function of $k$ for $\mdot$ ranging from $-0.01\,\msun$\ s$^{-1}$ (black lines) to $-1.0\,\msun$\ s$^{-1}$ (red lines). For fixed $\Gamma$, the lines overlap. The antesonic factors at each critical $k$ are marked with filled circles with color corresponding to $\mdot$. The antesonic factor at critical $k$ is linearly proportional to $\Gamma$. The horizontal dotted line marks the value for the isothermal case, $3/16=0.1875$.}
\label{fig:polyante}
\end{figure}

For the sake of completeness, we repeated the isothermal calculation with pressure-less free fall above the shock for a polytropic equation of state $P=k\rho^\Gamma$, where $k$ is the normalization factor and $\Gamma$ is the adiabatic index, which we vary from $1$ to $1.7$. We do not assume energy conservation through the shock so that $k$ is left as a free parameter, which we can vary to obtain its critical values. The results are qualitatively similar to those for isothermal accretion.  For fixed $\Gamma$, there exists a critical value of the constant $k$ above which there are no steady-state solutions with a bounding shock. As in the isothermal case, this critical value of $k$ corresponds to a critical condition that can be written as $\max\,(c_S^2/v_{\rm esc}^2) \approx 0.19\Gamma$ (compare with eq.~[4]), where $c_S$ is the adiabatic sound speed and the ``antesonic factor'' is linearly proportional to the adiabatic index of the flow. We illustrate the dependence of $\max\,(c_S^2/v_{\rm esc}^2)$ on $\Gamma$ in Figure~\ref{fig:polyante}. where we plot the antesonic ratio as a function of $k$ for a range of $\mdot$ and two values of $\Gamma$. We note that although we are able to write down a single local criterion for the antesonic condition in the adiabatic case, at the radial location where $\max\,(c_S^2/v_{\rm esc}^2)$ occurs, this condition should be interpreted as a global condition on the flow, and not a local one.  As in the isothermal case, the physics of the critical condition is determined by the ability of the flow to satisfy the jump conditions while simultaneously satisfying the Euler equations for steady-state accretion.

The physics of collapse, core bounce and shock stall dictate that there is a phase of steady-state accretion flow with a standing shock with time-changing $\mdot$ and $\ct$. If $\ct$ slowly increases at fixed $\mdot$, the shock radius advances until the critical value $\ct^{\rm crit}$ is reached. Physically, $\ct^{\rm crit}$ exists because as $\ct$ is increased the shock jump conditions simply cannot be satisfied given the properties of the flow upstream and the necessity of maintaining conservation of mass and momentum. What happens at $\ct^{\rm crit}$ depends on the properties of the flow upstream of the shock. In the case of Bondi flow, the profile can continuously transition to a shock-less B-type solution (accretion breeze) as shown by \citet{korevaar89}, \citet{velli94,velli01} and \citet{delzanna98}, because there is no jump in velocity at $\ct^{\rm crit}$. However, in the case relevant for supernovae, where the flow above the shock is in pressure-less free fall, the shock does not degenerate at $\ct^{\rm crit}$ and a catastrophic dynamical transition must occur, most likely to a supersonic wind, which can be identified with the supernova explosion \citep{burrows87,bg93,bhf95}. A similar conclusion was reached by \citet{yamasaki05}. It is likely that time-dependent hydrodynamical instabilities will modify the exact realization of this scenario. Nevertheless, our results from the isothermal accretion model are an important step towards understanding the mechanism of supernova explosions, and specifically the neutrino mechanism, as we detail below. Our isothermal model also provides a robust explosion criterion of broad applicability (the ``antesonic'' condition, Section~\ref{sec:coronal}).

\section{Steady-state accretion shock in core-collapse supernovae}
\label{sec:accretion_shock}

While isothermal flow calculations provide insight into the physics of steady-state accretion with shocks, we perform more involved calculations to assess the effects of individual parameters of the problem for supernova explosions. In this Section we first describe our equations and boundary conditions and then we proceed by discussing how the solution changes when approaching the critical neutrino luminosity. We then prove the correspondence between the critical neutrino luminosity and the critical sound speed from Section~\ref{sec:isot_acc}, and discuss properties of the critical solutions.

\subsection{Numerical setup}
\label{sec:num_setup}

We solve the time-independent Euler equations
\begin{eqnarray}
\frac{1}{\rho} \frac{\intd \rho}{\intd r} +\frac{1}{v} \frac{\intd v}{\intd r} + \frac{2}{r} &=& 0, \label{eq:mass_cons}\\
v\frac{\intd v}{\intd r} + \frac{1}{\rho}\frac{\intd P}{\intd r} &=& -\frac{GM}{r^2}, \label{eq:moment_equ}\\
\frac{\intd\varepsilon}{dr} - \frac{P}{\rho^2}\frac{\intd\rho}{\intd r} &=& \frac{\qdot}{v},\label{eq:energy_tot}
\end{eqnarray}
where $P$ is the gas pressure, $M$ is the mass within the radius of the neutrinosphere $\rnu$, $\varepsilon$ is the internal specific energy of the gas, and $\qdot = \mathcal{H} - \mathcal{C}$ is the net heating rate, a difference of heating $\mathcal{H}$ and cooling $\mathcal{C}$. These hydrodynamic equations couple through the equation of state (EOS), $P(\rho, T, \ye)$ and $\varepsilon(\rho, T, \ye)$, where $T$ is the gas temperature, to the equation for electron fraction $\ye$ 
\beq
v\frac{\intd\ye}{\intd r} = l_{\nue n} + l_{\posi n} - (l_{\nue n} + l_{\posi n} + l_{\nuebar p} + l_{\elec p})\ye \label{eq:el_frac},
\eeq
where $l_{\nue n}$, $l_{\posi n}$, $l_{\nuebar p}$, and $l_{\elec p}$ are reaction rates involving neutrons and protons. In Appendix~\ref{app:neutrino} we provide equations~(\ref{eq:mass_cons}--\ref{eq:energy_tot}) in the form of separated radial derivatives, which is useful for describing the connection between the critical luminosity and $\ct^{\rm crit}$ of Section~\ref{sec:isot_acc}. We also employ, for the first time in the context of calculations of the critical luminosity, gray neutrino radiation transport using the expression
\beq
\frac{\intd\lnu}{\intd r} = -4\pi r^2 \rho \qdot\label{eq:dldr},
\eeq
where $\lnu$ is the combined luminosity of electron neutrinos and antineutrinos, which are assumed to be equal everywhere. We do not consider neutrinos of other flavors. We assume that neutrinos have root-mean-square energy of $\epsilon_{\nu_e} = 13$\,MeV and antineutrinos $\epsilon_{\bar{\nu}_e}= 15.5$\,MeV \citep{thompson03}. Even this very simple approximation to radiation transport enables us to assess the relative importance of accretion and core neutrino luminosity for the supernova explosion. Our choice of equation of state, heating and cooling function, and reaction rates is a simplified version of \citet{schecketal06} and \citet{qian96}. More details are given in Appendix~\ref{app:neutrino}. 

Five equations~(\ref{eq:mass_cons}--\ref{eq:dldr}) are solved on an interval of radii between $\rnu$ and $\rs$, where $\rnu$ is the radius of the neutrinosphere and is fixed, and $\rs$ is the radius of the shock, which is left to vary. The matter inside of $\rnu$ is the PNS core and is characterized only by its mass $M$ and luminosity $\lcore$ and energy of neutrinos and antineutrinos that it isotropically emits. 

We thus need six boundary conditions to uniquely determine five functions: $\rho$, $v$, $T$, $\ye$ and $\lnu$, and shock radius $\rs$. We first demand that the flow has a fixed mass accretion rate $\mdot = 4\pi r^2 \rho v$ by applying this constraint at the inner boundary. We assume that the neutrino luminosity at the inner boundary is equal to the luminosity of the PNS core, $\lnu(\rnu) = \lcore$. At the outer boundary we apply the shock jump conditions\footnote{We do not include nuclear binding energy in our calculations. However, the pertinent effects are qualitatively assessed in Appendix~\ref{app:neutrino} and in \citet{yamasaki06}.}
\begin{eqnarray}
\rho v^2 + P &=& \rho^+ \vff^2,\label{eq:shock_jump_mom}\\
\frac{1}{2}v^2 + \varepsilon + \frac{P}{\rho} &=& \frac{1}{2} \vff^2,\label{eq:shock_jump_ene}
\end{eqnarray}
where $\vff = \sqrt{\Upsilon} v_{\rm esc}$ is the free fall velocity, and in accordance with analytically estimated accretion data of \citet{woosley02} supernova progenitor models available on-line\footnote{\url{http://www.stellarevolution.org/data.shtml}} we choose $\Upsilon = 0.25$. The quantity $\rho^+$ is the density just upstream of the shock and can be calculated from conservation of mass. We also assume that the matter entering the shock is composed of iron and hence $\ye = 26/56$ at the outer boundary. The last boundary condition comes from requirement that $\rnu$ is the neutrinosphere for electron neutrinos, which gives a constraint on the optical depth
\beq
\tau_\nu = \int_{\rnu}^{\rs} \kappa_{\nu_e} \rho\,\,\intd r = \frac{2}{3},
\label{eq:taudef}
\eeq
where $\kappa_{\nu_e}$ is the opacity to electron neutrinos. This condition is implemented by adding an extra variable corresponding to the optical depth and requiring that it is zero at the inner boundary and $2/3$ at the outer boundary.

The equations and boundary conditions above are solved using a relaxation algorithm \citep[p.~753]{nr} with logarithmic spacing of grid points \citep[see][and references therein]{thompson01}. The number of grid points was usually $\gtrsim 1000$. We refer to the problem outlined by equations~(\ref{eq:mass_cons}--\ref{eq:taudef}) and the physics described in Appendix~\ref{app:neutrino} as the ``fiducial calculation''. As this problem is complex, we occasionally resort to various simplifications to make some of our points more clear. We describe what exactly we modified with respect to the fiducial calculation at a various places in the text.

\subsection{Properties of the fiducial calculation}
\label{sec:prop_full}

\begin{figure*}
\plotone{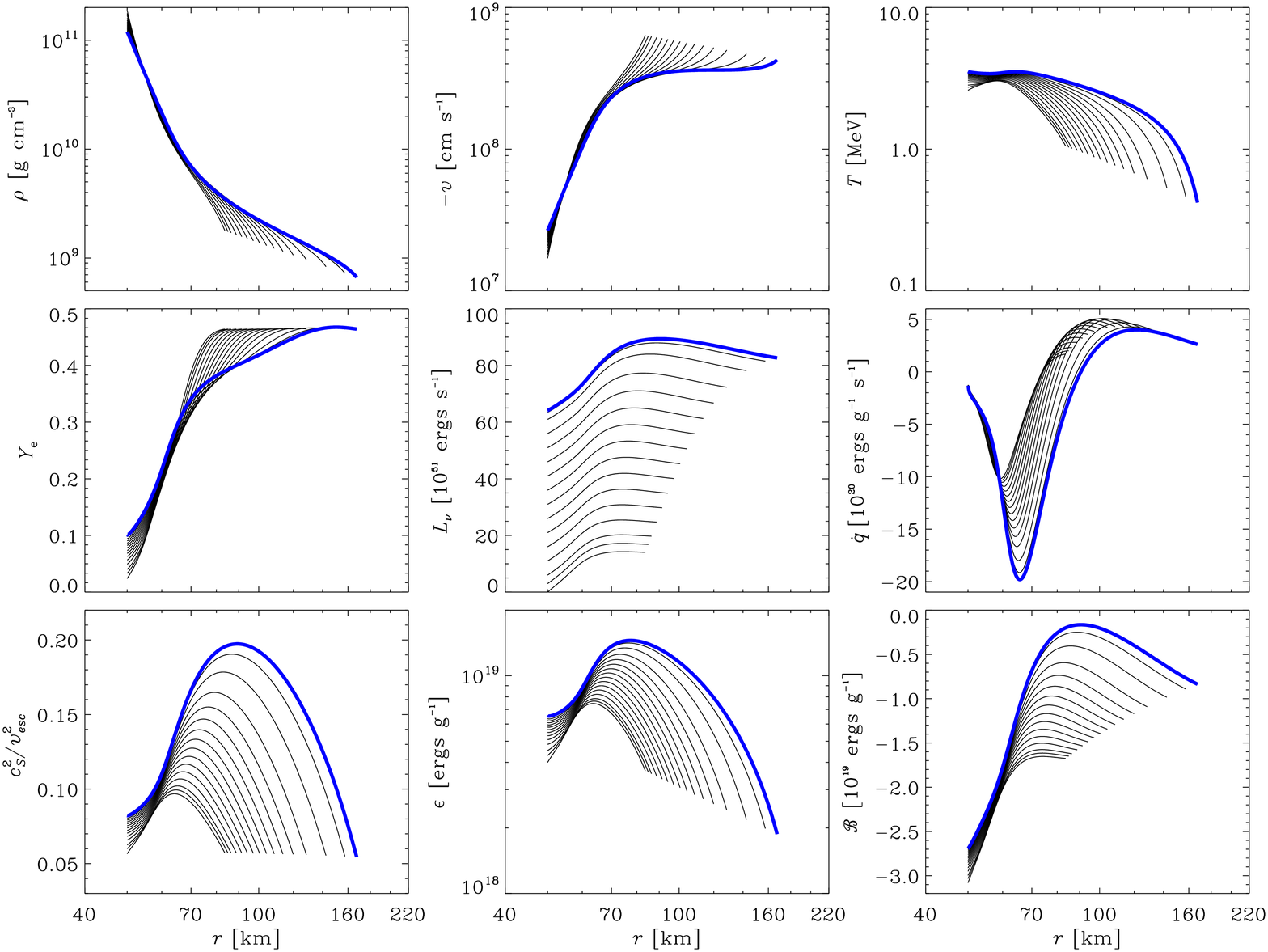}
\caption{Variables of interest as the core neutrino luminosity $\lcore$ approaches the critical value $\lcrit$ for $\mdot = -0.5\,\msun$\,s$^{-1}$, $M=1.4\,\msun$ and $\rnu = 50$\,km. Profiles of density $\rho$, velocity $v$, temperature $T$, electron fraction $\ye$, neutrino luminosity $\lnu$, net heating $\qdot$, adiabatic sound speed $c_S$, internal specific energy $\varepsilon$, and total specific energy $\mathscr{B}$ for $0 \leq \lcore \leq  \lcrit = 6.41\times 10^{52}$\,\ergsec are shown. The profiles at $\lcrit$ are shown with blue thick lines.}
\label{fig:profiles}
\end{figure*}

We include neutrino radiation transport and hence our solutions are parameterized by the core neutrino luminosity $\lcore$ and yield neutrino luminosity as a function of radius $\lnu(r)$. We confirm the existence of an upper limit to the core neutrino luminosity $\lcrit$ at fixed $\mdot$ \citep{bg93,yamasaki05,yamasaki06}. We determine $\lcrit$ by starting the relaxation code at a low value of $\lcore$ with an initial guess of power-law profiles of variables, and then increase $\lcore$ in small steps while using the converged solution from the previous step as an initial guess for the next step. We use bisection to trace $\lcrit$ to a specified relative precision. 

Figure~\ref{fig:profiles} shows profiles of thermodynamic quantities for $\lcore$ stepping from $0$ to $\lcrit$. Similar plots except for $\lnu(r)$ were published previously by \citet{yamasaki05,yamasaki06}, but we reproduce them here for the sake of completeness. Our results fall within the paradigm of  quasi-steady-state supernova structure \citep[e.g.][]{bruenn85,mayle88,janka01}. In particular, we find a  steep decrease in the density outside $\rnu$ that flattens when the dominant pressure support changes from an ideal gas of nucleons to a relativistic gas of electrons and positrons. The temperature profile is flat at small $r$, but then drops as the relativistic particles become more dominant. Still, the temperature changes only from approximately $3$\,MeV to $0.5$\,MeV, much less than the corresponding change by several orders of magnitude in density. This implies that our isothermal model is a legitimate rough approximation of the problem. We discuss in more detail the profiles of the adiabatic sound speed $c_S$, total specific energy in the form of Bernoulli integral\footnote{The importance of individual terms contributing to $\mathscr{B}$ is assessed in Appendix~\ref{app:toy_model}.}
\beq
\mathscr{B} = \frac{v^2}{2}+\varepsilon + \frac{P}{\rho} -\frac{GM}{r},
\label{eq:bernoul}
\eeq
and neutrino luminosity $\lnu$ that are to our knowledge not previously discussed in the context of the calculation of $\lcrit$. 

The total specific energy $\mathscr{B}$ is an increasing function of $\lcore$ in the whole region of interest. The maximum of $\mathscr{B}$ occurs at $\rgain$, which is defined as $\qdot(\rgain)=0$. Both $\rgain$ and $\mathscr{B}(\rgain)$ increase with increasing $\lcore$. We note that even at $\lcrit$ the energy is significantly negative in the entire region of interest. It is certainly not the case that positive $\mathscr{B}$ corresponds to explosion \citep[$\lcore=\lcrit$;][]{bhf95}.

The adiabatic sound speed $c_S$ was calculated using equation~(\ref{eq:app_cs}) and is plotted in the bottom left of Figure~\ref{fig:profiles} as a ratio to the local escape velocity $v_{\rm esc}(r)$. Similar to $\mathscr{B}$, we see that an increase of $\lcore$ results in a net increase of $c_S$ between $\rnu$ and $\rs$. The maximum of $c_S$ occurs at gradually larger radii for increasing $\lcore$, similar to the gain radius $\rgain$. However, the maximum of $c_S^2/v_{\rm esc}^2$ does not exactly coincide with $\rgain$; in fact, $\rgain$ is always about $1$ to $2\%$ larger for the particular calculation shown in Figure~\ref{fig:profiles}.

The middle center panel of Figure~\ref{fig:profiles} shows profiles of $\lnu(r)$. As expected, $\lnu$ grows in the cooling layer as a result of net neutrino and antineutrino production and decreases in the gain layer as a result of heating. How much does the accreting material cooling via neutrino emission contribute to the total neutrino flux emanating from below the shock? We define the accretion neutrino luminosity $\lacc$ as the difference between the neutrino luminosity at the shock and the luminosity of the core,
\beq
\lacc = \lnu(\rs)-\lcore.
\label{eq:lacc}
\eeq
The relative contribution of the accretion luminosity to the total neutrino luminosity decreases with growing $\lcore$. For $\lcore=0$, the accretion luminosity is the sole contributor to the total luminosity of about $1.4\times 10^{52}$\,\ergsec, but for the critical core luminosity of $6.41\times 10^{52}$\,\ergsec\ the accretion luminosity contributes $28\%$ to the total neutrino output. Clearly, the accretion luminosity is a sub-dominant, but important part of the total neutrino output. In Section~\ref{sec:crit_prop} we investigate how including the neutrino radiation transport and hence accretion luminosity affects $\lcrit$ for a range of $\mdot$.

\begin{figure*}
\plottwo{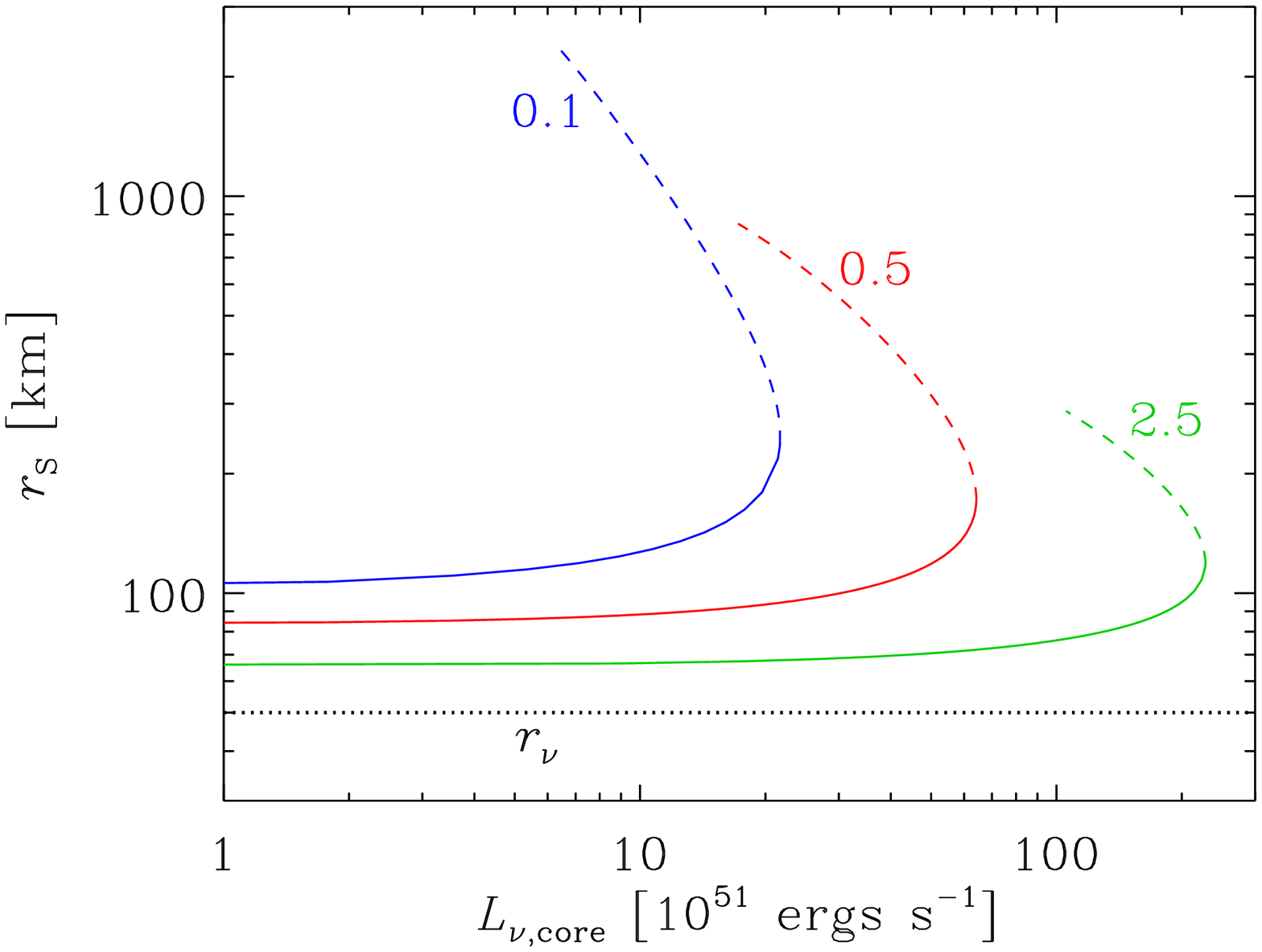}{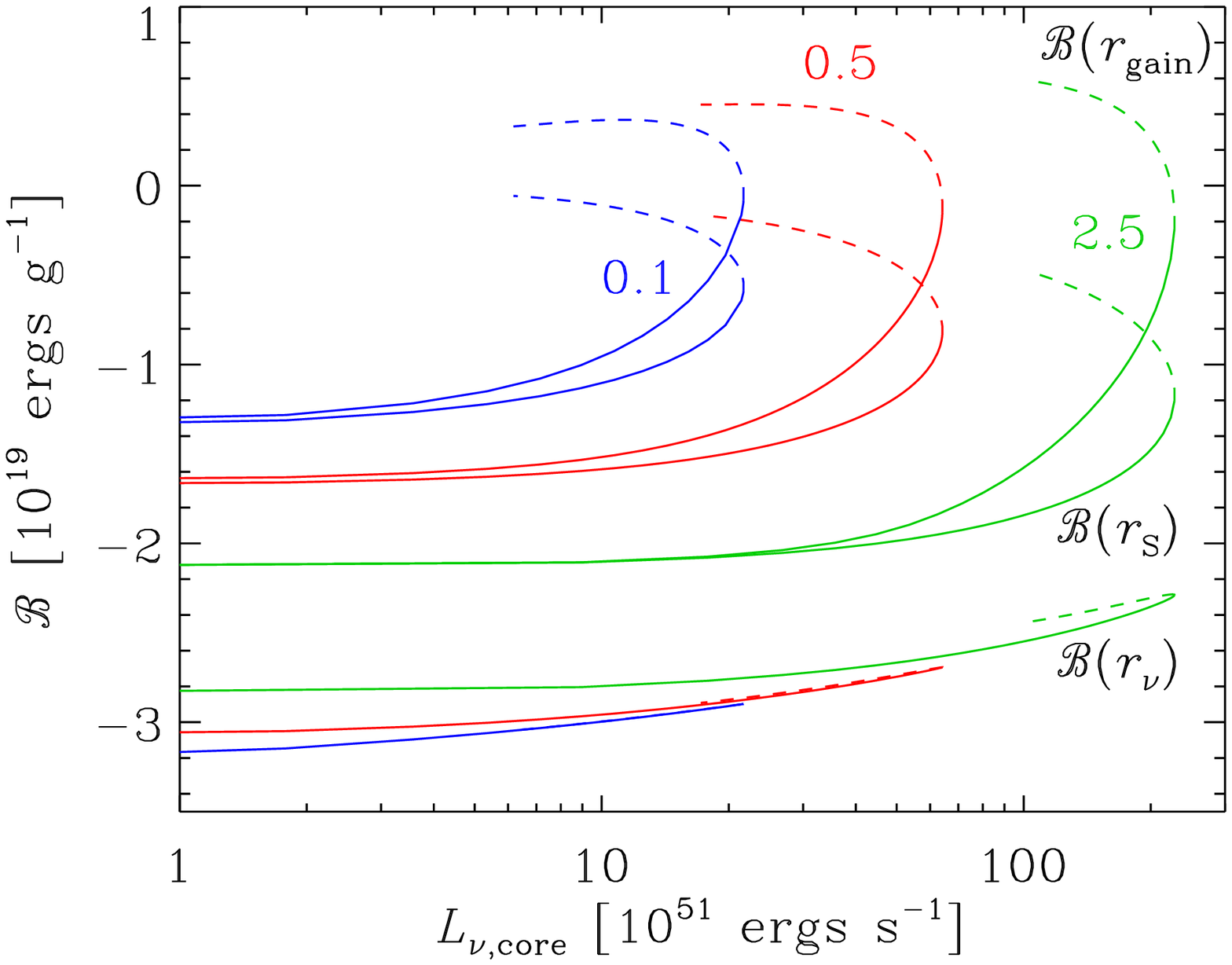}
\caption{Properties of the two branches of solutions for $M=1.4\,\msun$ and $\rnu =50$\,km for $\mdot = -0.1$ (blue), $-0.5$ (red) and $-2.5$\,$\msun\ s^{-1}$ (green). {\em Left}: Shock radius $\rs$ as a function of $\lcore$. $\lcrit$ occurs at $2.17\times 10^{52}$, $6.42 \times 10^{52}$, and $2.28 \times 10^{53}$\,\ergsec for the respective mass accretion rates. The black dotted line denotes $\rnu$. The solid and dashed parts of each line denote the lower (stable) and upper (unstable) solution branches, respectively. {\em Right}: Total specific energy $\mathscr{B}$ at $\rnu$ (bottom lines), $\rgain$ (top lines) and $\rs$ (middle lines). Color coding is the same as in the left panel.}
\label{fig:rshock}
\end{figure*}

We now turn our attention to another aspect of steady-state accretion shocks: the existence of two solutions with different shock radii for some values of $\lcore$, discovered by \citet{yamasaki05}. These authors also performed linear stability analysis and found that the upper solution branch with higher $\rs$ is unstable, the lower branch is stable, and both branches meet at a marginally stable point at $\lcrit$. We confirm the presence of the upper solution branch with our code. We switch to the upper solution branch by first calculating a sequence of models on the lower branch up to $\lcrit$. We take a solution close to $\lcrit$, manually scale up the radial coordinates by about $50\%$ and let the relaxation algorithm converge to a solution on the upper branch. 

We find that the profiles of thermodynamic variables for the upper solution branch are not fundamentally different from those of the lower branch presented in Figure~\ref{fig:profiles} and hence we do not present them here. We plot in Figure~\ref{fig:rshock} the shock radii and total specific energies at several radii of interest for both branches, and as a function of $\lcore$ for several values of $\mdot$. We see that the upper branch of solutions extends to larger shock radii, but ends at a finite radius and non-zero $\lcore$ (left panel, Figure~\ref{fig:rshock}). We specifically checked that this is not an artifact in our calculation. We note however, that for different input physics the upper branch can end at much larger radii, consistent with infinity in our code. We also find that the specific energy of the upper solution is always higher than that of the lower solution for the same $\lcore$ at all radii. In analogy with other physical systems that have two states with different energy, we can argue that the higher energy upper solution branch is unstable to transition, in agreement with \citet{yamasaki05} who find that the upper branch is indeed unstable to radial perturbations. We do not discuss the properties of the upper branch in greater detail, because it is unlikely that these solutions would be realized in a real physical system. However, the existence of the upper solution branch is analoguous to the upper solutions in Figure~\ref{fig:isot_acc}.

\subsection{Correspondence with isothermal accretion}
\label{sec:isot_corres}

It has been clearly established in 1D calculations that there exists an upper limit to the core neutrino luminosity $\lcrit$ that allows for a steady-state accretion shock \citep{bg93,yamasaki05,yamasaki06}, and this limit has been confirmed to exist in multi-dimensional dynamical supernova simulations \citep{murphy08,nordhaus10}, and in simulations of the stalled accretion shock \citep{ohnishi06,iwakami08}. To illustrate how the upper limit on the core neutrino luminosity is related to the limit on the sound speed in isothermal accretion discussed in Section~\ref{sec:isot_acc}, we consider the total specific energy of the system $\mathscr{B}$, which has the property that its radial derivative is proportional to the net deposited heating and cooling:
\beq
\frac{\intd \mathscr{B}}{\intd r} = \frac{4\pi r^2 \rho}{\mdot}\,\qdot.
\eeq
It is well established \citep[e.g.][]{janka01} and we confirm this in Appendix~\ref{app:toy_model}, that the contribution of the kinetic term $v^2/2$ to the total specific energy budget in equation~(\ref{eq:bernoul}) is negligible. At a fixed radius $r$ in the gain layer where $\qdot \approx \lcore/r^2$, the increase of $\lcore$ will result in an increase of enthalpy $h = \varepsilon + P/\rho$, and consequently, of the adiabatic sound speed, as $h \sim c_S^2$. We showed in Section~\ref{sec:prop_full} that $c_S$ as well as $\mathscr{B}$ at a given radius are increasing functions of $\lcore$ everywhere, not only in the gain layer. This correspondence between $\lcore$ and $c_S$ suggests that $\lcrit$ is a realization of $\ct^{\rm crit}$ (Section~\ref{sec:isot_acc}) in the context of the neutrino heating mechanism.

\begin{figure*}
\plotone{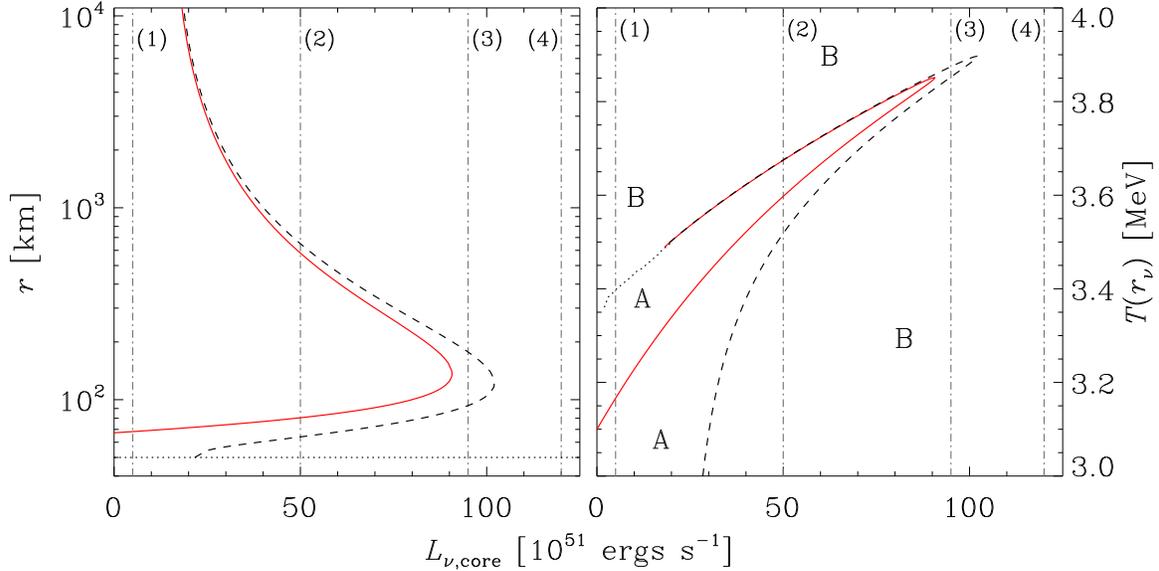}
\caption{Overview of the flow structure in the simplified supernova setup with $\lcrit = 9.07 \times 10^{52}$\,\ergsec. Vertical dash-dotted lines denoted by numbers mark values of $\lcore$ selected for detailed study in Figure~\ref{fig:sono_detail}. {\em Left}: Shock radius $\rs$ (red solid line) and sonic radius $r_{\rm sonic}$ (black dashed line) as a function of $\lcore$. {\em Right}: Temperature at the neutrinosphere $T(\rnu)$ for the solutions with a shock (red solid line) and with a sonic point (black dashed line). Regions of different solution types are marked with letters A and B, as in Figure~\ref{fig:isot_acc}. The dotted line is the separatrix between A- and B-type solutions where the sonic point lies at infinity.}
\label{fig:rshock_model}
\end{figure*}

As shown in Section~\ref{sec:isot_acc}, $\ct^{\rm crit}$ occurs when the flow cannot satisfy the shock jump conditions at any point in the accretion flow \citep[see also][]{yamasaki05}. We now prove that the same mechanism is responsible for the existence of $\lcrit$ in the supernova context. Essentially, we are trying to construct an equivalent of Figure~\ref{fig:isot_acc} for non-isothermal flows and more realistic physics. In order to make this feasible we consider a simplification of the problem by setting $\intd \ye /\intd r = 0$ and $\intd \lnu/\intd r = 0$ (eqs. [\ref{eq:el_frac}] and [\ref{eq:dldr}]) and setting the heating and cooling terms to $\mathcal{H} = 1.2 \times 10^{-18}\,{\rm cm^2\ g^{-1}} \lcore/r^2$ and $\mathcal{C} = 2 \times 10^{18}\,(T/{\rm MeV})^6$\,\ergsec\ g$^{-1}$, respectively. We also modify the boundary conditions by requiring $\rho(\rnu) = 3\times 10^{10}$\,g\ cm$^{-3}$ instead of fixing $\tau_\nu$. With these changes, the problem simplifies considerably, because $\rho(\rnu)$ and $v(\rnu)$ are now fixed and the only variables left are $T(\rnu)$ and $\rs$, which are determined by the two shock jump conditions. Consequently, if we know the value of $T(\rnu)$, the flow profile is now uniquely determined as an initial value problem.

We want to show how the solution for the flow structure with a shock relates to all possible accretion solutions. For a given $\lcore$ and $T(\rnu)$ we calculate the A-type solutions (those reaching $\mathscr{M}=-1$ in Figure~\ref{fig:isot_acc}) by setting $v^2=c_S^2$ at the outer boundary, which gives a continuum of solutions for a range of $T(\rnu)$. Solutions going through the sonic point are obtained by setting the numerator and denominator of the momentum equation~(\ref{eq:app_momentum}) simultaneously to zero at the outer boundary. The accretion breezes (B-type solutions in Figure~\ref{fig:isot_acc}) are obtained from the A-type solutions by setting $v$ to a specific fraction of $c_S$ at a fixed radius corresponding to $v=c_S$ in an A-type solution. By varying this fraction, we get a continuum of flows with different $T(\rnu)$. We extend the breezes to larger radii with an initial-value integrator of stiff equations based on algorithms from \citet{nr} and using the results from our relaxation algorithm as a starting point. We verified that for B-type solutions (subsonic accretion breezes) the ratio $|v/c_S|$ is always less than unity.

In the left panel of Figure~\ref{fig:rshock_model} we show the position of the shock radius $\rs$ (solid) and the sonic point $r_{\rm sonic}$ (dashed) as a function $\lcore$. In the right panel, we show the values of $T(\rnu)$ for the same solutions as in the left panel. We see that $r_{\rm sonic}$ exhibits qualitatively similar behavior to $\rs$, except that for low values of $\lcore$ there is no sonic point with a finite $r_{\rm sonic}$. Furthermore, the position of the shock is well separated from the sonic point, except for the tip of the upper solution branch, where these two appear to merge, similar to what happens in isothermal accretion in Figure~\ref{fig:isot_acc}. The same trend is seen in $T(\rnu)$ (the only free parameter of the problem for fixed $\lcore$), which suggests that the upper branches of shock and sonic point indeed coincide at some $\lcore$. 

\begin{figure*}
\plottwo{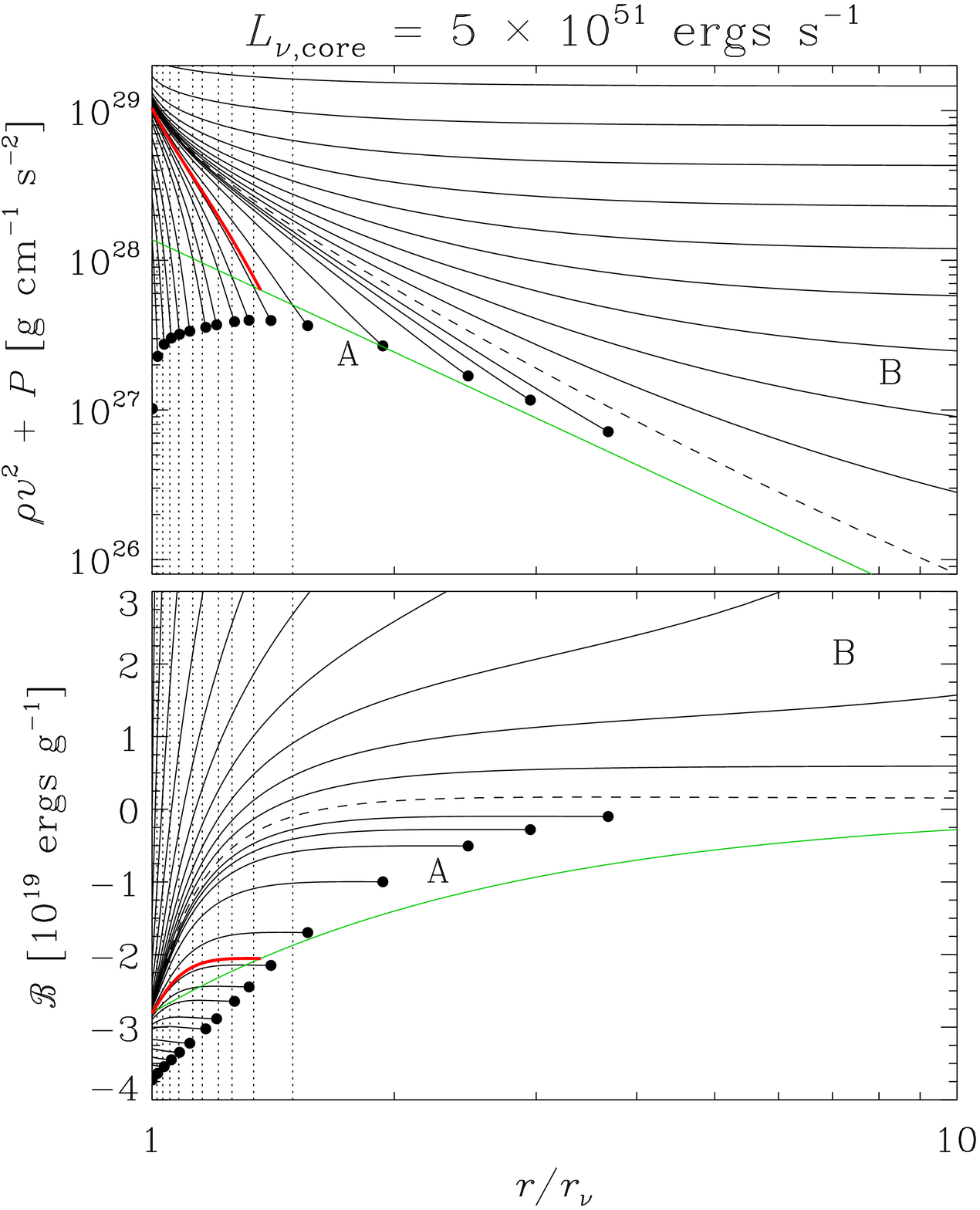}{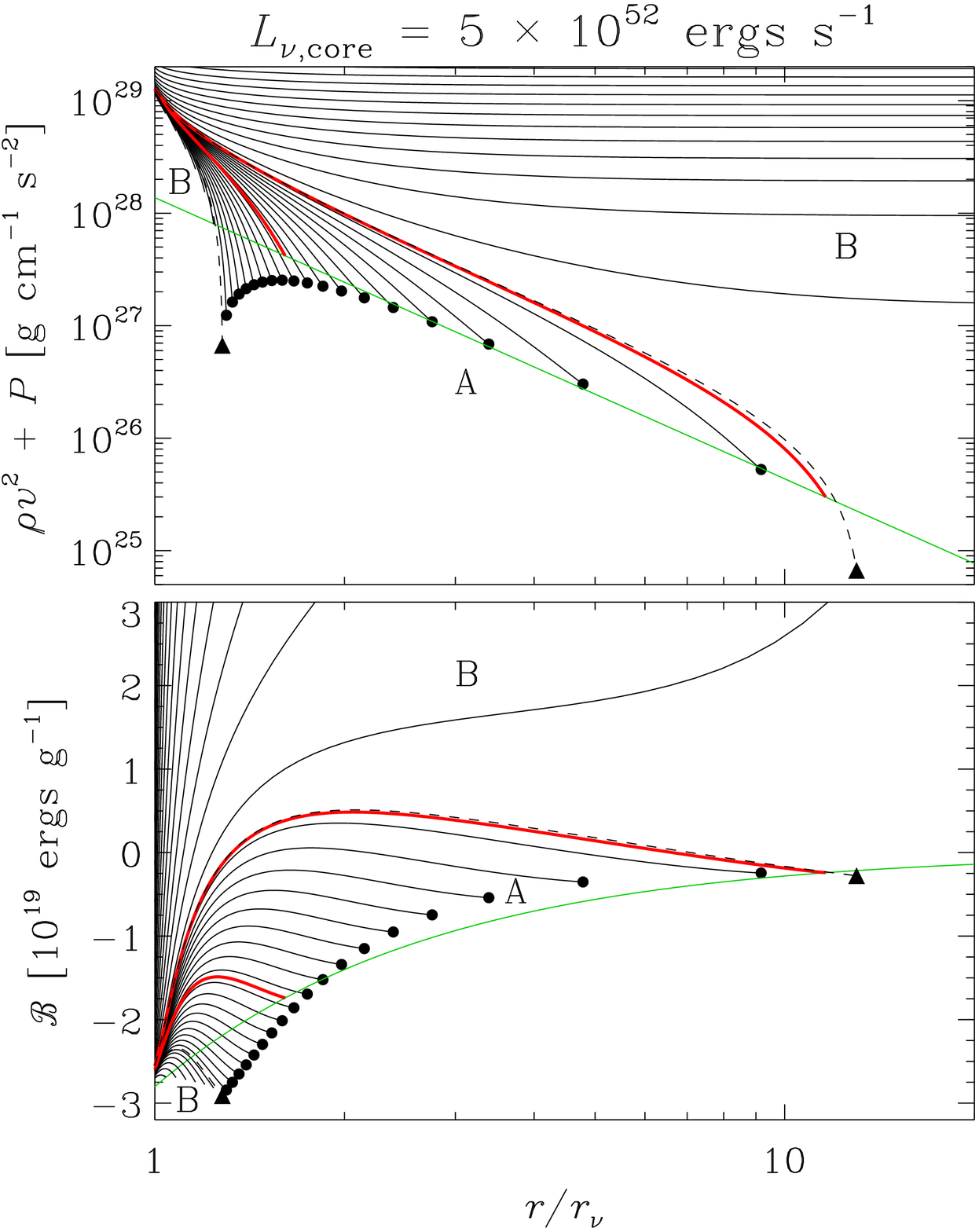}
\plottwo{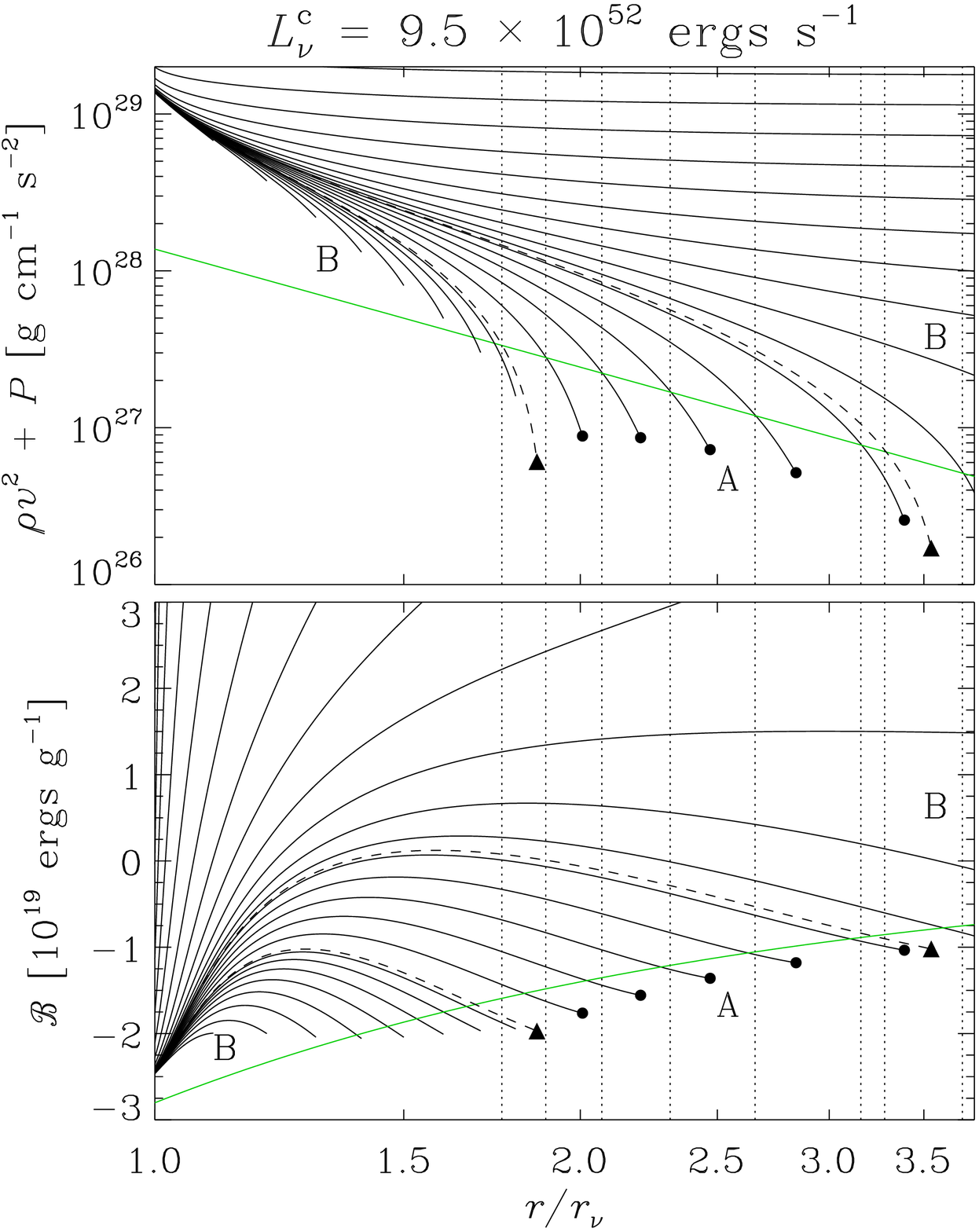}{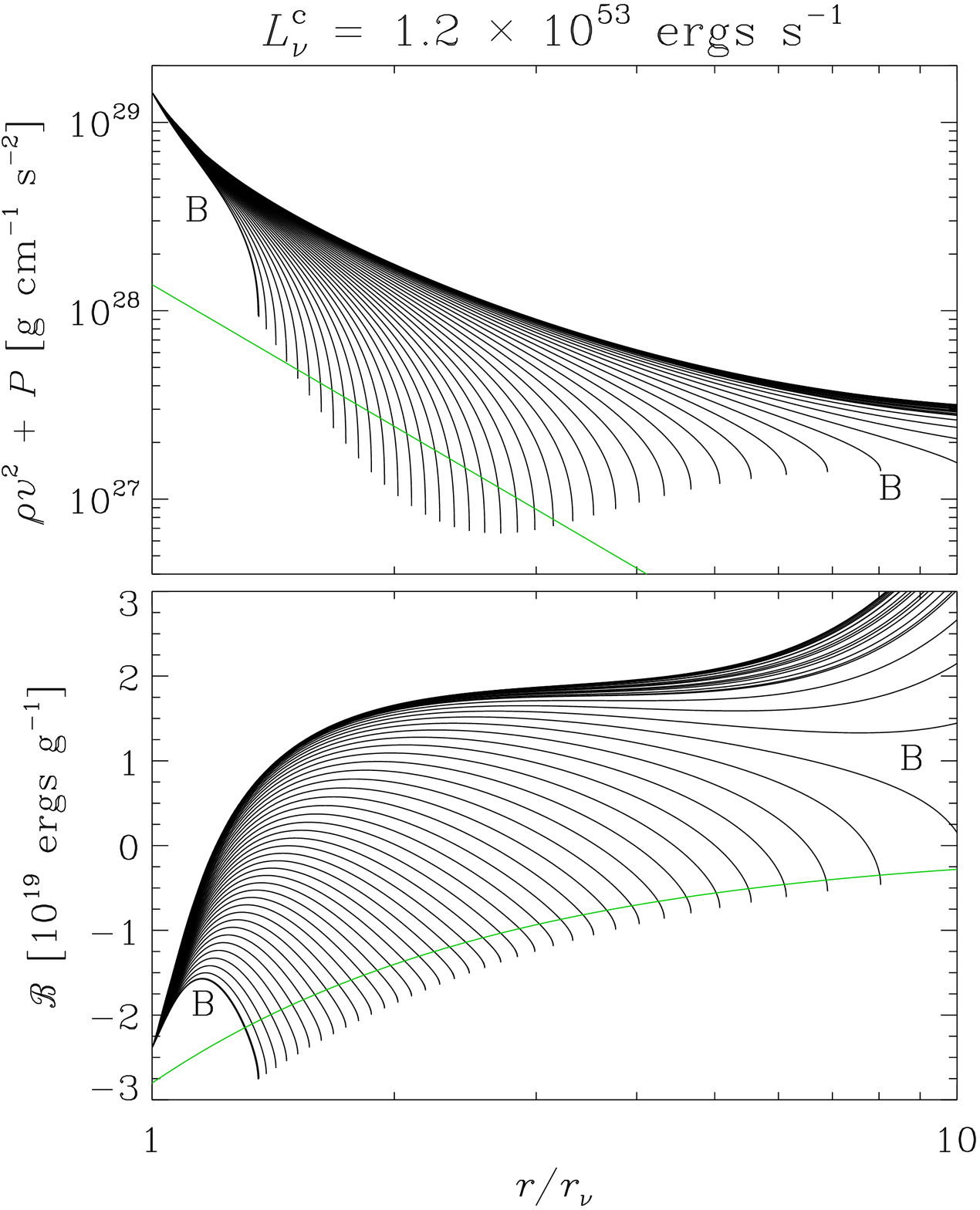}
\caption{Detailed study of the structure of flows in the simplified supernova setup summarized in Figure~\ref{fig:rshock_model}. The four panels show flow structure at different values of $\lcore$ corresponding to different situations: single shock radius and no sonic point (upper left), two shock radii and two sonic-point solutions (upper right), no shock solution and two sonic-point solutions (lower left), and neither shock or sonic-point solution (lower right). In each panel the flows that satisfy the shock jump conditions are shown with red solid lines, flows going through the sonic point (filled triangle) are shown with black dashed lines, and all remaining flows are shown with solid black lines. A-type solutions going through $v^2=c_S^2$ end with filled circles. Regions of different solution types are marked with letters A and B, as in Figure~\ref{fig:rshock_model}. Green solid lines mark the shock jump conditions. Additionally, the dotted vertical lines in the upper left and lower left panels mark radii of intersection of the flows with the momentum shock jump condition.}
\label{fig:sono_detail}
\end{figure*}

In order to explore the situation in more detail, we selected four values of $\lcore$ marked by vertical dot-dashed lines to illustrate the four configurations visible in Figure~\ref{fig:rshock_model}: (1) low $\lcore$ with one solution with finite shock radius and no finite sonic radius, 
(2) intermediate $\lcore$ where two solutions are present for both the shock and the sonic point, (3) $\lcore > \lcrit$ but with two sonic point solutions still present, and (4) high $\lcore$ with no solution for either the shock or the sonic point. In Figure~\ref{fig:sono_detail} we plot flow solutions for each of these luminosities (black solid lines) with shock jump conditions overplotted (green solid lines). The Figure should be interpreted in a way similar to Figure~\ref{fig:isot_acc}. We found that plotting the Mach number of the flows only obfuscates the issue because the lines cross in a complicated manner; instead we plot momentum $\rho v^2+P$ and specific energy $\mathscr{B}$. In this parameter space the different flow lines generally do not intersect and the shock jump conditions are a function of only the radius as can be seen from the equations~(\ref{eq:shock_jump_mom}) and (\ref{eq:shock_jump_ene}).

The upper left part of Figure~\ref{fig:sono_detail} shows flow profiles for low $\lcore$, where there is only one solution for the shock radius (far left vertical dash-dotted line in Figure~\ref{fig:rshock_model}). Each solid line corresponds to a different value of $T(\rnu)$. For low $T(\rnu)$, we find that the flows are of type A. That is, they reach a point where $v^2=c_S^2$ (not a sonic point; shown with black dots in Figure~\ref{fig:sono_detail}). For increasing $T(\rnu)$, the intersection of the flow with the shock jump conditions moves to larger radii. A-type solutions extend to  $T(\rnu) \simeq 3.32$\,MeV, where the point where $v^2=c_S^2$ moves to infinite radius. For yet larger $T(\rnu)$ we are able to find only B-type solutions --- subsonic accretion breezes --- with $|v/c_S| < 1$ everywhere. For a single value of $T(\rnu)$ the profiles of momentum and energy of the flow intersect with the shock jump conditions at the same radius, which means that only at this $T(\rnu)$ is a solution with a shock possible (red line). The vertical dotted lines mark radii of intersection of momentum of the flow with the momentum shock jump condition. We then plot these radii in the energy panel and we see that for flows lying below the red line the intersection of $\mathscr{B}$ with the appropriate shock jump condition occurs at systematically smaller radii than what is required by the momentum profiles. Conversely, for curves above the red line the intersection occurs at larger radii. Thus, the momentum and energy at the shock can only be conserved simultaneously for a single $T(\rnu)$ at $\lcore=5\times 10^{51}$\,\ergsec, which then defines not only $T(\rnu)$, but also $\rs$.

The upper right panel of Figure~\ref{fig:sono_detail} displays flow profiles for intermediate $\lcore$, which allows for two solutions for the shock and the sonic point (vertical dash-dotted line labelled (2) in Figure~\ref{fig:rshock_model}). The topology of the solutions for this $\lcore$ differs from the previous case. The A-type solutions exist only for a small range of temperatures ($3.517\,{\rm MeV} < T(\rnu) < 3.675\,{\rm MeV}$). There are two transonic solutions going through two distinct sonic points at finite but vastly different radii, which separate A-type solutions from the B-type solutions, (which occur for $T(\rnu) < 3.517$\,MeV and $T(\rnu) > 3.675$\,MeV). Due to severe stiffness of the equations, we were not able to follow the B-type solutions to very large radii. In contrast to the upper left panel of Figure~\ref{fig:sono_detail} at lower $\lcore$, the flow profiles intersect the shock jump conditions at just two inner boundary temperatures shown by the two solid red lines. The upper solutions are analoguous to those obtained in the isothermal model in Figure~\ref{fig:isot_acc}.

The lower left part of Figure~\ref{fig:sono_detail} explores the region with $\lcore > \lcrit$ where two solutions for the sonic point are still present. The topology of the solutions is similar to the previously discussed case with $\lcore = 5 \times 10^{52}$\,\ergsec, except that the A-type solutions exist for an even smaller range of inner boundary temperatures ($3.852\,{\rm MeV} < T(\rnu) < 3.875$\,MeV). However, in this case, there are no simultaneous intersections of the momentum and energy profiles of the flows with the shock jump conditions. This is explicitly shown by the vertical dotted lines, which mark several radii of intersection of the momentum profiles of the flows with the momentum shock jump condition. Clearly, the intersections of the energy profiles with the energy shock jump conditions occur at systematically smaller radii. This means that at this $\lcore$ no flow profile can connect to the free-falling matter upstream, because it is not possible to simultaneously satisfy conservation of mass, momentum and energy at the shock. Although the mismatch in radii decreases as we go to the higher $T(\rnu)$ of B-type solutions, they eventually cease to cross the energy shock jump condition altogether. Overall, as implied by Figure~\ref{fig:rshock_model}, for $\lcore = 9.5\times 10^{52} $\,\ergsec\ it is impossible to find a flow solution that simultaneously conserves momentum and energy at the shock.

For the sake of completeness we show in the lower right panel of Figure~\ref{fig:sono_detail} the flow profiles for high $\lcore$, where there is no possible solution with a shock or a sonic point. All flows displayed are B-type solutions with $|v/c_S| < 1$. Despite the fact that we were not able to extend some of the solutions to large radii due to stiffness of the equations, $v/c_S$ of the solutions systematically converged to zero at large radii.

With the $\lcore$--$T(\rnu)$ space systematically explored we can now label regions of A and B solutions in Figure~\ref{fig:rshock_model} in analogy with Figure~\ref{fig:isot_acc}. We note that the more physically realistic simulation displayed in Figure~\ref{fig:rshock} has exactly the same behavior except that sonic points exist at finite radii all the way to $\lcore=0$. The upper shock solution branch, however, ends as indicated in Figure~\ref{fig:rshock}. As can be seen from Figure~\ref{fig:rshock_model}, the position of the shock when $\lcore=\lcrit$ does not align with the position of the sonic point and hence a smooth transition to a B-type solution is not possible, unlike the results of \citet{korevaar89}, \citet{velli94,velli01} and \citet{delzanna98}. With the same logic as in Section~\ref{sec:isot_acc} we argue that at $\lcrit$ the structure of the accretion flow catastrophically changes: if $\lcore > \lcrit$ the shock moves outwards dynamically and a neutrino-driven wind with a sonic point below the shock is established. However, multi-dimensional time-dependent hydrodynamical instabilities might modify the realization of this process.

\subsection{Properties of the critical solutions}
\label{sec:crit_prop}

\begin{figure*}
\plotone{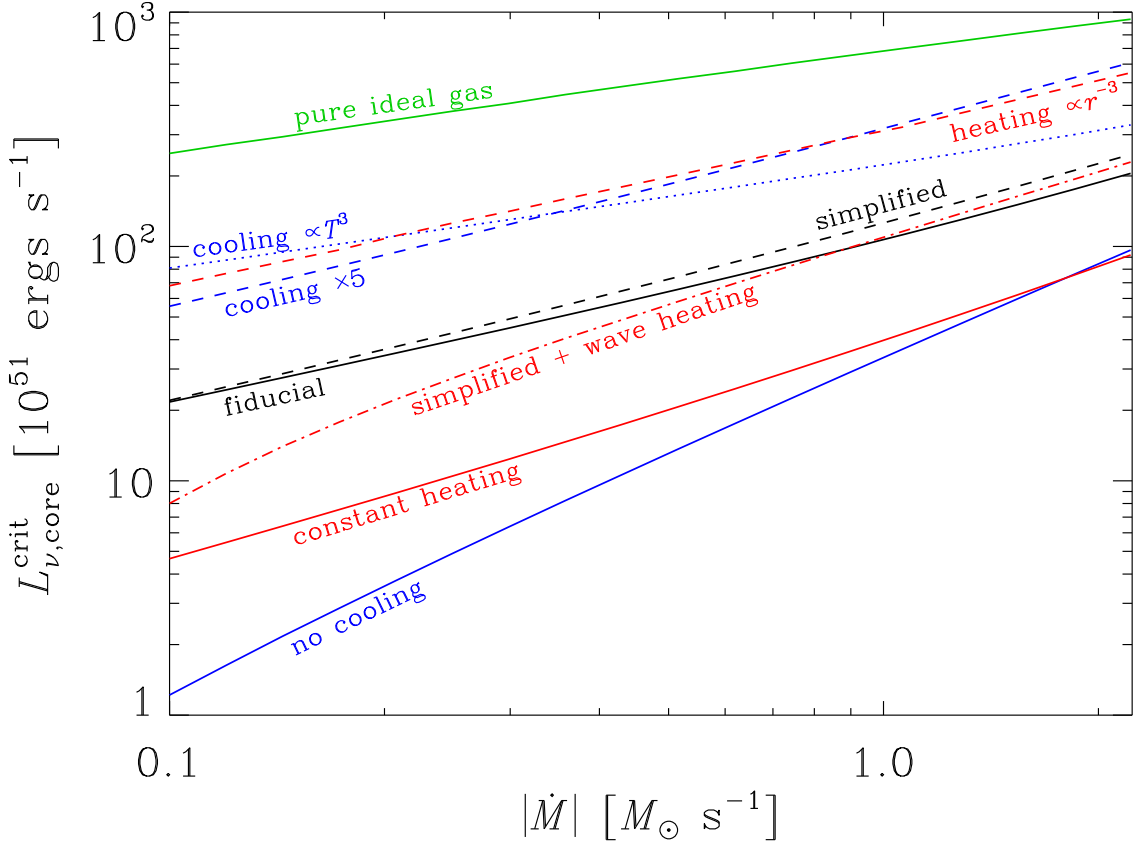}
\caption{Effects of different physics on the critical curve for $M=1.4\,\msun$, $\rnu=50$\,km. The solid black line shows the fiducial model, while the black dashed line denotes the model with simplified physics (see text for details), which was then subject to additional changes. We consider no cooling, $\mathcal{C} = 0$ (solid blue), cooling increased by a factor of $5$, $\mathcal{C} = 5\mathcal{C}_{\rm s}$ (blue dashed), cooling proportional to $T^3$, $\mathcal{C}= 8.19\times 10^{-11}\,T^3$ (blue dotted), heating proportional to $r^{-3}$, $\mathcal{H} = \mathcal{H}_{\rm s}\times (\rnu/r)$ (red dashed), constant heating rate $\mathcal{H} =  1.2\times 10^{-18}\lcore/\rnu^2$ (solid red), neutrino and Alfv\'en waves heating, $\mathcal{H} = \mathcal{H}_{\rm s} + 10^{21}\,{\rm ergs\ s^{-1}\ g^{-1}}\,(\rnu/r)^2 \exp[(1-r/\rnu)/5]$ \citep[dash-dotted red;][]{metzger07}, and EOS in the form of pure ideal gas (solid green).
}
\label{fig:diff_phys}
\end{figure*}

In Figure~\ref{fig:diff_phys} we present the critical core neutrino luminosity $\lcrit$ versus the mass accretion rate, the critical curve, as calculated with the fiducial physical model as described in Section~\ref{sec:num_setup}. Before we dive into properties of the critical curve, we investigate whether it is a robust phenomenon from the point of view of input physics. To this end, we simplify the problem by setting the electron chemical potential to zero, shutting off neutrino radiation transport ($\intd \lnu /\intd r = 0$), electron fraction evolution ($\intd \ye / \intd r = 0$), and choosing simplified heating and cooling prescription with $\mathcal{H}_{\rm s}=1.2\times 10^{-18}\lcore/r^2$ and $\mathcal{C}_{\rm s}=8.19\times 10^{-43}\,T^6$, respectively. The critical curve for this modification is shown in Figure~\ref{fig:diff_phys} with a dashed black line. With respect to the simplified version of the calculation we make additional individual changes to see how the critical curve changes: we modify cooling by multiplying it by a factor of $5$, making it proportional to $T^3$ instead of $T^6$, or shutting it off altogether. We modified heating by making it proportional to $r^{-3}$, constant at all radii or adding to $\mathcal{H}_{\rm s}$ heating motivated by Alfv\'en wave energy deposition \citep{metzger07}. We also changed the EOS to be that of a pure ideal gas. We also tried an EOS of pure relativistic particles, but we were able to converge to solution only by relaxing the condition on optical depth and replacing it with fixed density at $\rnu$. We thus do not show the critical curve here.

\begin{figure*}
\plottwo{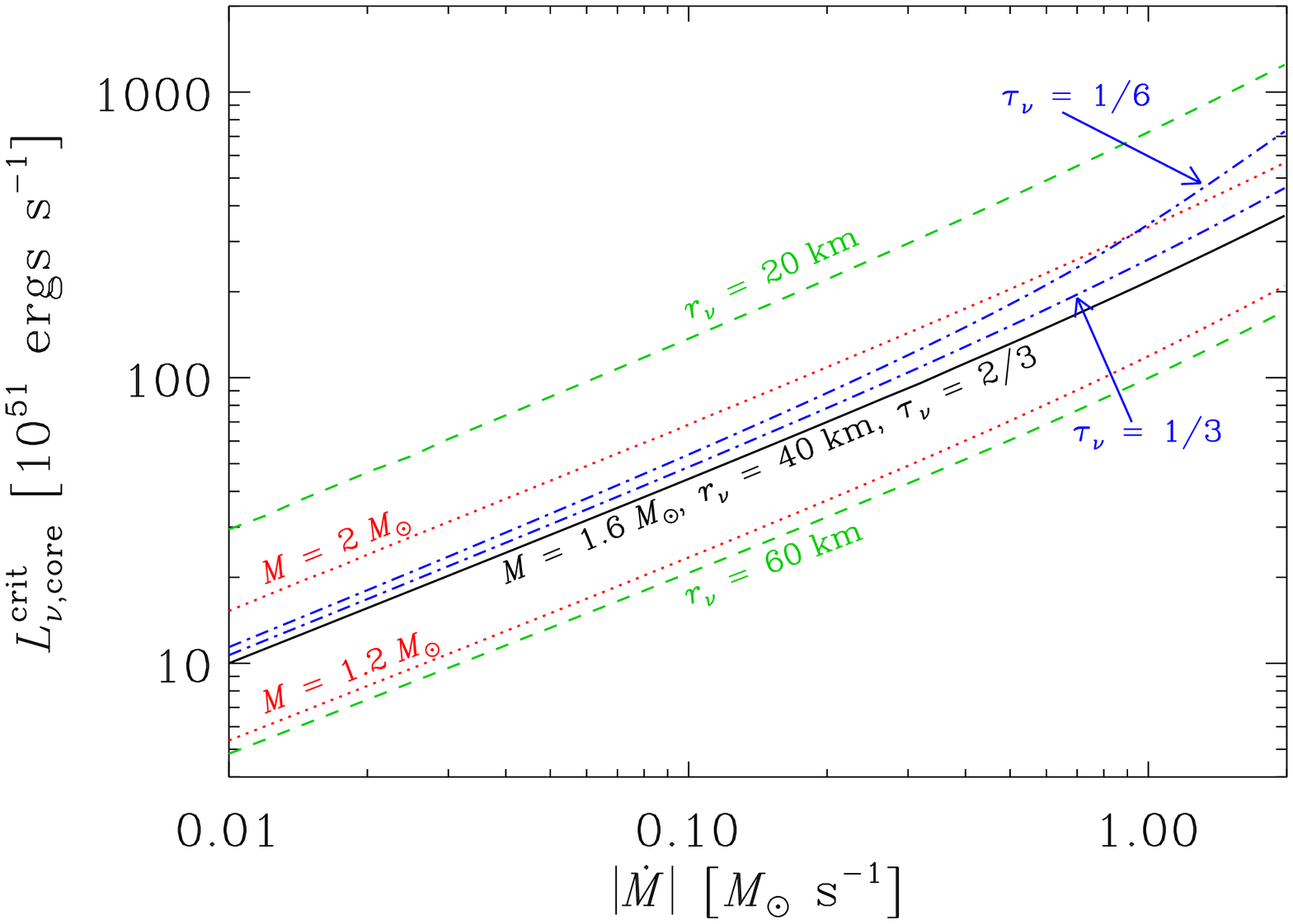}{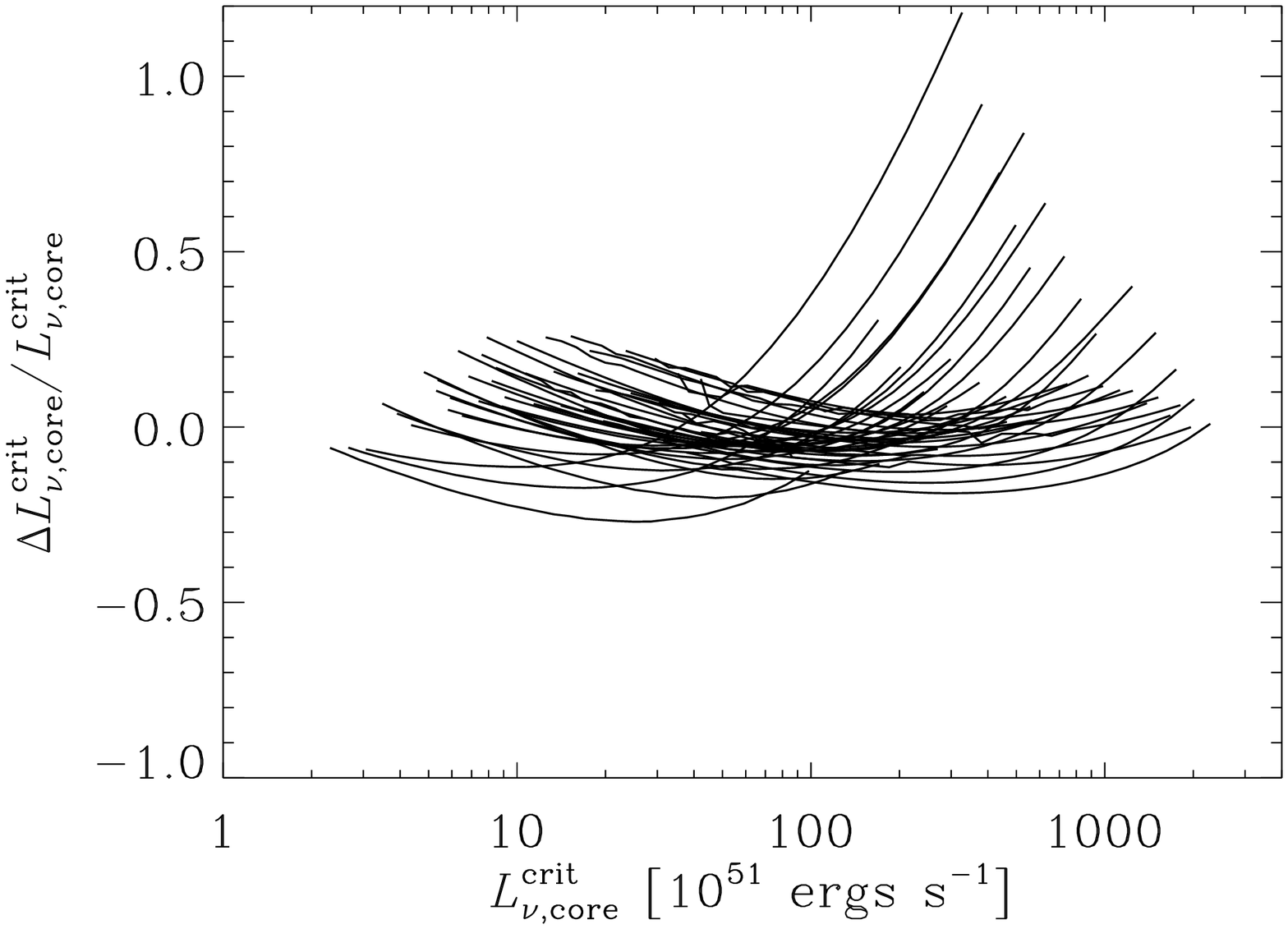}
\caption{{\em Left}: Curves of critical core neutrino luminosity as a function of $\mdot$. The black solid line is for $M=1.6\,\msun$, $\rnu=40$\,km and $\tau_\nu=2/3$. Red dotted lines are for the same parameters except $M=2.0\,\msun$ (upper line) or $1.2\,\msun$ (lower line). Green dashed lines are for $\rnu =20$\,km (upper line) or $60$\,km (lower line). Blue dash-dotted lines are for $\tau_\nu =1/3$ (lower line) or $1/6$ (upper line). {\em Right}: Relative residuals from the fitting of equation~(\ref{eq:lcrit_empir}) to the ensemble of calculated critical luminosities. Each line corresponds to a sequence of $\mdot$ from $-0.01$ to $-2\,\msun$ s$^{-1}$ with other parameters fixed.}
\label{fig:crit_curve}
\end{figure*}

Figure~\ref{fig:diff_phys} shows that even drastic changes to the functional form of the heating and cooling functions and EOS consistently yield critical values of $\lcore$. Naturally, the exact value of $\lcrit$ at a given $\mdot$ changes considerably. However, it is interesting that we do not see wildly discrepant slopes of individual lines in Figure~\ref{fig:diff_phys}. Specifically, changing the form of the heating affects the slope only very slightly, while the form of cooling has a much more pronounced effect, except when simply multiplied by a constant factor. This can be explained by the fact that cooling couples directly to the temperature, and the temperature gradient is directly responsible for the hydrostatic equilibrium of the region when relativistic gas dominates.

In order to obtain dependencies of $\lcrit$ on the parameters for the fiducial problem (solid line in Figure~\ref{fig:diff_phys}), we determine critical core luminosities on a grid of parameter values chosen as $M \in \{1.2, 1.4, 1.6, 1.8, 2.0\}\,\msun$, $\rnu \in \{20, 40, 60\}$\,km, $\tau_\nu \in \{1/6, 1/3, 2/3\}$, and 30 values of $\mdot$ logarithmically spaced between $-0.01\,\msun\ {\rm s^{-1}}$ and $-2\ \msun\ {\rm s^{-1}}$. Although in principle the neutrinosphere radius $\rnu$ and core luminosity $\lcore$ are connected by physics of neutrino diffusion out of the PNS, we consider them separately here. The critical curves for sample values of parameters are shown in the left panel of Figure~\ref{fig:crit_curve}. In addition to the well-known dependence of $\lcrit$ on $\mdot$, we see that $\lcrit$ increases with increasing $M$ and decreasing $\rnu$ and $\tau_\nu$. Lower values of $\tau_\nu$ also increase the curvature of the lines in the $\mdot$--$\lcrit$ space. In order to quantify these observations we fit to all our results a combination of power laws to get
\begin{eqnarray}
\lcrit &=& 8.18 \times 10^{53}\,{\rm ergs\ s}^{-1}\ \ \tau_\nu^{-0.206} \left( \frac{M}{\msun}\right)^{1.84} \times \nonumber\\
&\times& \left( \frac{\mdot}{\msun\ {\rm s}^{-1}}\right)^{0.723} \left( \frac{\rnu}{10\,{\rm km}}\right)^{-1.61}
\label{eq:lcrit_empir}
\end{eqnarray}

The relative differences between calculated and fitted values of $\lcrit$ are plotted in the right panel of  Figure~\ref{fig:crit_curve}. On the first sight, the multiple power-law fit of equation~(\ref{eq:lcrit_empir}) is not perfect. However, for most of the data the power-law fit is good within $30\%$ of $\lcrit$. There are systematic trends in the residuals as a function of $\mdot$ in the sense that all lines are curved upward: at low and high $\mdot$ the value of $\lcrit$ is slightly higher than what would correspond to the mean power law fitted to the data. We can understand this on the basis of our isothermal accretion model and its critical curve presented in Figure~\ref{fig:isot_crit}, which also exhibits an upward curvature of the critical curves, which is a manifestation of the nearly-hydrostatic exponential density structure of the flow (see Section~\ref{sec:isot_acc}).

In the fit of equation~(\ref{eq:lcrit_empir}) we did not include the dependence on the remaining parameters of our calculation, including the incoming electron fraction $\ye(\rs)$, and energy of the neutrinos $\epsilon_{\nu_e}$ and antineutrinos $\epsilon_{\bar{\nu}_e}$. We performed a simple check to see how changes in these parameters affect $\lcrit$. We find that changing $\ye(\rs)$ from $0.45$ to $0.5$ shifts the critical luminosity by at most $1\%$. As suggested by the physics of the neutrino heating, $\lcrit$ scales as $\epsilon_{\nu_e}^{-2}$. We confirm this scaling to within $2\%$ by varying $\epsilon_{\nu_e}$ from $8$\,MeV to $14$\,MeV while keeping the ratio $\epsilon_{\nu_e}/\epsilon_{\bar{\nu}_e}$ fixed at $13/15.5 \simeq 0.839$.

One of the uncertainties in core-collapse supernova simulations and neutron star physics in general is the EOS of dense matter. For example, a softer high-density nuclear EOS generically leads to smaller PNS radii at earlier times after collapse and bounce, but also leads to higher average neutrino energies,
which might be favorable for explosion \citep[e.g.][]{baron85a,baron85b,sumiyoshi05,janka05,mareketal09}. Can we put some constraints on the EOS using our results on $\lcrit$? In more realistic simulations, $\lcore$, $\rnu$ and the temperature of emitted neutrinos $T_\nu \sim \epsilon_{\nu_e}$ are connected by the physics of diffusion. In the simplest case, we can assume that these quantities follow the black-body law
\beq
\rnu^2 = \frac{4\lcore}{7\pi \sigma T_\nu^4},
\label{eq:rnu_bb}
\eeq
where $\sigma$ is the Stefan-Boltzmann constant. In addition to equation~(\ref{eq:lcrit_empir}) we found that $\lcrit \sim \epsilon_{\nu_e}^{-2}$ to a very good precision. Eliminating the dependence on $\epsilon_{\nu_e}$ using equation~(\ref{eq:rnu_bb}), we find 
\beq
\lcrit \propto \tau_\nu^{-0.14} M^{1.23}\mdot^{0.482}\rnu^{-0.41}.
\eeq
Therefore, larger neutrinosphere radii suggest lower $\lcrit$, when all other quantities are equal. Thus, stiff equations of state that yield higher $\rnu$ are potentially preferred. We note that this result is enabled by superlinear dependence of $\lcrit$ on $\rnu$ in equation~(\ref{eq:lcrit_empir}). On the other hand, soft EOS leads to a faster contraction of the PNS, higher core luminosities and neutrino energies, which can be favorable for the explosion too \citep{schecketal06,schecketal08,marek09,mareketal09}.

\begin{figure}
\plotone{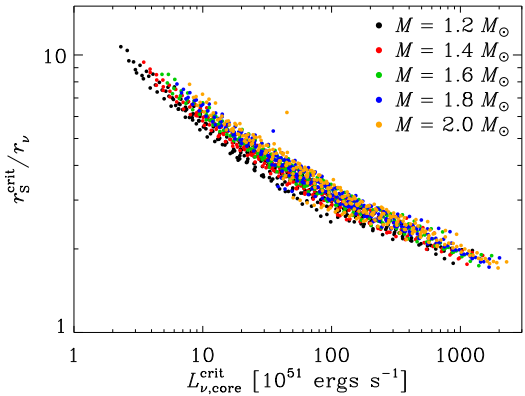}
\caption{Critical shock radii relative to $\rnu$ as a function of $\lcrit$ for calculations presented in Figure~\ref{fig:crit_curve}. Different colors correspond to different values of $M$.}
\label{fig:crit_rshock}
\end{figure}

Finally, we briefly mention the behavior of shock radii at $\lcrit$, $\rs^{\rm crit}$. In Figure~\ref{fig:crit_rshock}, we plot $\rs^{\rm crit}/\rnu$ as a function of $\lcrit$ for data from Figure~\ref{fig:crit_curve}. We see that the points lie close to a line described approximately as $\rs^{\rm crit}/\rnu \propto ({\lcrit})^{-0.26}$. The weak dependence on $\lcrit$ can probably be attributed to changes in the efficiency of cooling as a function $\lcrit$ and hence $\mdot$ (see Section~\ref{sec:lcrit_multid} for a related discussion). There are no noticeable residuals as a function of $\rnu$, $M$, and $\tau_\nu$.

\subsection{Interpretation of $\lcrit$ and the importance of $\lacc$}
\label{sec:lcrit_interpretation}

As we have demonstrated in Section~\ref{sec:crit_prop}, the critical luminosity depends on $\mdot$, $M$ and $\rnu$. A natural combination of these variables to have the dimension of luminosity is $GM\mdot/\rnu$, an expression for power released by accretion of material. Although the powers of individual quantities are somewhat different than those in equation~(\ref{eq:lcrit_empir}), one might wonder whether these expressions for the critical luminosity are intimately related in the sense that $\lcrit$ is directly proportional to $\lacc$. This interpretation is especially tempting, because dynamical supernova models seem to be on the verge of explosion over a wide range of $\mdot$, which would suggest that a significant fraction of the neutrino luminosity is supplied by cooling of the accreting gas. Any change in $\mdot$ yields a proportional change of $\lacc$, which again somehow proportionally changes the critical luminosity while keeping it higher than $\lacc$. This picture is, however, incorrect. 

\begin{figure}
\plotone{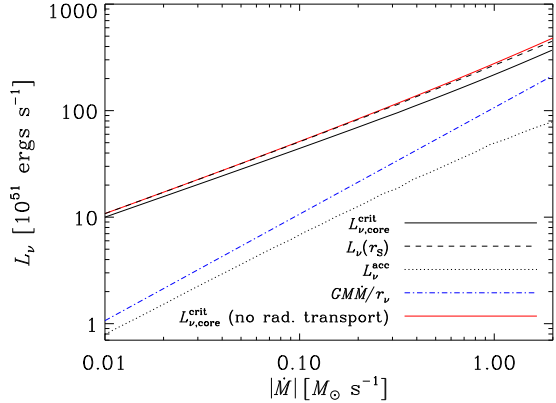}
\caption{Effect of radiation transport and accretion luminosity on the critical curve. The black solid line shows critical core neutrino luminosity $\lcrit$ for $M=1.6\,\msun$ and $\rnu=40$\,km, the black dashed line is the neutrino luminosity at the shock $\lnu(\rs)$, and the black dotted line is their difference, the accretion luminosity $\lacc$. The blue dash-and-dotted line is the maximum accretion luminosity, $GM\mdot/\rnu$. The red solid line is a critical curve for the same parameters except that neutrino radiation transport was shut off.}
\label{fig:crit_acc}
\end{figure}

To illustrate why, we plot in Figure~\ref{fig:crit_acc} $\lcrit$ along with the neutrino luminosity leaving the shock $\lnu(\rs)$ at $\lcrit$, and their difference, $\lacc$. We also plot $GM\mdot/\rnu$, corresponding to the maximum achievable $\lacc$ for a cold accretion flow. We see that $\lacc$ is always a small fraction of $\lcrit$, and that even the maximum possible value of $\lacc$ falls short of the necessary luminosity by a factor of several as found also by \citet{bg93}. Clearly, the major source of neutrinos powering the explosion is the core of the PNS. However, the accretion luminosity is not entirely unimportant. We plot in Figure~\ref{fig:crit_acc} $\lcrit$ for a calculation with the same parameters but with neutrino radiation transport shut off. Including the radiation transport in the calculation has a positive effect of lowering the critical core luminosity by about $8\%$ at $\mdot=-0.01\,\msun$ s$^{-1}$ and by about $23\%$ at $\mdot=-2\,\msun$ s$^{-1}$, respectively. 

Instead of regarding equation~(\ref{eq:lcrit_empir}) as a manifestation of accretion luminosity, we can understand it in the terms of the critical condition on the sound speed in isothermal accretion given in equation~(\ref{eq:isot_cond}). The dependencies of equation~(\ref{eq:lcrit_empir}) can be then explained in simple terms. Increasing the mass of the PNS core makes the escape velocity higher and velocity profile steeper, and more heating is then necessary to get a mild enough velocity profile that it can no longer connect to the shock jump conditions (Section~\ref{sec:isot_acc}). Increasing $\rnu$ shifts the region of interest to weaker gravitational potential with lower escape velocity and hence lower neutrino luminosity is required to get a mild velocity profile that again does not connect to the shock jump conditions. Decreasing $\tau_\nu$ with all other parameters fixed means less neutrinos absorbed and clearly a higher critical neutrino luminosity is required. 

\begin{figure*}
\plotone{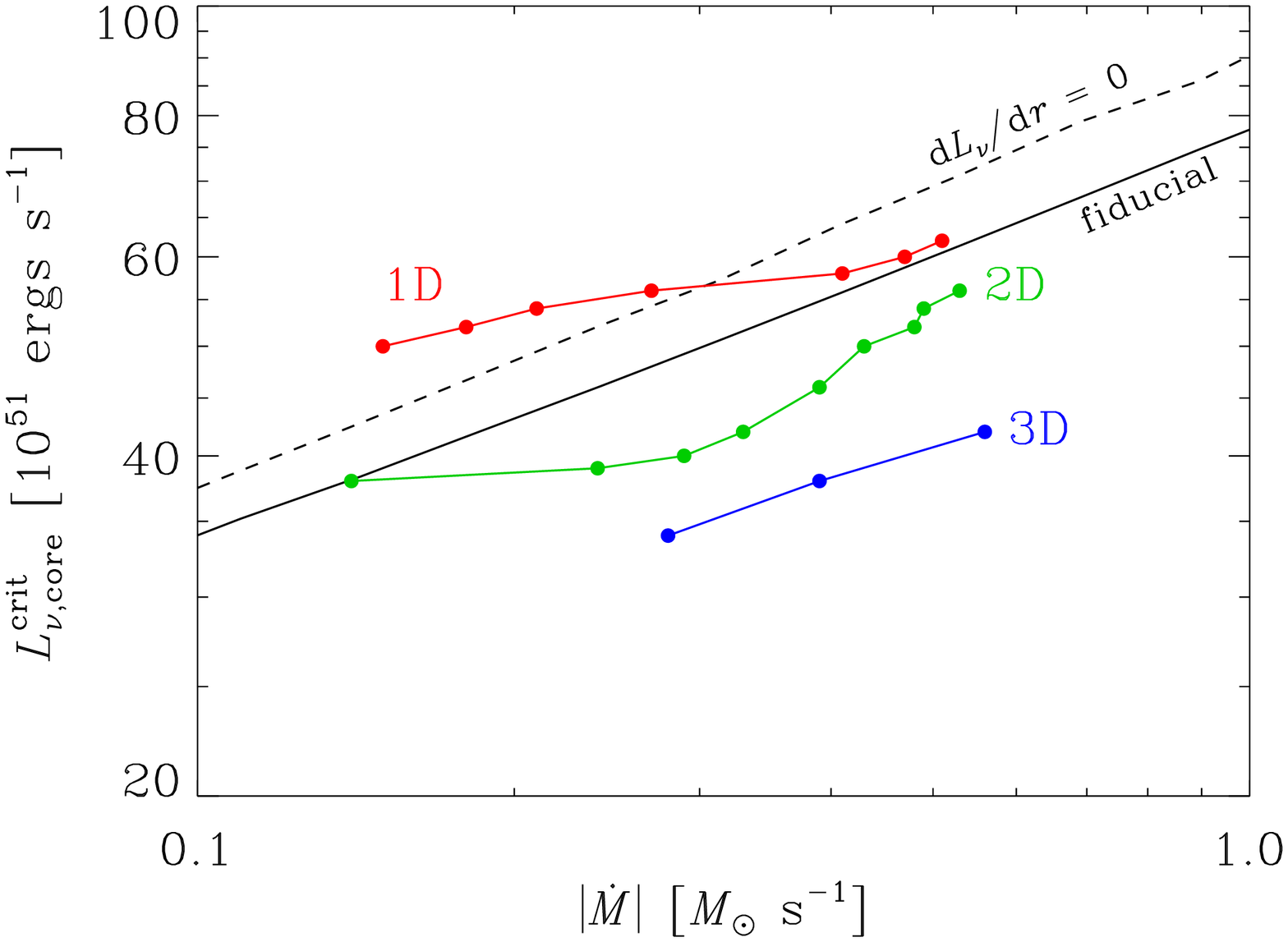}
\caption{Comparison of our calculations with the results of \citet{nordhaus10} (red, green and blue lines with points). The black line is our fiducial model with $M=1.4\,\msun$ and $\rnu$ calculated from equation~(\ref{eq:rnu_bb}), while the dashed line is for our calculation with $\intd \lnu /\intd r = 0$ (eq.~[\ref{eq:dldr}]). Heating from the accretion luminosity reduces $\lcrit$ by an amount comparable to going from 1D to 2D or from 2D to 3D.}
\label{fig:nordhaus_comp}
\end{figure*}

To illustrate this in greater detail, we can write $\qdot/v \sim \intd c_S^2/\intd r \sim c_S^2/\rnu$ and hence $c_S^2 \sim h \sim \qdot\rnu /v \sim \rho_\nu\rnu \lcore /\mdot$, where $h$ is the enthalpy and $\rho_\nu $ is the density at the neutrinosphere. Plugging this expression as $\ct^{\rm crit}$ in equation~(\ref{eq:isot_cond}), we obtain $\lcrit \sim GM\mdot/(\rho_\nu \rnu^2)$. The product $\rho_\nu\rnu$ is proportional to the optical depth $\tau_\nu$, because most of opacity to the neutrinos is concentrated close to the neutrinosphere. Thus, by relating $\ct^{\rm crit}$ of the isothermal accretion model through Euler equations to $\lcrit$ we arrive through a simple estimate to essentially the same expression for $\lcrit$ as in the dimensional analysis in the beginning of this Section. The power-law dependence of these estimates is of course different than in the empirical fit in equation~(\ref{eq:lcrit_empir}) due to nonlinear nature of the problem.

Several authors have produced critical curves based on either steady-state calculations like our own, or hydrodynamic simulations. Through steady-state calculations, \citet{bg93} obtained a critical curve that is best fit by a power law, $\lcrit \sim \dot{M}^{0.43}$. This power-law slope is lower by about $0.3$ than what we measure. This difference can be reconciled by realizing that \citet{bg93} assumed that the neutrinosphere radius $r_\nu$ is determined by $\lcore$ and neutrino temperature $T_\nu$ through a black-body law. If we substitute equation~(\ref{eq:rnu_bb}) to equation~(\ref{eq:lcrit_empir}) to eliminate $\rnu$, we get
\beq
\lcrit \propto \tau_\nu^{-0.114} M^{1.02} \mdot^{0.401} \epsilon_{\nu_e}^{0.68}.
\eeq
The power-law slope in $\mdot$ is about $0.40$, quite close to \citet{bg93}.

In Figure~\ref{fig:nordhaus_comp} we compare our results to the critical curves of \citet{nordhaus10} based on hydrodynamic simulations. In order to make our calculations compatible with theirs, we couple $\rnu$ to $\lcore$ through equation~(\ref{eq:rnu_bb}) with $T_\nu = 4.5$\,MeV. We keep $M$ fixed at $1.4\,\msun$, which makes the slope of our critical curves slightly milder than that of \citet{nordhaus10}: for a given progenitor model lower, but fixed, $\lcore$ means that the explosion occurred later when $M$ is higher, because the PNS had a longer time to gain mass, and hence $\lcrit$ is higher. We see that our results are in good agreement with \citet{nordhaus10} and that including heating from accretion luminosity reduces $\lcrit$ by an amount comparable to going from 1D to 2D or from 2D to 3D. 

\section{Explosion conditions}
\label{sec:conditions}

Our goal in this section is to find out how well some of the conditions proposed in the supernova literature for reviving the stalled accretion shock work in diagnosing $\lcrit$ within the context of steady-state calculation. Similar work was performed by \citet{murphy08} with dynamical simulations. Our code does not allow for time evolution, but we can directly and exactly compare how a given explosion condition relates to the derived steady-state value of $\lcrit$ at various values of $\mdot$. 

\subsection{Positive acceleration condition of \citet{bethewilson85}}

\citet{bethewilson85} were the first to analyze the delayed neutrino mechanism. They argued that shock recession creates a steeper pressure profile that in turn overcomes gravitation and gives an outward acceleration, $\ddot{r} > 0$. In the language of our steady-state model, $\ddot{r} = v (\intd v/\intd r) >0$ is the statement of their critical explosion condition. From Figure~\ref{fig:profiles} we clearly see that below the shock the flow decelerates everywhere, except maybe for a small range of radii, where $v$ is basically constant. We also find through an investigation of the velocity profiles that having $\intd v/\intd r = 0$ somewhere in the flow does not correlate with approaching the critical curve at different $\mdot$. In the discussion of our toy model in Section~\ref{sec:toy_model} and in Appendix~\ref{app:toy_model} we show that velocity and acceleration terms are subdominant in the momentum and energy equation of the problem and that $\lcrit$ does not depend on these terms. From this evidence we conclude that the vanishing acceleration criterion of \citet{bethewilson85} does not coincide with the critical neutrino luminosity.

\subsection{Explosion condition of \citet{janka01}}

\begin{figure}
\plotone{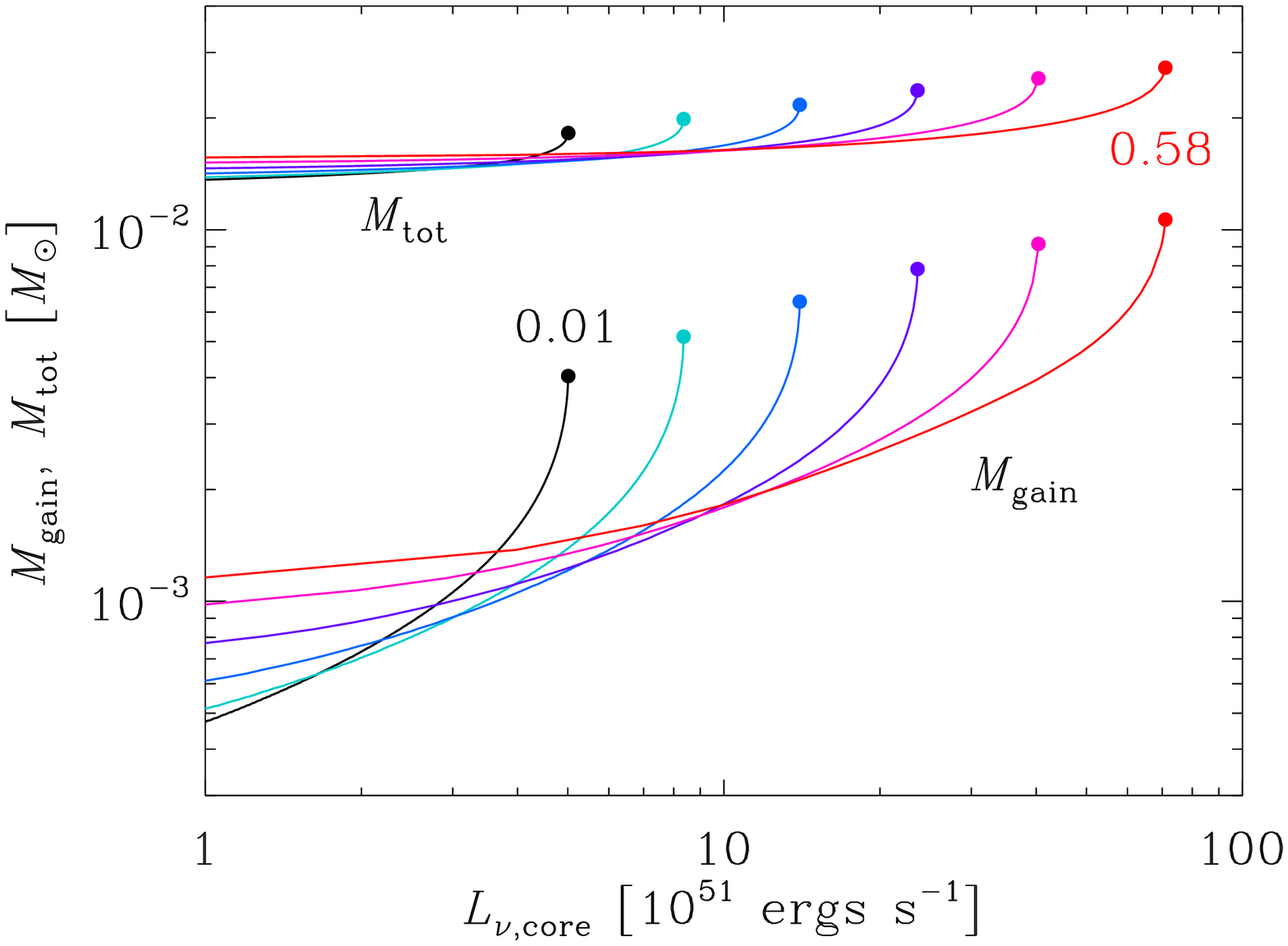}
\plotone{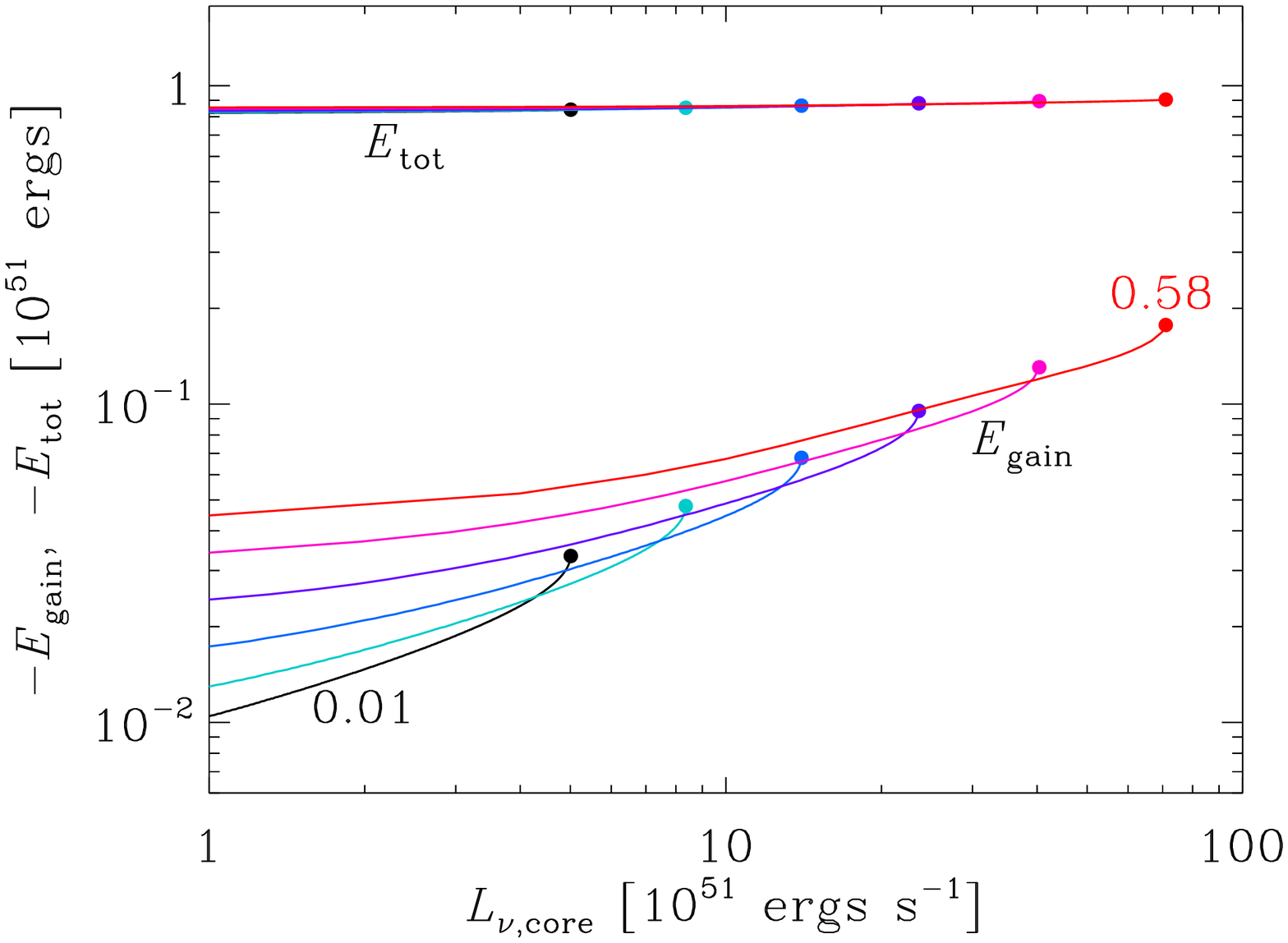}
\caption{{\em Top}: Mass in the gain layer $M_{\rm gain}$ (lower set of lines) and the total mass between the neutrinosphere and the shock $M_{\rm tot}$ (upper set of lines) as a function of $\lcore$. Filled circles at the end of the lines denote position of $\lcrit$. Colors denote variations in the mass-accretion rate going from $|\mdot| = 0.01\,\msun\ \invs$ (black) to $|\mdot| = 0.58\,\msun\ \invs$ (yellow) in uniform logarithmic steps. {\em Bottom}: Energy \'a la \citet{janka01} in the gain layer $E_{\rm gain}$ (lower set of lines) and the energy between the neutrinosphere and the shock $E_{\rm tot}$ (upper set of lines). The meaning of symbols and colors is the same as in the upper panel.}
\label{fig:janka}
\end{figure}

\citet{janka01} presented a quasi-time-dependent analytic analysis of the conditions necessary for revival of a stalled accretion shock. He found that the shock experiences both expansion and outward acceleration when both the mass and energy in the gain region grow with time. These conditions define two critical lines in the $\lcore$--$\mdot$ plane that enclose a region favorable for explosion. As the approach of \citet{janka01} is completely different from that of \citet{bg93} and also ours, his critical condition is also very different. As can be seen in Figures~4 and 5 of \citet{janka01}, the critical luminosities are decreasing function of increasing $|\mdot|$, unlike the plots in \citet{bg93}, \citet{yamasaki05,yamasaki06}, and our Figure~\ref{fig:crit_acc}, where the critical luminosities increase with increasing $|\mdot|$.

In Figure~\ref{fig:janka}, we plot as a function of $\lcore$ and $\mdot$ the mass in the gain layer, $M_{\rm gain} = \int_{\rgain}^{\rs} 4\pi r^2\rho\,\intd r$, total mass between the neutrinosphere and the shock, $M_{\rm tot}$, energy in the gain layer, $E_{\rm gain} = \int_{\rgain}^{\rs} 4\pi r^2\rho (\varepsilon - GM/r)\,\intd r$, and the total energy between the neutrinosphere and the shock $E_{\rm tot}$. We see that for a model with fixed mass accretion rate the masses grow with increasing $\lcore$ as $\lcrit$ is approached. While the total energy at any specific radius in the gain layer increases with growing $\lcore$ (see Figure~\ref{fig:profiles} for a plot of the related Bernoulli integral), the total energy within that region decreases, because the gain layer gets physically larger with increasing $\lcore$.

The critical conditions of \citet{janka01} demand that the mass and energy in the gain region increase with time. As there is no time dependence in our steady-state models and thus this condition is not directly testable, we might emulate the time dependence with a sequence of steady-state models. If we fix $\mdot$ and increase $\lcore$, following either of the lines in Figure~\ref{fig:janka}, $M_{\rm gain}$ grows and $E_{\rm gain}$ decreases. On the other hand, if we fix $\lcore$ to, say, $5\times 10^{51}$\,\ergsec\ and decrease $|\mdot|$, which corresponds to moving vertically from red through blue to black lines in Figure~\ref{fig:janka}, $M_{\rm gain}$ first drops and then increases. Similarly, $E_{\rm gain}$ first increases and then just before reaching the critical curve it decreases. Thus, from the point of view of our steady-state models, the conditions of \citet{janka01} are not equivalent to the critical neutrino luminosity.

\subsection{Advection time vs.\ heating time}

The ratio of advection to heating time $t\adv/t\heat$ has been extensively used as a metric of hydrodynamic simulations \citep[e.g.][]{thompson00,thompson_murray01,thompson05,buras06b,schecketal08}. \citet{murphy08} found that this condition is only a rough diagnostic of the critical condition for explosion.

\begin{figure}
\plotone{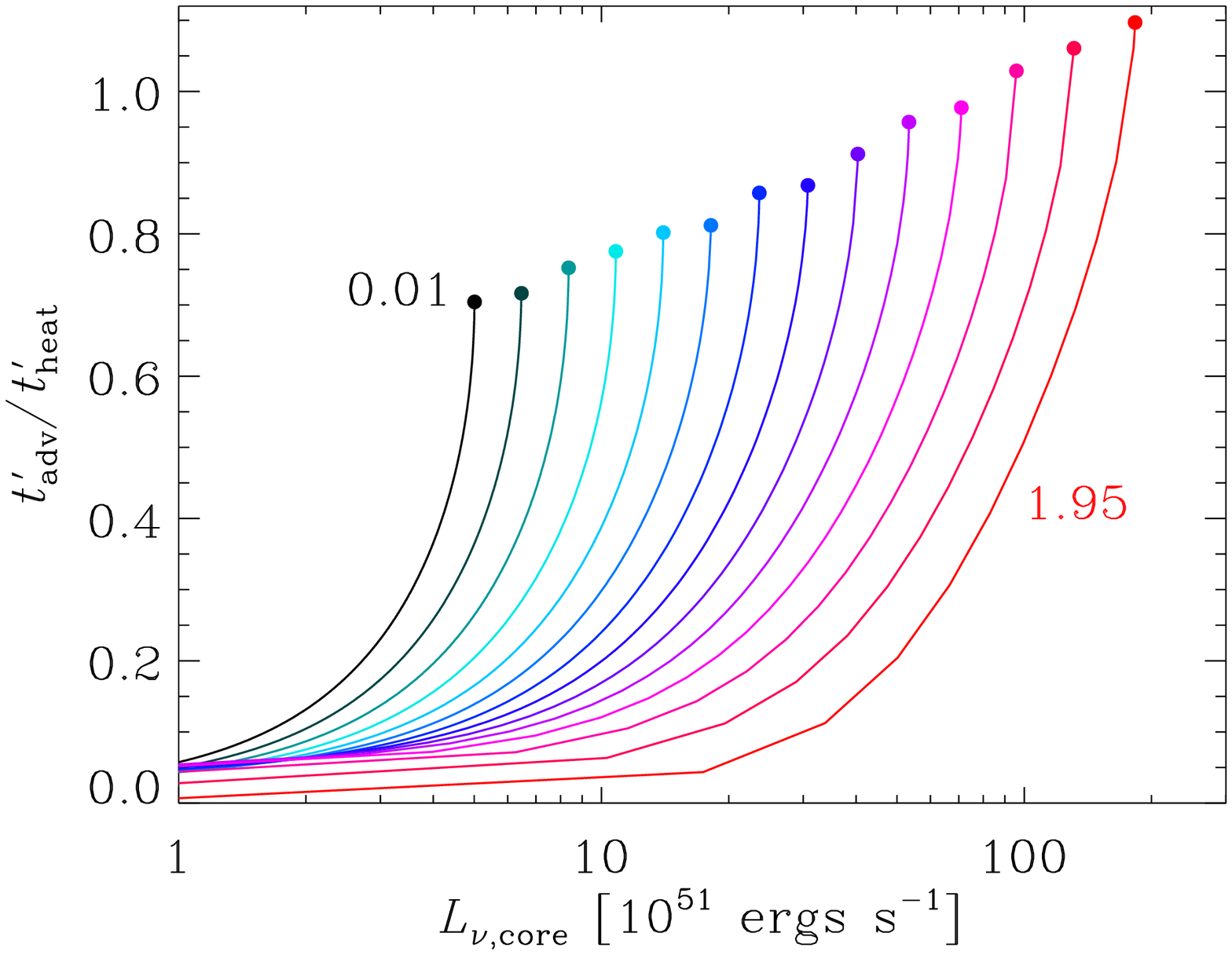}
\caption{Ratio of advection and heating times as defined by \citet{murphy08}. Each line shows the ratio as a function of core luminosity 
$\lcore$ and fixed $\mdot$. The color coding corresponds to different values of $|\mdot|$ going lef to to right from $0.01\,\msun$\ s$^{-1}$ (black) to $1.95\,\msun$\ s$^{-1}$ (red). The filled circles at the end of the lines mark the value at $\lcrit$.}
\label{fig:tadv_theat1}
\end{figure}

\begin{figure}
\plotone{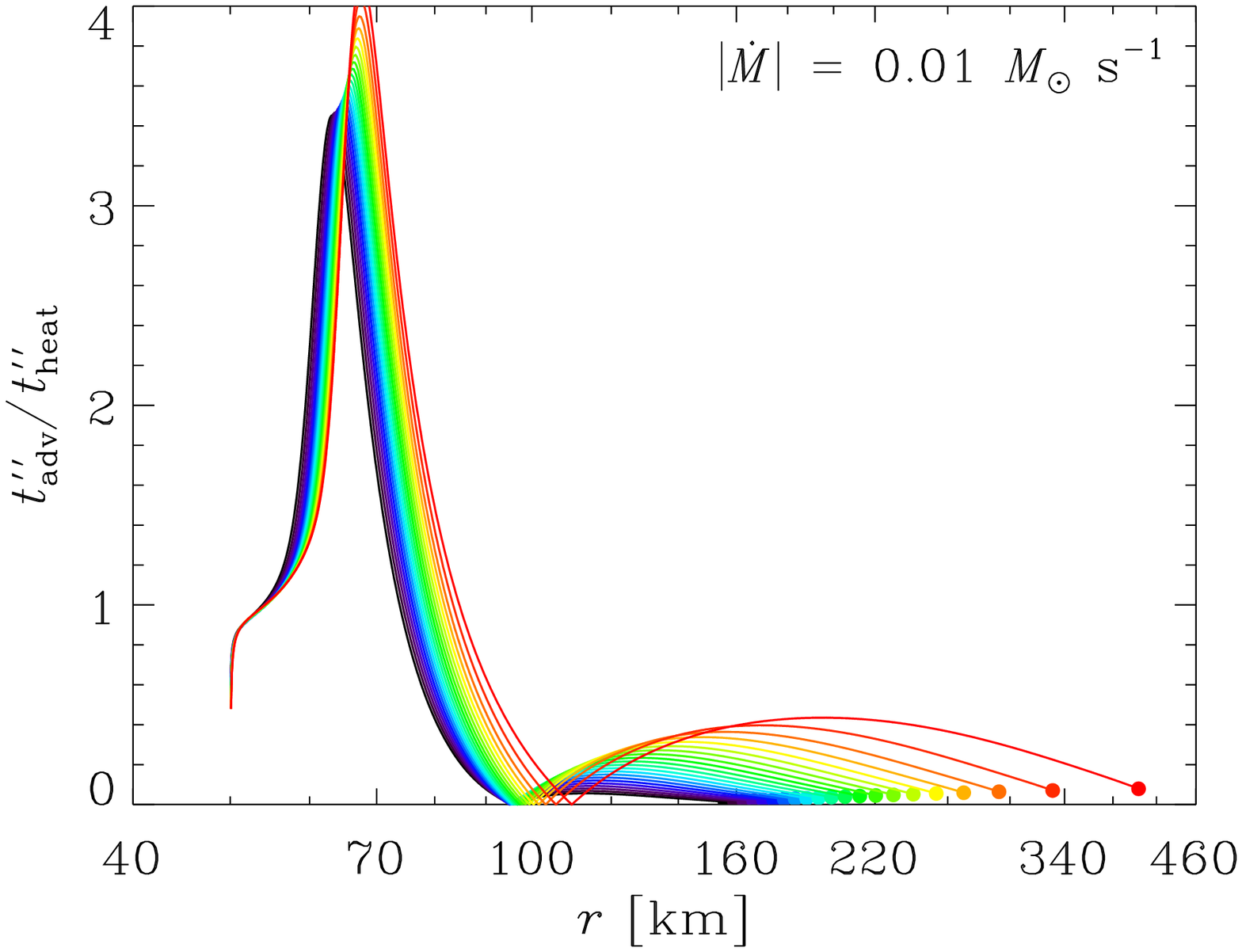}
\plotone{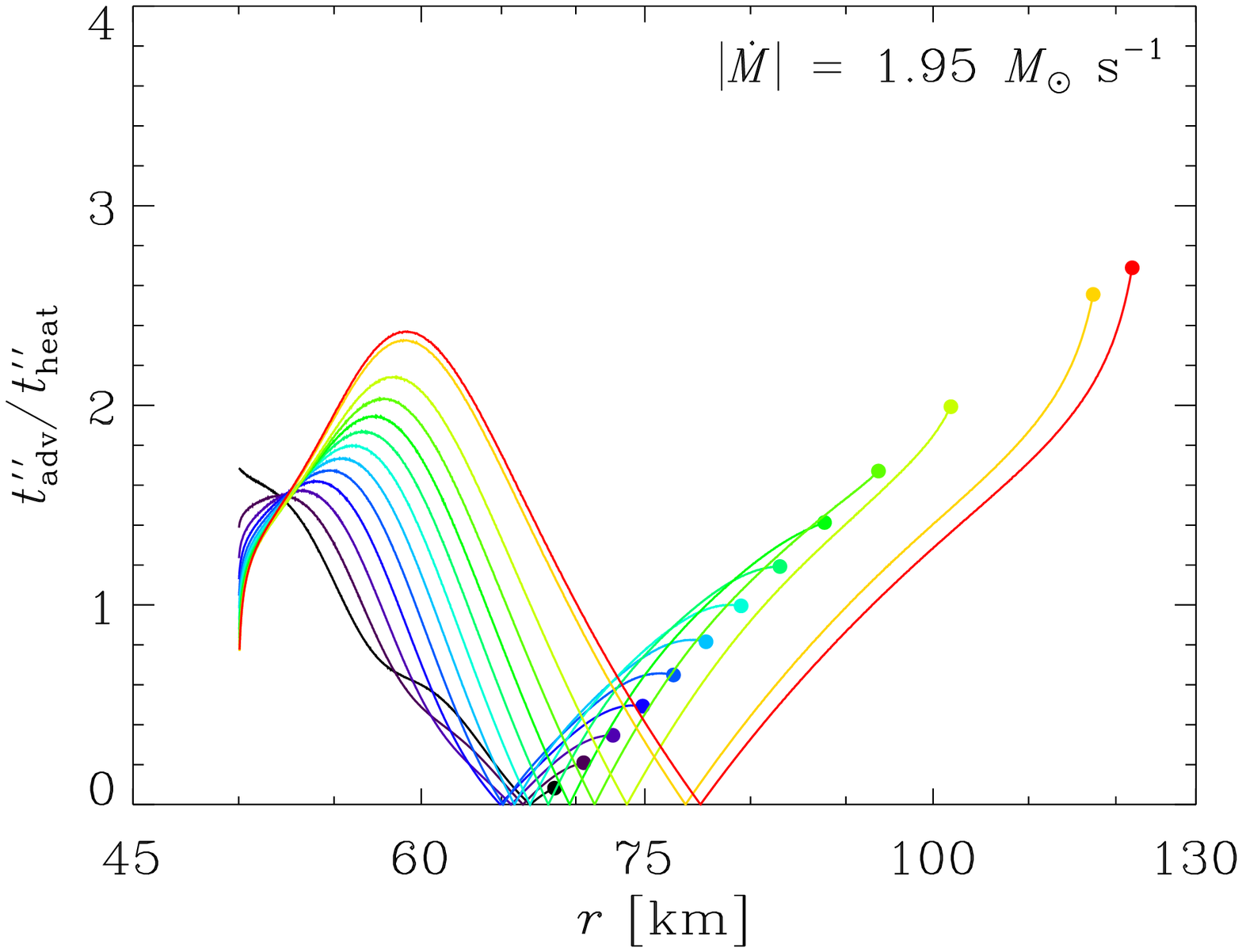}
\caption{Ratio of advection and heating times as defined by \citet{thompson05} for two values of $\mdot$. Each line corresponds to a radial profile of the ratio for a given luminosity going from $1\times 10^{51}$\,\ergsec\ (black) to the critical value (red). Filled circles at the end of the lines mark the value just inside of the shock. }
\label{fig:tadv_theat2}
\end{figure}

Here, we specifically test two definitions of advection and heating times. The first is an integral condition on the gain layer due to \citet{murphy08}
\beq
\addtocounter{equation}{1}
t'\adv = \int_{\rgain}^{\rs} \frac{\intd r}{|v|}, \tag{\arabic{equation}a}\label{eq:tadva}
\eeq
\beq
\addtocounter{equation}{1}
t'\heat = \frac{\int_{\rgain}^{\rs} 4\pi r^2 \rho\, \varepsilon\,\intd r}{\int_{\rgain}^{\rs} 4\pi r^2 \rho\, \qdot\,\intd r}. \tag{\arabic{equation}a}\label{eq:theata}
\eeq
We plot in Figure~\ref{fig:tadv_theat1} values of the ratio $t'\adv/t'\heat$ for a range of mass accretion rates. To a factor of $\sim\!2$ the condition $t'\adv/t'\heat \sim 1$ at $\lcrit$. However, the specific value of the ratio changes from about $0.7$ to almost $1.2$ over the range of mass accretion rates -- more than a $50\%$ increase. Clearly, this definition does not exactly correspond to the critical luminosity.

The second definition of advection and heating times is due to \citet{thompson05}, who defined the times locally as
\beq
\addtocounter{equation}{-1}
t''\adv = \left|\frac{r_P}{v}\right|, \tag{\arabic{equation}b}\label{eq:tadvb}
\eeq
\beq
\addtocounter{equation}{1}
t''\heat = \left|\frac{P/\rho}{\qdot}\right|, \tag{\arabic{equation}b}\label{eq:theatb}
\eeq
where $r_P = (\intd \ln P / \intd r)^{-1}$ is the pressure scale-height at a given radius. In Figure~\ref{fig:tadv_theat2} we plot the ratio $t''\adv/t''\heat$ as a function of radius for two values $\mdot$. For the smaller $\mdot$, the ratio spikes to high values below the gain radius, while it stays consistently below unity above it, even though the critical luminosity was reached. Contrary to this, the higher $\mdot$ calculation tolerates $t''\adv/t''\heat > 1$ in the gain layer for a number of sub-critical luminosities. 

Neither definition of advection and heating times yields the desired exact correspondence with $\lcrit$. We tried to adjust the definitions in many ways, but failed to produce a condition that would be valid over a wide range of parameters (e.g., $\mdot$, $M$, and $\rnu$).

\subsection{Antesonic condition}
\label{sec:coronal}

\begin{figure*}
\plotone{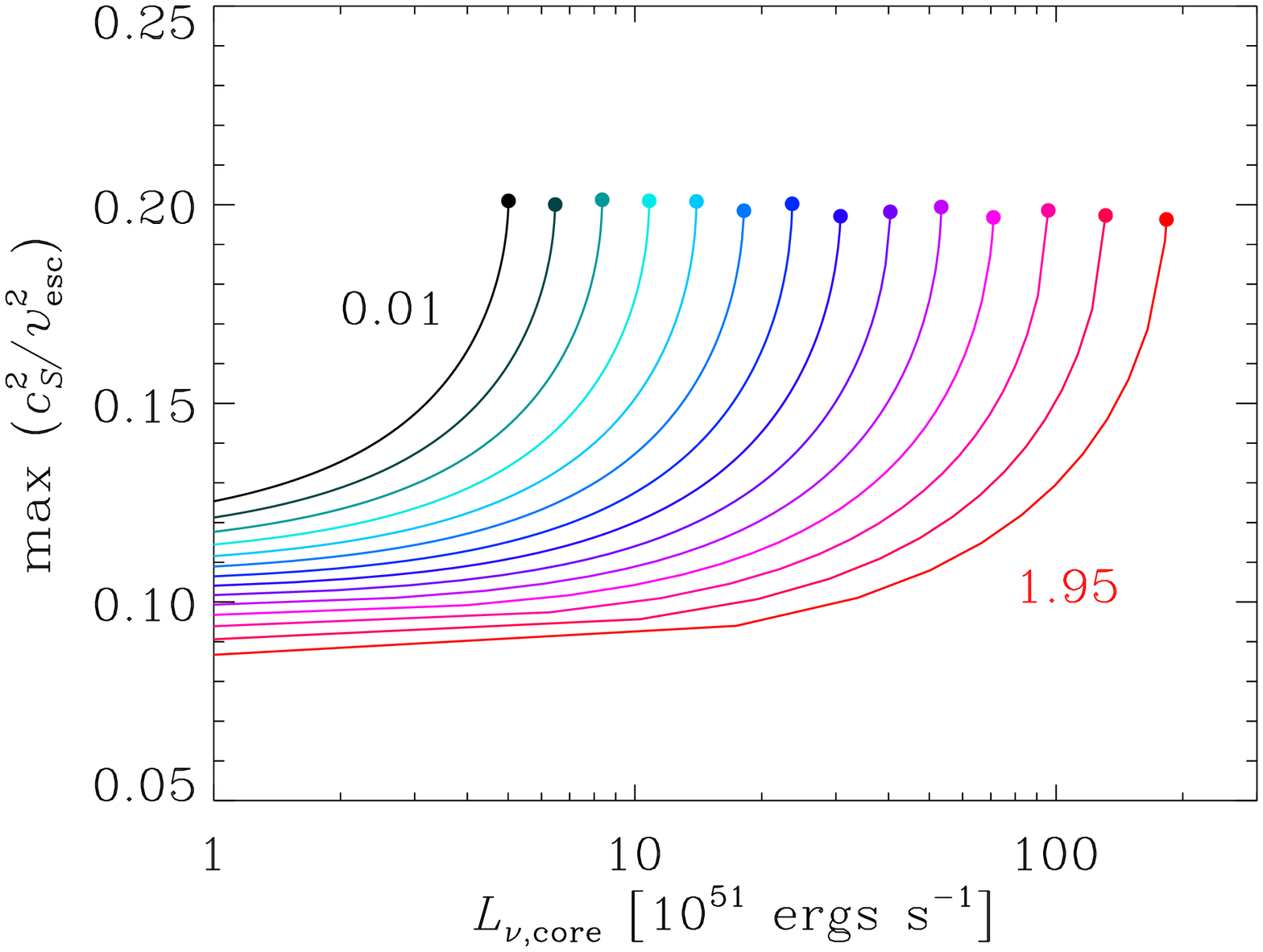}
\caption{The antesonic condition. Maximum value of the ratio of the adiabatic sound speed $c_S$ to the escape velocity $\vesc$ as a function of $\lcore$ (individual lines). Colors distinguish between different mass accretion rates and have the same meaning as in Figure~\ref{fig:tadv_theat1}. Filled circles mark $\lcrit$. As discussed in Section~\ref{sec:isot_acc} in the case of polytropic accretion flows, although the critical condition plotted here appears to be a local condition on the sound speed in the accretion flow, our analysis in Section~\ref{sec:isot_acc} and in Figures~\ref{fig:rshock_model} and \ref{fig:sono_detail} shows that the quantity $\max\,(c_S^2/v_{\rm esc}^2)$ is a merely a scalar metric for solution space of Euler equations and thus it is a global condition.}
\label{fig:soundspeed}
\end{figure*}

In Section~\ref{sec:isot_acc} we showed that the antesonic condition $(\ct^{\rm crit})^2/v_{\rm esc}^2 = 0.1875$ is equivalent to the critical condition for isothermal accretion with pressure-less free fall at the shock, and it can be generalized to $\max\, (c_S^2/v_{\rm esc}^2) \approx 0.19\Gamma$ for a polytropic EOS. Does a similar condition hold also in the more realistic calculations? In order to test this, we studied the maximum value of the ratio of the adiabatic sound speed $c_S$ to the escape velocity, $\max\, (c_S^2/v_{\rm esc}^2)$, similarly to the polytropic case. We plot in Figure~\ref{fig:soundspeed} how this value changes with core luminosity and mass accretion rate. We find that the value of this parameter is surprisingly constant at  $\max\, (c_S^2/v_{\rm esc}^2) \approx 0.20$ over almost three orders of magnitude in $\mdot$ at $\lcore=\lcrit$. Furthermore, we took all $1350$ critical luminosities used to construct Figure~\ref{fig:crit_curve} and calculated a histogram of values of $\max\, (c_S^2/\vesc^2)$ at $\lcrit$. We find that the histogram can be very well fit with a Gaussian with a maximum at $0.193$ and with a width of $0.009$, that is only $5\%$ of its value! While from Figure~\ref{fig:soundspeed} we observe in critical values a slight trend with $\mdot$, we also see from the same Figure that the numerical noise contributes significantly to the total scatter\footnote{Although critical luminosities can be determined very precisely with our code, other parameters like critical shock radii are determined less precisely. This is because $|\partial \rs / \partial \lcore|  = \infty$ at the critical luminosity, as can be inferred from Figure~\ref{fig:rshock}. Hence, a small uncertainty in $\lcrit$ is dramatically amplified in $\rs$ and related quantities, as can be noted in Figures~\ref{fig:tadv_theat1} and \ref{fig:soundspeed}.}.

How does the value of $\max\, (c_S^2/v_{\rm esc}^2)$ depend on input physics? We determined $\max\, (c_S^2/v_{\rm esc}^2)$ for data of Figure~\ref{fig:diff_phys} and found that there is some variability. For constant heating (solid red line) the ratio is about $0.15$, while for no cooling (solid blue line) and heating proportional to $r^{-3}$ (red dashed line) the mean value is about $0.26$. For all other cases of different physics, the mean value ranges from $0.19$ to $0.22$, quite close to the result from our fiducial calculation. Furthermore, we also find that increasing $\Upsilon$, the fraction of free-fall velocity of the incoming matter (eq.~[\ref{eq:shock_jump_ene}]), from $0.25$ to unity increases the value of the ratio from $0.19$ to about $0.22$. We found in Section~\ref{sec:isot_acc} that for polytropic EOS the antesonic ratio is linearly proportional to adiabatic index $\Gamma$. With this information at hand, the close agreement between $\max\, (c_S^2/v_{\rm esc}^2)$ measured in our fiducial calculation and the value in isothermal accretion can be regarded as a coincidence. The fiducial calculation is not exactly isothermal ($\Gamma > 1$), which compensates for lowering $\Upsilon$ to $0.25$. Even so, despite the fact that $\lcrit(\mdot)$ changes dramatically with the large changes to the input physics considered in Figure~\ref{fig:diff_phys}, $\max\, (c_S^2/\vesc^2)$ is virtually constant $\sim\! 0.2$.

Another question is how does the radius at which $c_S^2/v_{\rm esc}^2$ peaks relate to other significant radii of the problem like the gain and shock radii? In the isothermal calculation this critical radius coincides with the shock radius, because at this position the escape velocity is smallest. In the full problem this is clearly not the case as shown in the lower left panel of Figure~\ref{fig:profiles}. Instead we find that radius of maximum $c_S^2/v_{\rm esc}^2$ nearly coincides with the gain radius, but that they are offset by as much as $2\%$ in both directions. For the case of no cooling, the maximum occurs at the neutrinosphere, and there is no gain radius. 
As the radius of maximum $c_S^2/v_{\rm esc}^2$ closely tracks the gain radius in the fiducial calculation, their difference might not be noticeable at all in simulations.

\citet{bhf95} proposed a similar ``coronal'' condition for the explosion, where $T$ of the substantial amount of matter needs to be higher than the local escape temperature. However, we have demonstrated that the transition from accretion to wind does not require sound speeds above the escape speeds. Instead, the sound speed vs.\ escape speed condition is only a manifestation of the inability to satisfy the shock jump conditions. Indeed, as Figure~\ref{fig:profiles} shows, the Bernoulli integral is negative at explosion as noted also by \citet{bhf95}.

\section{Analytic toy model}
\label{sec:toy_model}

In this Section we discuss an analytical toy model that retains the important properties of the full numerical steady-state solution (as characterized by \citealt{bg93} and \citealt{yamasaki05,yamasaki06}), namely the existence of $\lcrit$ and two branches of solutions with different shock radii. 
We are able to derive explicit analytic expressions for $\lcrit$ and $\rs$ (eqs.~\ref{eq:toy_rshock_sol} \& \ref{eq:toy_lcrit}). The details of the construction of the toy model are given in Appendix~\ref{app:toy_model}. Here we discuss the most crucial points of the model and the results we obtain.

\subsection{Construction of the toy model}

\begin{figure*}
\center{\includegraphics[width=0.8\textwidth]{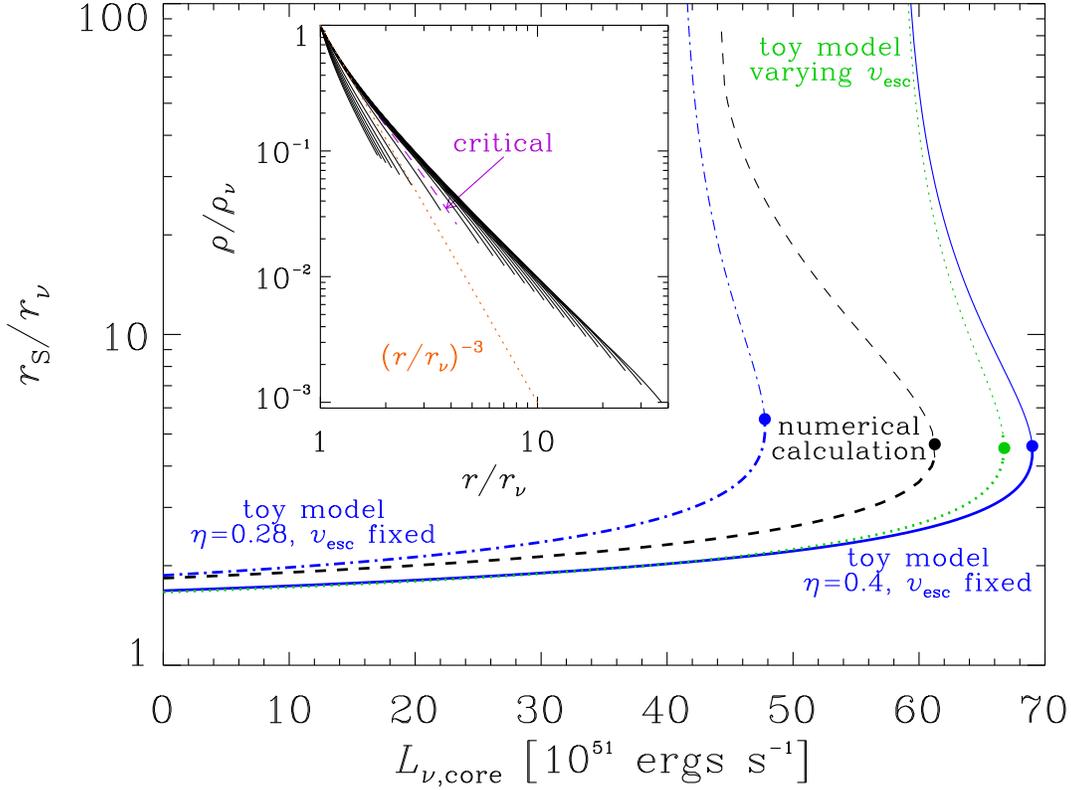}}
\caption{Shock radii $\rs$ as a function of $\lcore$ for the toy model described in Section~\ref{sec:toy_model} and Appendix~\ref{app:toy_model}. The solid blue line shows the analytic solution given by equation~(\ref{eq:toy_rshock_sol}), the dotted green line is a numerical solution to equation~(\ref{eq:rshock_pol}) where the dependence of $\vesc$ on $\rs$ was included, and the dashed black line is the numerical solution without fixing the density profile. All have $\eta=0.4$ (eq.~[\ref{eq:cooling_eff}]). The dash-dotted blue line shows the solution to equation~(\ref{eq:toy_rshock_sol}) with $\eta=0.28$. Critical points are marked with filled circles, and the upper and lower solution branches are shown with thin and thick lines, respectively. The inset plot shows density profiles from the numerical solution for a range of  $\lcore$. In order of increasing shock radii, the solid lines are profiles separated by $10^{52}\,$\ergsec\ up to $\lcrit$ (dashed purple line). Solid lines for higher shock radii (upper solution branch) are for $\lcore$ decreased in steps of $10^{51}$\,\ergsec\ down to the minimum for the upper solution branch. The dotted orange line shows the density profile assumed in the analytic toy model, $\rho/\rho_\nu = (r/\rnu)^{-3}$.}
\label{fig:toy_rshock}
\end{figure*}

Our toy model is based on the assumption of hydrostatic equilibrium, conservation of energy, and the fact that $\lcrit$ occurs when the shock jump conditions cannot be simultaneously matched to any solution of the Euler equations (Section~\ref{sec:isot_acc}). We find that we can realistically model $\lcrit$ using the solution to simple algebraic equations. Specifically, we obtain a quadratic equation for $\rs$, which gives $\lcrit$ when the discriminant vanishes.  The cost of this simplicity is that we must introduce a number of approximations, albeit reasonable, and are forced to make our model somewhat internally inconsistent. Therefore, our toy model produces only qualitative agreement with the numerical solution. Even this approach, however, enables us to grasp the basic elements of the physics involved and to extend them to more complicated and more complete models (see Appendix~\ref{app:toy_model}).

A primary simplification of the model is that we take the heating and cooling functions to depend only on radius $r$: $\mathcal{H}(r) = a\lcore/r^2$ and $\mathcal{C}(r) = b/r^4$. While $\mathcal{H}$ is a good approximation,  $\mathcal{C}$ lacks the important self-regulatory feature of realistic cooling, which is proportional to $T^6$, that an excess in the internal energy can be quickly radiated away. In other words, when no heating is present, the material cannot radiate away via neutrino cooling more energy than its total internal energy content. To quantify this constraint on the cooling function $\mathcal{C}(r)$, we assume that a cold parcel of matter falls from large distance through the shock to the neutrinosphere at $\rnu$. When no heating is present, the parcel will radiate a fraction of its gravitational potential energy
\beq
\int_{\rnu}^\infty \frac{4\pi r^2\rho}{\mdot}\mathcal{C}(r)\,\intd r = \eta\frac{GM}{\rnu},
\label{eq:cooling_eff}
\eeq
where $\eta$ is the cooling efficiency. If the sole source of neutrino flux is the accretion flow itself (ie.\ there is no $\lcore$), then $\eta\leq1$. In the presence of heating we still consider equation~(\ref{eq:cooling_eff}) to be valid, but with a different (and potentially higher) value of $\eta$. A condition similar to equation~(\ref{eq:cooling_eff}) was employed by \citet{fernandez09a} to limit excessive cooling due to discreteness effects. 

\begin{figure}
\plotone{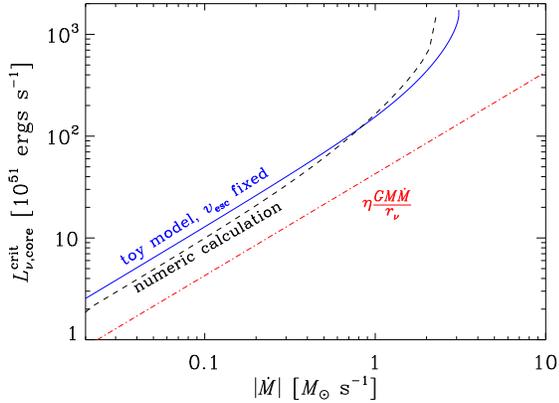}
\caption{Comparison of $\lcrit$ from the numerical solution to eqs.~(\ref{eq:pres_balance_simple}) and (\ref{eq:energy_balance_simple}) (dashed black line) to the analytic solution for the toy model provided in eq.~(\ref{eq:toy_lcrit}) (solid blue line; only the solution with lower $\lcrit$ is shown).  The dash-dot red line shows $\eta GM|\mdot|/\rnu$ ($\eta=0.4$; compare with Fig.~\ref{fig:toy_rshock}).}
\label{fig:toy_lcrit}
\end{figure}

Shock radii as a function of $\lcore$ and $\lcrit$ as a function of $\dot{M}$ are shown in Figures~\ref{fig:toy_rshock} and \ref{fig:toy_lcrit},
respectively, at two levels of approximation.  In both, the results from the numerical solution to equations
(\ref{eq:pres_balance_simple}) and (\ref{eq:energy_balance_simple}) are labelled ``numerical'' and the results
from a truly analytic toy model are labelled ``toy.''  We find that equations
(\ref{eq:toy_rshock_sol}) and (\ref{eq:toy_lcrit}) from the toy model, which provide explicit analytic solutions
for $\rs$ and $\lcrit$, reproduce the qualitative behavior of the more
complete solution remarkably well.   

\subsection{Relation of $\lcrit$ to $\lacc$}

Is it possible to reach $\lcrit$ with just $\lacc$? 
Within the framework of the toy model, the answer to the question of whether $\lcrit$ is always higher than $\eta GM|\mdot|/\rnu$ depends on the ratio of the term in the square brackets in equation~(\ref{eq:toy_lcrit_small_mdot}) (which is always positive and larger than unity) to the denominator, $2\pi\rnu\rho_\nu a$, an upper limit of the optical depth $\tau_\nu$ for our fixed density profile. For small $\eta$ we can thus write (see eq.~\ref{eq:toy_lcrit})
\beq
\lcrit \approx \frac{\eta}{\tau_\nu}\frac{GM|\mdot|}{\rnu},
\label{eq:toy_lcrit_tau}
\eeq
where the exact factor in front of the right-hand side is comparable to, but always higher than unity. For parameters given in Table~\ref{tab:toy_coeff} we get $\tau_\nu \approx 2\pi\rnu\rho_\nu a \doteq 0.36$ and $\lcrit > \eta GM|\mdot|/\rnu$. We confirmed this result for all $0 < \eta \leq 1$ by inspecting graphs similar to Figure~\ref{fig:toy_lcrit}.

From equation~(\ref{eq:toy_lcrit_tau}) we see that if $\tau_\nu \sim 1$, then even if the whole region between the neutrinosphere and the accretion shock is immersed in a neutrino flux corresponding to the maximum integrated neutrino cooling $\eta GM\mdot/\rnu$, the critical luminosity cannot be attained. Furthermore, if $\eta \gtrsim \tau_\nu$ then the critical luminosity is even higher than the maximum gravitational energy release rate. In reality, the accretion luminosity available for absorption will be substantially lower than in the ideal scenario described, and hence sufficient core neutrino luminosity independent of the accretion flow is required to reach an explosion. One could rightfully argue that the optical depth is not constrained within our toy model and that at high $\tau_\nu$ some effect might come to play that lowers the critical luminosity below the accretion luminosity. However, as demonstrated by our numerical solution to the more detailed version of the problem described in Section~\ref{sec:crit_prop} and depicted in Figure~\ref{fig:crit_acc}, $\lcrit$ is in fact much higher than the maximum possible accretion luminosity.

\subsection{$\lcrit$ as a function of dimension}
\label{sec:lcrit_multid}

\begin{figure}
\plotone{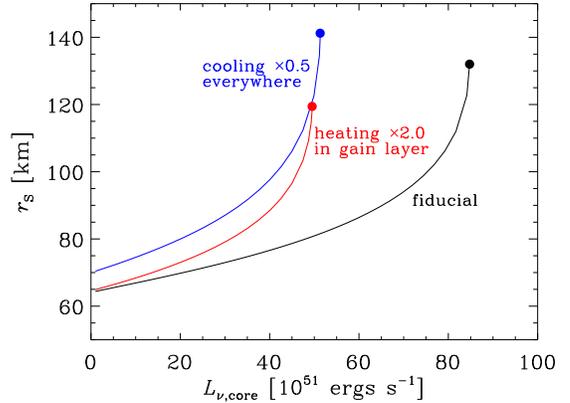}
\caption{Illustration of effects of modified heating and cooling on $\rs$ (solid lines) and $\lcrit$ (filled circles). Decreasing cooling (blue) and increasing heating (red) both act to decrease $\lcrit$, and to increase $r_{\rm S}$ at fixed $\lcore$ with respect to the fiducial calculation (black).
However, lower cooling efficiency at $\lcrit$ increases $r_{\rm S}$, whereas higher heating decreases $r_{\rm S}$ at $\lcrit$.}
\label{fig:multid}
\end{figure}

\citet{ohnishi06} and \citet{iwakami08} investigated the shock radii and stability of 2D and 3D accretion flows with heating and cooling based on steady-state models of \citet{yamasaki05}. They found that going from 1D to 2D increases the shock radii even when the region is convectively stable, and that this effect is more pronounced for higher $\lcore$. They also noticed that their $\lcore = 6.0\times 10^{52}$\,\ergsec\ simulation exploded even though it was stable in spherical symmetry and steady-state calculations. The common opinion shared by \citet{ohnishi06} and \citet{iwakami08} is that the SASI is responsible for the shock revival, essentially lowering $\lcrit$. However, we have shown in our toy model (Figure~\ref{fig:toy_rshock} and in more involved calculations, Figure~\ref{fig:diff_phys}) that decreasing the cooling efficiency $\eta$ lowers $\lcrit$, as displayed in equation~(\ref{eq:toy_lcrit_tau}). In order to illustrate this point more clearly, we plot in Figure~\ref{fig:toy_rshock} shock radii for somewhat lower cooling efficiency $\eta$. We see that not only is $\lcrit$ decreased, but also that  the shock radii consistently increase over the whole range of allowed core luminosities and that at fixed luminosity the increase of $\rs$ is both relatively and absolutely more prominent. For lower $\eta$ the shock radius at $\lcrit$ increases only modestly, as can be seen from Figure~\ref{fig:toy_rshock}.

To put this on more quantitative grounds, \citet{ohnishi06} found non-exploding dynamical 1D models for neutrino luminosities up to $\lcore = 6.5\times 10^{52}$\,\ergsec, while the models in 2D did not explode up to $\lcore = 5.5\times 10^{52}$\,\ergsec. These numbers can be identified as $\lcrit$ in 1D and 2D, respectively.  Interpreted using equation (\ref{eq:toy_lcrit_tau}), these values of $\lcrit$ suggest that $\eta$ decreased by about $15\%$ in going from 1D to 2D. However, from the physics of the toy model and our derivation of equations (\ref{eq:toy_lcrit_tau}) and (\ref{eq:toy_rshock_sol}),
we find that changes in $\eta$ are always accompanied by changes in $\rs$ at $\lcrit$, such that the ratio of two critical shock radii $\rs^{\rm crit}$ for two cooling efficiencies $\eta_1$ and $\eta_2$ at their respective $\lcrit$ is
\beq
\frac{\rs^{\rm crit}(\eta_1)}{\rs^{\rm crit}(\eta_2)} = \sqrt{\frac{\eta_2}{\eta_1}}.
\label{eq:toy_rshock_crit_eta}
\eeq
This equation implies that a decrease in $\eta$ of $15\%$ should correspond to an increase in $\rs$ at $\lcrit$
by $8-9\%$, in good agreement with the larger shock radii found by \citet{ohnishi06} in their 2D simulations (see their Fig.~4).

Similarly, \citet{nordhaus10} found that in 3D $\lcrit$ drops to about $60\%$ of the 1D value (see our Figure~\ref{fig:nordhaus_comp}). What modification of the physics would be necessary to explain such a decrease? As shown in Figure~\ref{fig:multid}, we find within our fiducial model that reducing cooling by a factor of two yields the desired drop of $\lcrit$. Simultaneously, the entropy in the gain layer at fixed $\lcore$ increases by about $30\%$ and $\rs$ at $\lcrit$ increases by about $5$ to $7\%$. We obtain similar changes in $\lcrit$ and entropy if we increase the net heating $\qdot$ by a factor of two in the gain layer, thus simulating in a crude way increased energy deposition as might be provided by convection. However, in this case $\rs^{\rm crit}$ {\em decreases} by about $8$ to $10\%$, although at fixed $\lcore$ $\rs$ is larger. \citet{yamasaki06} performed a calculation similar to ours, which included convection in a phenomenological way, that essentially flattens the entropy gradient, and they found that while $\lcrit$ decreased substantially, the shock radii also decreased.

We note here, that $\eta$ is the only free parameter within our toy model and therefore the effects of all unknown physics get projected into $\eta$. Even so, the decrease in the critical luminosities and the increase of shock radii as a function of dimension observed in simulations \citep[e.g.][]{ohnishi06,iwakami08,murphy08,nordhaus10} may occur because higher dimensional flows have less efficient cooling,  and not because of SASI or any other instability. This is supported by \citet{nordhaus10} who did not see in their 3D simulations as vigorous SASI activity as in their 2D simulations, yet $\lcrit$ was lower than in 1D. Furthermore, while both decreased cooling and increased heating give lower $\lcrit$, higher entropy, and higher $\rs$ at fixed $\lcore$, the latter gives lower $\rs$ at $\lcrit$, which we consider incompatible with results of hydrodynamical simulations. Finally, while it is somewhat difficult to disentangle the effects of modified heating and cooling, we think that the cooling is responsible for the reduction in $\lcrit$ because the temperature dependence of cooling offers greater prospects for modifying the structure of the flow, providing lower $\lcrit$, higher $\rs$, and higher entropy. Looking at Figure~\ref{fig:diff_phys}, we see that with no cooling $\lcrit$ can be more than order of magnitude smaller at low $|\mdot|$. As the usefulness of our 1D calculations and the toy model are quite limited for discussing multi-dimensional effects, our work can provide predictions and scaling relations of various quantities that will be useful in understanding the results of future simulations.

\section{Discussion and summary}
\label{sec:summary}


We studied spherically-symmetric accretion flows in order to understand the physics of the critical neutrino luminosity, $\lcrit$, which separates steady-state accretion from explosion --- a formulation of the neutrino mechanism of supernovae. Our results can be summarized as follows.\\

\noindent 1.~We investigated isothermal accretion flows with fixed mass accretion rate $\mdot$ and variable sound speed $\ct$. In Figure~\ref{fig:isot_acc} we showed that there is a maximum, critical sound speed $\ct^{\rm crit}$ which allows for a steady-state shock standoff accretion shock. For $\ct > \ct^{\rm crit}$ the flow cannot conserve mass and momentum simultaneously at the shock and it must move outward on a dynamical timescale to establish a wind solution. For the shock jump conditions relevant for the supernova problem we showed that there is no smooth transition to an outflow solution. We studied the ratio $\ct^2/\vesc^2$ and we find that it is constant at $\ct^{\rm crit}$ and stays always below the value for the sonic point. Thus, for the first time, we derive the ``antesonic'' condition for explosion: if $\ct^2$ exceeds $\simeq 0.19 \vesc^2$ (eq.~[\ref{eq:isot_cond}]), there is no solution for a steady-state accretion flow with a standoff accretion shock, and explosion necessarily results. The physical realization of this transition is likely modified by time-dependent hydrodynamical instabilities.\\

\noindent 2.~The mechanism of $\ct^{\rm crit}$ in the isothermal model does not depend on any particular heating mechanism, and we explicitly showed in Section~\ref{sec:isot_corres} that it is directly analogous to $\lcrit$ observed in more complete steady-state calculations. Specifically, in Figure~\ref{fig:sono_detail} we investigated and classified the structure of accretion flows in the supernova problem as a function of the core neutrino luminosity $\lcore$ with respect to the position of the sonic point and the constraints imposed by the shock jump conditions. We found that the critical solution does not correspond to the sonic point, but it is always marginally close (Figure~\ref{fig:rshock_model}). We showed that the maximum of the ratio $c_S^2/\vesc^2$ at $\lcrit$ is close to $\simeq 0.19$ (Figure~\ref{fig:soundspeed}) and stays constant to within $5\%$ over a wide range of $\mdot$, and masses and radii of the PNS. Thus, our ``antesonic'' condition corresponds directly to $\lcrit$ and is thus superior in monitoring the approach to $\lcrit$ when compared to other heuristic conditions proposed in the literature (Section~\ref{sec:conditions}). For example, the ratio of the advection to the heating time in the accretion flow is not constant along the critical curve (Figures \ref{fig:tadv_theat1} \& \ref{fig:tadv_theat2}). Instead, the physics of $\lcrit(\dot{M})$, and thus the neutrino mechanism of supernovae itself, is identical to the physics of the antesonic condition.\\

\noindent 3.~Extending the previous works of \citet{bg93} and \citet{yamasaki05,yamasaki06} we studied the thermodynamical profiles of the flows for $\lcore$ ranging from $0$ to $\lcrit$ (Figure~\ref{fig:profiles}). We confirm that there are two solutions with different shock radii and different energies for a given $\lcore$ (Figure~\ref{fig:rshock}), which merge at $\lcrit$. We also found that the critical curve robustly exists for a wide range of microphysics, although the exact normalization and slope vary (Figure~\ref{fig:diff_phys}). Finally, we quantified the dependence of $\lcrit$ over a wide range of $\mdot$, and mass and radius of the PNS, providing a useful power-law fit as a function of the key parameters of the problem (Figure~\ref{fig:crit_curve} and eq.~[\ref{eq:lcrit_empir}]). Our results imply that larger PNS radius decreases $\lcrit$, potentially favoring stiff high-density nuclear equations of state (see Section 3.4; eq.~[18]). However, faster contractions of the PNS in soft EOS models can be favorable for explosion by increasing neutrino luminosities and energies.\\

\noindent 4.~We include in our calculations a simple approximation to gray neutrino radiation transport (Figure~\ref{fig:profiles}). We found that the neutrino cooling of the flow above the neutrinosphere, the accretion neutrino luminosity, $\lacc$, lowers $\lcrit$ by a small but significant amount (Figure~\ref{fig:crit_acc}). Thus, it is the core, not the accretion flow, which is responsible for the transition to the explosion. Specifically, we show by analysis of our numerical results and by calculations within our toy model that $\lcrit$ is always higher than the maximum available accretion power $GM\mdot/\rnu$ (Figure~\ref{fig:toy_lcrit} and eq.~[\ref{eq:toy_lcrit_tau}]). Even though $\lcrit$ is always $>\!\!\lacc$, we found in Figure~\ref{fig:nordhaus_comp} that the inclusion of heating by accretion luminosity decreases $\lcrit$ by an amount comparable to the reduction caused by going from 1D to 2D, or from 2D to 3D. \\

\noindent 5.~Our numerical calculations and toy model imply (but do not prove) that the reduction of $\lcrit$ seen in recent 2D and 3D simulations likely arises from an overall reduction in the cooling efficiency allowed by the accretion flows in multi-dimensions (Section~\ref{sec:lcrit_multid} and Figure~\ref{fig:toy_rshock}), and not from from additional heating caused by convection or the presence of shock oscillations.  In particular, while both a reduction in cooling and an increase in heating increase the entropy in the gain layer and decrease $\lcrit$, a reduction in cooling always occurs with an increase in $\rs(\lcrit)$, while an increase in heating results in a decrease in $\rs(\lcrit)$. Because the simulations imply that $\rs(\lcrit)$ is somewhat larger in 2D and 3D with respect to 1D, our equation (\ref{eq:toy_rshock_crit_eta}), derived from the toy model, implies that a decrease in the cooling efficiency of the flow is the cause of the observed decrease in $\lcrit$.  However, more work on this issue is clearly warranted.\\

\section*{Acknowledgements}

This work is supported in part by an Alfred P. Sloan Foundation Fellowship and by NSF grant AST-0908816. We thank Brian Metzger, Christopher Kochanek, Kris Stanek, and John Beacom for discussions and encouragement. We thank Adam Burrows, Hans-Thomas Janka, and Jeremiah Murphy for critical reading of the text.

\appendix
\section{Euler equations, equation of state and neutrino physics}
\label{app:neutrino}

Using thermodynamic expansion of $\varepsilon$ and $P$ we express the radial derivatives of $v$, $\rho$ and $T$ by inverting equations~(\ref{eq:mass_cons}--\ref{eq:energy_tot}). The radial derivatives are
\begin{eqnarray}
\frac{\intd \rho}{\intd r} &=& -\frac{\rho}{2r}\left(\frac{4v^2 - v_{\rm esc}^2}{v^2-c_S^2}\right) + \frac{\qdot}{v}\frac{K}{v^2-c_S^2}  +  \frac{W}{v^2-c_S^2}\left(\frac{\intd \ye}{\intd r}\right),\\
\frac{\intd v}{\intd r} &=& -\frac{v}{2r}\left(\frac{v_{\rm esc}^2 - 4c_S^2}{v^2-c_S^2}\right) -  \frac{\qdot}{\rho}\frac{K}{v^2-c_S^2} - \frac{v}{\rho}\frac{W}{v^2-c_S^2}\left(\frac{\intd \ye}{\intd r}\right),\label{eq:app_momentum}\\
\frac{\intd T}{\intd r} &=& -\frac{\rho Z}{2rC_V}\left(\frac{4v^2 - v_{\rm esc}^2}{v^2-c_S^2}\right)+\frac{\qdot}{vC_V}\left(\frac{v^2-\ct^2}{v^2-c_S^2}\right) -\nonumber\\
& & -\frac{1}{C_V}\left(\frac{\intd \ye}{\intd r}\right) \left[\left(\frac{v^2-\ct^2}{v^2-c_S^2}\right)\left(\frac{\partial \varepsilon}{\partial \ye}\right)_{\rho,T}-\frac{Z}{v^2-c_S^2} \left(\frac{\partial P}{\partial \ye}\right)_{\rho,T} \right].
\end{eqnarray}
Combinations of thermodynamic variables in these equations are defined as
\begin{eqnarray}
C_V &=& \left(\frac{\partial \varepsilon}{\partial T}\right)_{\rho,\ye},\\
\ct^2 &=& \left(\frac{\partial P}{\partial \rho}\right)_{T,\ye},\\
K &=& \frac{1}{C_V}\left(\frac{\partial P}{\partial T}\right)_{\rho, \ye},\\
W &=&  \left( \frac{\partial P}{\partial \ye}\right)_{\rho,T}   - K \left(\frac{\partial \varepsilon}{\partial \ye}\right)_{\rho,T},\\
Z &=&  \frac{P}{\rho^2} - \left(\frac{\partial \varepsilon}{\partial \rho}\right)_{T,\ye},\\
c_S^2 &=& c_T^2 +KZ.\label{eq:app_cs}
\end{eqnarray}
Our form of thermodynamic equations is the same as the Newtonian version of equations of \citet{thompson01}, except that we include dependency on $\ye$ and we do keep $Z$ and $K$ separate, because they are related only for a special case of equation of state.

Our equation of state includes relativistic electrons and positrons with chemical potential and nonrelativistic free protons and neutrons \citep{qian96}. In our neutrino physics we use prescriptions of heating, cooling, opacity and reaction rates for charged-current processes\footnote{We also modified the heating and cooling functions to include electron-positron annihilation and creation, which increased $\lcrit$ in the fiducial calculation by about $2$ to $5\%$.} with neutrons and protons from \citet{schecketal06}. We explicitly calculate the degeneracy parameter of electrons and positrons, but we assume that degeneracy parameter of (anti)neutrinos is zero. We also assume that the energies of neutrinos and antineutrinos do not change between $\rnu$ and $\rs$.

While we do not include nuclear binding energy in our energy shock jump condition (eq.\ [\ref{eq:shock_jump_ene}]), we can assess its effects on $\rs$ and $\lcrit$ with the help of Figure~\ref{fig:sono_detail}. Nuclear binding energy would appear as a positive term on the left-hand side of equation~(\ref{eq:shock_jump_ene}), thus the total energy $\mathscr{B}$ just downstream of the shock would be smaller at a given radius. This corresponds to shifting the green line down in the energy panels of Figure~\ref{fig:sono_detail}. This would lead to smaller shock radii, because the common intersection for the momentum and energy jump conditions would occur at smaller radius. Similarly, $\lcrit$ would increase, because the common intersection could occur at higher $\lcore$. For example, consider the energy panel in lower left part of Figure~\ref{fig:sono_detail}. If we shift the energy shock jump condition (green line) down by a small amount, it will intersect the flow profiles (black lines) exactly at the position of the similar intersection in the momentum panel (vertical dotted lines). Thus, a solution with a shock is possible and $\lcrit$ increases. By moving the green line down even more, the intersection would move to lower radii, thus decreasing $\rs$. We verified that these qualitative assessments are correct by a sample numerical calculation. Finally, we point out that nuclear binding energy was included in similar calculations of \citet{yamasaki06}.

\section{Construction of the toy model}
\label{app:toy_model}

\begin{figure*}
\plottwo{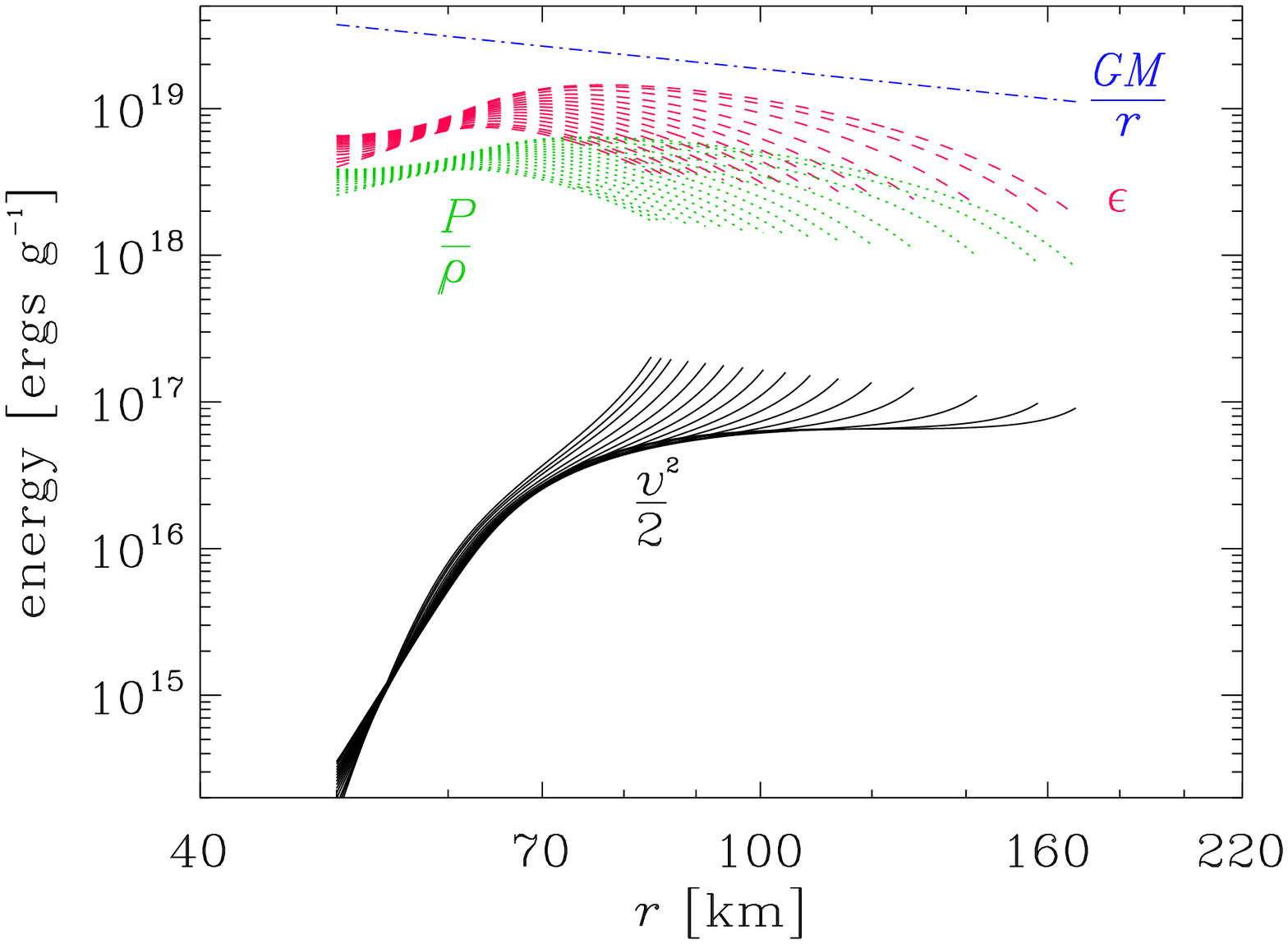}{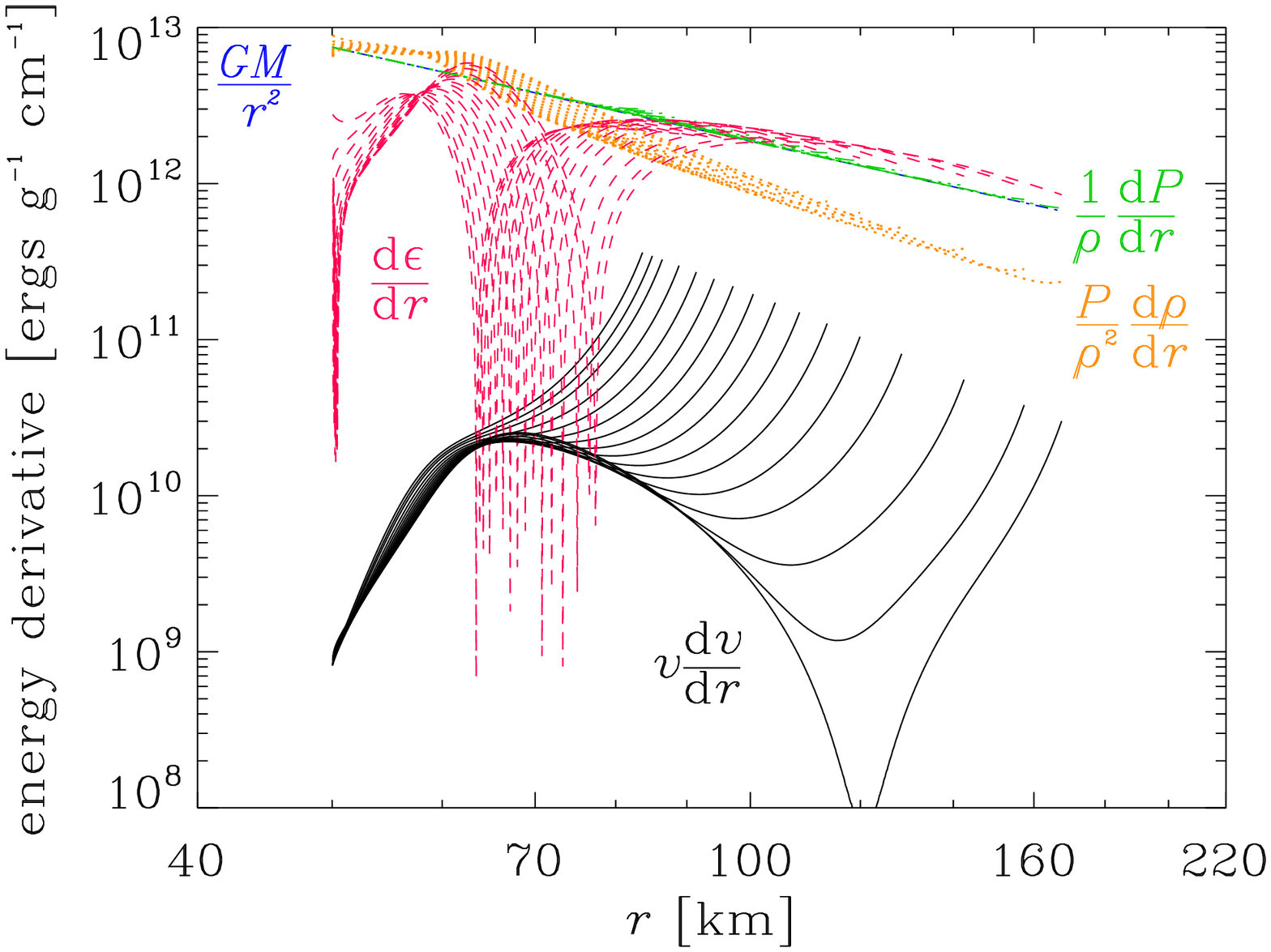}
\caption{{\em Left}: Absolute values of individual components of the Bernoulli integral for models from Figure~\ref{fig:profiles}: kinetic energy $v^2/2$ (solid black), internal energy $\varepsilon$ (dashed red), pressure $P/\rho$ (dotted green) and gravitational potential energy $GM/r$ (dash-dotted blue). {\em Right}: absolute values of the radial derivative of the profiles in the top panel: $v(\intd v/\intd r)$ (solid black), $\intd \varepsilon / \intd r$ (dashed red), $(\intd P/\intd r)/\rho$ (long-dashed green), $P/\rho^2 (\intd \rho / \intd r)$ (dotted orange), and $GM/r^2$ (dash-dotted blue).}
\label{fig:bernoul_comp}
\end{figure*}

It is well known \citep[e.g.][]{janka01} that the region between the accretion shock and the neutrinosphere is very close to hydrostatic equilibrium. Figure~\ref{fig:bernoul_comp} shows individual components of the Bernoulli integral (equation~[\ref{eq:bernoul}]) along with their radial derivatives. We see that contributions of both the kinetic energy term $v^2/2$ to the total energy budget and of the $v(\intd v/\intd r)$ term to the momentum equilibrium are negligible. Thus, we can formulate the steady-state problem without the $v(\intd v/\intd r)$ term in the momentum equilibrium equation and neglect the $v^2$ terms on the left-hand side of the shock jump conditions. In this setting, the flow is hydrostatic and velocity can be formally defined as $v=\mdot/(4\pi r^2 \rho)$. We also set $\intd \ye / \intd r = 0$ and $\intd \lnu / \intd r = 0$, which implies $\tau_\nu \ll 1$. With these simplifications the equations~(\ref{eq:mass_cons}--\ref{eq:energy_tot}) can be reformulated as
\begin{eqnarray}
P(r_1)-P(r_2) &=& - \int^{r_1}_{r_2} \frac{GM\rho(r', \rs)}{r'^2}\intd r' \label{eq:pres_balance_simple}\\
h(r_1)-h(r_2) &=&   \frac{GM}{r_1} - \frac{GM}{r_2} + \int^{r_1}_{r_2} \frac{4\pi r'^2 \rho(r', \rs)}{\mdot}\,\qdot(r',\lcore)\, \intd r',\label{eq:energy_balance_simple}
\end{eqnarray}
where $h = \varepsilon +P/\rho$ is the specific enthalpy of the flow and we explicitly state that the density profile $\rho(r,\rs)$ depends on the shock radius $\rs$. 

As boundary conditions we choose the two shock jump conditions assuming that the matter up the shock is in free fall, specifically $h(\rs) = GM/\rs$ and $P(\rs) = \vesc \mdot/ (4\pi \rs^2)$, and we fix the inner boundary density to $\rho(\rnu) = \rho_\nu$. To make use of the boundary conditions, we set $r_1=\rnu$ and $r_2=\rs$ in equations~(\ref{eq:pres_balance_simple}) and (\ref{eq:energy_balance_simple}). We choose the equation of state of relativistic particles which satisfies $\varepsilon = 3P/\rho$ and hence $h = 4P/\rho$. As discussed in Section~\ref{sec:toy_model}, we choose $\qdot=\mathcal{H}-\mathcal{C}$ with heating and cooling functions defined as
\begin{eqnarray}
\mathcal{H} &=& \frac{a\lcore}{r^2},\\
\mathcal{C} &=& \frac{b}{r^4}.
\end{eqnarray} 
While $\mathcal{H}$ is a good approximation,  $\mathcal{C}$ lacks the important self-regulatory feature of realistic cooling, which is proportional to $T^6$, that an excess in the internal energy can be quickly radiated away. In other words, when no heating is present, the material cannot radiate away via neutrino cooling more energy than its total internal energy content. To quantify this constraint on the cooling function $\mathcal{C}(r)$, we assume that a cold parcel of matter falls from large distance through the shock to the neutrinosphere at $\rnu$. When no heating is present, the parcel will radiate a fraction of its gravitational potential energy
\beq
\int_{\rnu}^\infty \frac{4\pi r^2\rho}{\mdot}\mathcal{C}(r)\,\intd r = \eta\frac{GM}{\rnu},
\eeq
where $\eta$ is the cooling efficiency. If the sole source of neutrino flux is the accretion flow itself (ie.\ there is no $\lcore$), then $\eta\leq1$. In the presence of heating we still consider equation~(\ref{eq:cooling_eff}) to be valid, but with a different (and potentially higher) value of $\eta$. A condition similar to equation~(\ref{eq:cooling_eff}) was employed by \citet{fernandez09a} to limit excessive cooling due to discreteness effects.

We verified with our relaxation code that our choices of equation of state and heating and cooling along with an assumption of hydrostatic equilibrium do not fundamentally change the phenomenology of the problem. Namely, even in this highly simplified version of the problem the critical neutrino luminosity and the two branches of solutions are still present.

We construct our toy model by evaluating the momentum balance (equation~[\ref{eq:pres_balance_simple}]) and enthalpy conservation (equation~[\ref{eq:energy_balance_simple}]) between the neutrinosphere and the shock and then solving for $\rs$. In order to have an analytically tractable result we fix the density profile to be $\rho(r) = \rho_\nu (r/\rnu)^{-3}$. However, as the problem is fully determined and the density profile is an integral part of the solution, an ad hoc prescription of the density profile violates one of the boundary conditions. We choose to keep the boundary conditions on enthalpy and pressure, $h(\rs) = GM/\rs$ and $P(\rs) = \mdot\vesc/(\pi \rs^2)$, and to violate the related condition on density $\rho(\rs) = \mdot \vesc / (\pi GM \rs)$. This internal inconsistency means that our toy model will produce only qualitative agreement with the corresponding numerical solution. However, our point is to illustrate basic ingredients and physical principles of steady-state accretion, not to provide any numeric values.

Integrating equations~(\ref{eq:pres_balance_simple}--\ref{eq:energy_balance_simple}) and applying the boundary conditions we get a polynomial equation for the shock radius
\beq
\frac{2\pi \rnu^3 \rho_\nu }{\mdot} \left(\frac{a\lcore }{\rnu^2} - \frac{b}{2\rnu^4}\right) \left(\frac{\rs}{\rnu}\right)^4 + \left( {\frac {\mdot\vesc}{\pi\rho_\nu\rnu^2}}-2 {\frac {\pi \rho_\nu \rnu^3}{\mdot}  \frac{a\lcore}{\rnu^2}} \right) \left(\frac{\rs}{\rnu}\right)^2-\frac{GM}{\rnu}+{\frac {\pi \rho_\nu b}{\rnu\mdot}}=0,
\label{eq:rshock_pol}
\eeq
where we neglect the dependence of the free-fall velocity $\vesc$ on the shock radius $\rs$ to get an analytic solution. We will discuss how this approximation changes the toy model results further below. With this assumption, equation~(\ref{eq:rshock_pol}) is solved for the shock radius yielding

\begin{eqnarray}
\left(\frac{\rs}{\rnu}\right)^2  &=& \left(\frac{2a\lcore}{\rnu^2} - \frac{b}{\rnu^4}\right)^{-1} \times  \left[ \frac{a\lcore}{\rnu^2} -\frac{\vesc\mdot^2}{2\pi^2 \rho_\nu^2\rnu^5}\right. \pm \nonumber \\ 
& \pm & \left. \sqrt{\qdot(\rnu)^2  + \frac{2GM\mdot}{\pi\rho_\nu\rnu^4}\left(\frac{a\lcore}{\rnu^2} - \frac{b}{2\rnu^4}\right) - \frac{\vesc\mdot^2}{\pi^2\rho_\nu^2\rnu^5}\frac{a\lcore}{\rnu^2} + \frac{\vesc^2\mdot^4}{4\pi^4\rho_\nu^4\rnu^{10}} }  \right].
\label{eq:toy_rshock_sol}
\end{eqnarray}
While it is not possible to readily provide obvious physical interpretation to the individual terms in equation~(\ref{eq:toy_rshock_sol}), we can say that they are combination of heating and cooling terms that include coefficients $a$ and $b$, which measure the strength of heating and cooling, the gravitational potential energy term proportional to $GM/\rnu$, and the ram pressure term that is proportional to $\vesc\mdot$.

Our solution for the square of the shock radius has two critical points. The first critical point occurs when the denominator vanishes, $\lcore = b/2a\rnu^2$. In this case one of the solutions stays finite and the other goes from negative (and hence unphysical) to infinite. This corresponds to the minimal neutrino luminosity that allows two solutions for the shock radius\footnote{One might be tempted to interpret $\int_{\rnu}^\infty 4\pi r^2 \rho\qdot/\mdot \intd r=0$ as an equivalent and general condition for the minimum luminosity that gives two solutions for the shock radius. However, this is a mere coincidence arising from our choice of density profile and equation of state that cancels several terms in equations~(\ref{eq:pres_balance_simple}) and (\ref{eq:energy_balance_simple}) when $\rs \rightarrow \infty$.}. Note that this happens for a finite and positive value of $\lcore$.

The second critical point occurs when the discriminant vanishes. At this point the two solution branches merge and there is no solution for parameter combinations that give a negative discriminant. Thus, the neutrino luminosity corresponding to this critical point can be readily identified as the critical luminosity $\lcrit$. The equation for the critical curve can be obtained by setting the discriminant in equation~(\ref{eq:toy_rshock_sol}) to zero and solving the quadratic equation for the neutrino luminosity
\beq
a\lcrit = \frac{b}{\rnu^2} + \frac{GM|\mdot|}{\pi\rho_\nu \rnu^2} + \frac{\vesc\mdot^2}{2\pi^2\rho_\nu^2\rnu^3} \pm \frac{GM|\mdot|}{\pi\rho_\nu \rnu^2}\sqrt{1 - \frac{\pi \rho_\nu b}{GM\mdot} + \frac{\vesc b}{\rnu (GM)^2} - \frac{\vesc\mdot}{\pi\rho_\nu\rnu GM}},
\label{eq:toy_lcrit}
\eeq
where the physically relevant critical luminosity is the smaller of the two solutions. To explain this result, let us keep for the moment only the first two terms on the right-hand side. Merging the terms with $a$ and $b$ to $\qdot^{\rm crit}$ and realizing that $\mdot = 4\pi \rnu^2 \rho_\nu v(\rnu)$, we can write $\qdot^{\rm crit}\rnu/v(\rnu) \sim GM/\rnu$, and $c_S^2 \sim \qdot\rnu/v$ (similar to the discussion in Section~\ref{sec:crit_prop}). Combining these expressions shows that $(c_S^{\rm crit})^2 \sim GM/\rnu$, which is basically equivalent to the critical condition for isothermal accretion as given by equation~(\ref{eq:isot_cond}). The additional terms on the right-hand side arise because within the framework of the toy model we were able to calculate the critical condition {\em exactly}.

We note at this point that the analytic and relatively short form of equations~(\ref{eq:rshock_pol}--\ref{eq:toy_lcrit}) is a result of a particular choice of the density profile and the heating and cooling functions. It is indeed possible to repeat the derivations with different power-law indices for heating and cooling and even with a density profile prescription that satisfies the outer boundary condition on density, which would make the toy model internally consistent. The final polynomial equation would be of higher order and generally without analytic solution at all. Even without an analytic solution it is possible to determine the two critical points using Descartes' rule of signs (appearance of the second solution branch for the shock radius) and setting the polynomial discriminant to vanish (maximum neutrino luminosity that gives steady-state solution). However, we decide not to discuss these more complicated toy models as the lengthy equations are not more illuminating than the simplest version we present here.

We set $b = \eta GM|\mdot|/\pi\rho_\nu$ as discussed in Section~\ref{sec:toy_model} and rewrite equation~(\ref{eq:toy_lcrit}) as
\beq
\lcrit = \frac{GM|\mdot|}{\rnu} \frac{1}{\pi\rnu\rho_\nu a} \left[ 1+\eta - \frac{\vesc\mdot}{2\pi\rnu\rho_\nu GM} \pm \sqrt{\left(1+\eta\right)\left(1 - \frac{\vesc\mdot}{\pi\rnu\rho_\nu GM} \right)}  \right],
\label{eq:toy_lcrit_eta}
\eeq
where the physically relevant solution is the one that gives lower critical luminosity. 

With values of other parameters of the problem given in Table~\ref{tab:toy_coeff} we can test how our toy model compares with our numerical solution with the same input physics. Figure~\ref{fig:toy_rshock} shows the shock radius $\rs$ as a function of $\lcore$ for $\mdot = -0.5\,\msun\ \invs$. We compare the analytic solution given by equation~(\ref{eq:toy_rshock_sol}) (solid line) to the numerical solution of the simplified problem (dashed line). With the dotted line we also plot the numerical solution to equation~(\ref{eq:rshock_pol}), which differs from the analytic solution by accounting for the fact that $\vesc$ is a function of the shock radius $\rs$. Both analytic and numerical solution to equation~(\ref{eq:rshock_pol}) agree very well within each other and hence we prove the assumption of constant $\vesc$ is justified. The agreement between the numerical solution without fixed density profile and the analytic solution is, however, only qualitative. From the inset plot, which shows the density profiles of the numerical solution, we see that the density profiles are to first order power laws with the slope dependent on the shock radius. This explains why there is a mismatch of $\sim 8\times 10^{51}$\,\ergsec\ in the critical luminosity between the numerical and analytic solution. There is also a slight curvature to the numerically obtained density profiles that results in these shock radii always being slightly higher than the analytic ones. Despite the obvious inadequacies we are able to very well reproduce the numerical shock radii with our analytic model.

Figure~\ref{fig:toy_lcrit} shows the comparison of the analytic critical curve given by equation~(\ref{eq:toy_lcrit_eta}) with the numerical solution to the simplified problem. We see a very good agreement both in the absolute value and in the slope for $|\mdot| \lesssim 1\,\msun\ \invs$. At higher $|\mdot|$, we see an upturn in both the analytic and numerical critical curves. In fact, the analytic critical curve ends when the term inside the square root vanishes. We were not able to get a smoothly extending numerical critical curve above $\sim 2.5\,\msun\ \invs$, which suggests that it is absent or at least substantially changed at very high mass-accretion rates. However, we did not investigate this issue further as we do not see anything as nearly dramatic in the realistic critical curves presented in Section~\ref{sec:crit_prop} and it is thus an artifact of our simplistic assumptions in the toy model. Furthermore, the mass-accretion rates where the critical curve ends are fairly high and steady-state approximation ceases to be valid in this area of parameter space.

\begin{table}
\begin{center}
\caption{Values of parameters used in the toy model.}
\label{tab:toy_coeff}
\begin{tabular}{ccl}
\hline\hline
Parameter & Value & Description \\
\hline
 $M$ & $1.2\,\msun$ & neutron star mass \\
 $\rnu$  & $40$\,km & neutrinosphere radius\\
 $\rho_\nu$ & $3 \times 10^{10}\,{\rm g\ cm^{-3}}$ & density at the neutrinosphere\\
 $\vesc$ & $-7.31\times 10^9\,{\rm cm}\ \invs $ & velocity just outside the shock\\
 $\eta$ & $0.4$ & cooling efficiency\\
 $a$  & $6.45 \times 10^{-19}\,{\rm cm^2\ g^{-1}}$ & heating function coefficient\\
 $b$ & $\eta\frac{GM|\mdot|}{\pi\rho_\nu}$ & cooling function coefficient\\
\hline
\end{tabular}
\end{center}
\end{table}

To make a direct comparison with the numerical results of the full fiducial model presented in Section~\ref{sec:crit_prop} we expand the square-root term in equation~(\ref{eq:toy_lcrit_eta}) in the limit of low mass-accretion rate ($|\mdot| \lesssim 1\,\msun\ \invs$) and low cooling efficiency ($\eta \lesssim 1$) to get
\beq
\lcrit \approx \eta\frac{GM|\mdot|}{\rnu}\frac{1}{2\pi\rnu\rho_\nu a} \left[1 + \frac{\eta}{4}\left(1 +\frac{1}{\eta}\frac{\vesc\mdot}{\pi\rnu\rho_\nu GM}\right)^2 \right],
\label{eq:toy_lcrit_small_mdot}
\eeq
which is correct up to second order. The term in the square brackets is always positive and larger than $1$. The denominator can be interpreted as an upper limit on the optical depth $\tau_\nu$, because $\tau_\nu \approx \int_{\rnu}^\infty 4\pi\rho a \intd r = 2\pi\rnu\rho_\nu a$ for our fixed density profile.

We conclude with an explanation of derivation of equation~(\ref{eq:toy_rshock_crit_eta}). The discriminant in equation~(\ref{eq:toy_rshock_sol}) vanishes at $\lcrit$, which allows us to plug in the expression for $\lcrit$ from equation~(\ref{eq:toy_lcrit_small_mdot}). Under the same assumptions that were used to derive equation~(\ref{eq:toy_lcrit_small_mdot}), we can calculate that the critical shock radius $\rs^{\rm crit}$ does not depend on the particular $\lcrit$, but it is proportional to $\eta^{-1/2}$.

\clearpage

\end{document}